\documentclass[useAMS,usenatbib]{mn2e}

\usepackage{epsfig,color}
\usepackage{natbib}
\bibpunct{(}{)}{;}{a}{}{,}
\def\spose#1{\hbox to 0pt{#1\hss}}
\def\ltsimm{\mathrel{\spose{\lower 3pt\hbox{$\sim$}}
        \raise 2.0pt\hbox{$<$}}}
\def\gtsimm{\mathrel{\spose{\lower 3pt\hbox{$\sim$}}
        \raise 2.0pt\hbox{$>$}}}
\def\Mdot{\hbox{${\dot M}$}}

\def\km{{\rm\thinspace km}}
\def\cm{{\rm\thinspace cm}}

\def\s{{\rm\thinspace s}}
\def\yr{{\rm\thinspace yr}}
\def\g{{\rm\thinspace g}}
\def\kmps{\hbox{${\rm\km\s^{-1}\,}$}}

\def\erg{{\rm\thinspace erg}}

\def\Hz{{\rm\thinspace Hz}}

\def\ster{{\rm\thinspace ster}}
\def\ergps{\hbox{${\rm\erg\s^{-1}\,}$}}
\def\Rsol{\hbox{${\rm\thinspace R_{\odot}}$}}
\def\Msol{\hbox{${\rm\thinspace M_{\odot}}$}}

\def\Msolpyr{\hbox{${\rm\Msol\yr^{-1}\,}$}}
\def\pcm{\hbox{${\rm\cm^{-1}\,}$}}
\def\pcm2{\hbox{${\rm\cm^{-2}\,}$}}
\def\pcm3{\hbox{${\rm\cm^{-3}\,}$}}
\def\ergpscm3Hz{\hbox{${\rm\ergps\cm^{-3}\Hz^{-1}\,}$}}
\def\ergpscm3Hzster{\hbox{${\rm\ergps\cm^{-3}\Hz^{-1}\ster^{-1}\,}$}}
\def\gpcm3{\hbox{${\rm\g\cm^{-3}\,}$}}
\def\ergpcm2{\hbox{${\rm\erg\cm^{-2}\,}$}}
\def\ergpcm3{\hbox{${\rm\erg\cm^{-3}\,}$}}
\def\phpscm2{\hbox{${\rm photons\s^{-1}\cm^{-2}\,}$}}

\title[Radio emission models of colliding wind binaries]{3D models of radiatively driven colliding winds in massive O+O star binaries - II. Thermal radio to sub-mm emission} 
\author[J.~M.~Pittard] {J. M. Pittard\thanks{E-mail:
jmp@ast.leeds.ac.uk}\\School of Physics and Astronomy, The
University of Leeds, Leeds LS2 9JT, UK\\ }

\begin{document}

\date{Accepted 6th Aug 2008 Received 4th Aug 2008; in original form 4th July 2008}

\pagerange{\pageref{firstpage}--\pageref{lastpage}} \pubyear{2009}

\maketitle

\label{firstpage}

\begin{abstract}
In this work the {\em thermal} emission over cm to sub-mm wavelengths
from the winds in short-period O+O-star binaries is investigated
(potential {\em non-thermal} emission is presently ignored). The
calculations are based on three-dimensional hydrodynamical models
which incorporate gravity, the driving of the winds, orbital motion of
the stars, and radiative cooling of the shocked plasma. The {\em
thermal} emission arises from the stellar winds and the region where
they collide. We investigate the flux and spectrum from a variety of
models as a function of orbital phase and orientation of the observer,
and compare to the single star case. The emission from the wind-wind
collision region (WCR) is strongly dependent on its density and
temperature, being optically thick in radiative systems, and optically
thin in adiabatic systems.  The flux from systems where the WCR is
highly radiative, as investigated for the first time in this work, can
be over an order of magnitude greater than the combined flux from
identically typed single stars. This excess occurs over a broad range
of wavelengths from cm to sub-mm. In contrast, when the WCR is largely
adiabatic, a significant excess in the thermal flux occurs only below
100\,GHz.

In circular systems with (near) identical stars the observed
variability in synthetic lightcurves is typically less than a factor
of 2. Eccentric systems may show order of magnitude or greater flux
variability, especially if the plasma in the WCR transitions from an
adiabatic to a radiative state and vice-versa - in such cases the flux
can display significant hysteresis with stellar separation.
We further demonstrate that clumping can affect the variability of radio
lightcurves.

We investigate the spectral index of the emission, and often find
indices steeper than $+0.6$. Synthetic images display a variety of
morphologies, with the emission sometimes resembling an intertwined
``double-helix''.  We conclude by comparing our results to observations. 
The predictions made in this paper are of interest to future
observations with the next generation of radio and sub-mm telescopes,
including the EVLA, e-MERLIN, ALMA, and the SKA, and future upgrades
to the VLBA.
\end{abstract}


\section{Introduction}
\label{sec:intro}
Luminous stars of OB and Wolf-Rayet (WR) spectral types drive fast,
dense, ionized winds with mass-loss rates that are high enough to
significantly affect their evolution and their surroundings.  The
winds emit free-free radiation over a wide range of wavelengths, from
cm emission in the radio band, to $\mu$m emission in the
near-infra-red. Since the bremsstrahlung opacity is proportional to
the square of the wavelength, the observed emission originates above a
characteristic radius that increases with wavelength (typical values
are $\sim 50-100\,R_{*}$ at 20\,cm (1.5\,GHz), and $\sim 2-5\,R_{*}$
at 1\,mm (300\,GHz)). At wavelengths where the wind is optically
thick, the thermal emission can be used to determine stellar mass-loss
rates. This has mostly been done using radio data
\citep*[e.g.][]{Abbott:1980,Abbott:1981,Bieging:1989,Leitherer:1995,Contreras:1996,Scuderi:1998,Benaglia:2001,Schnerr:2007},
since the wind can be assumed to be at terminal velocity, and the
interpretation of the flux is not strongly dependent on details of the
ionization conditions \citep{Wright:1975,Panagia:1975}. Wind
asphericity typically does not alter radio derived mass-loss rates by
more than a factor of two \citep{Schmid-Burgk:1982}.

The free-free flux is expected to behave like $S_{\nu} \propto
\nu^{\alpha}$, where the spectral index $\alpha \approx +0.6$ at cm
wavelengths \citep{Wright:1975}. However, the spectral index of
``thermal'' sources\footnote{We repeat that in this work we
concentrate solely on the {\em thermal} emission from the winds of
early-type stars, though observations reveal that there is also a
significant population of {\em non-thermal} emitters which appears to
be linked to the presence of a companion star
\citep{Dougherty:2000,DeBecker:2007} and the resulting collision of
the winds \citep[see][and references therein, for models of the
non-thermal emission]{Pittard:2006}. The non-thermal emission from the
hydrodynamical models presented in \citet{Pittard:2009a}, which are
the basis for the current work, will be investigated in future
papers.} at shorter (mm and far-infrared) wavelengths increases to
around $+0.7$ to $+0.9$
\citep*[e.g.][]{Williams:1990,Leitherer:1991,Altenhoff:1994,Nugis:1998}.
Several possible causes for this steepening have been suggested,
including a decrease of the wind temperature or the ionization state
with radius or an increase of the wind speed with radius
\citep{Schmid-Burgk:1982,Leitherer:1991}.  However, since temperature
gradients only weakly affect the flux through the gaunt factor, while
the wind speed changes relatively little between the cm and mm
emitting regions, these effects are unable to explain the difference
between theoretical and observed spectral indices between cm and mm
wavelengths. \citet{Schmid-Burgk:1982} also noted that deviations from
the expected behaviour ``should not easily be blamed'' on
non-spherical winds. This led both \citet{Williams:1990} and
\citet{Leitherer:1991} to suggest a gradient in ionization as the most
likely cause.

Another possibility involves a radial variation of the degree of
clumping within the wind. For instance, an excess of mm emission over
that expected from scaling the radio emission to shorter wavelengths
can be caused by a decline in the degree of clumping between the
regions of the wind emitting at mm and cm wavelengths
\citep[][]{Runacres:1996,Blomme:2002}. \citet{Nugis:1998} have also
argued for models which combine clumping and a highly ionized outer
wind region.  A recent review of small-scale structure in massive star
winds can be found in \citet*{Puls:2008}. 

At very high frequencies, the photospheric emission, which is rapidly rising, 
begins to dominate the emission from the wind. This transition occurs at
roughly the same frequency as the whole wind becomes optically thin.
The intrinsic thermal spectrum from the wind only then flattens, as the 
emission then arises from essentially a blackbody of constant projected linear
radius $R_{\rm c}$ at all frequencies greater than the critical value
given by Eq.~12 of \citet{Wright:1975}. However, the dominance of the
photospheric emission means that observationally the emission retains
a strong wavelength dependence.

The presence of a companion star may also change the observed thermal
flux, its spectral index, and the inferred mass-loss rates. First,
there are now two stars, rather than one, and the winds of both will
contribute to the observed emission.  Moreover, the interaction of the
winds may also produce significant free-free
emission. \citet{Stevens:1995} investigated the total thermal flux
from a wide colliding winds binary (CWB) as a function of the wind
momentum ratio, and discovered that it could be about 50 per cent
higher than that expected from the more massive wind alone.  In more
recent work, \citet{Pittard:2006} investigated the emission from an
adiabatic WCR, where the plasma in the WCR remains at high temperature
as it flows out of the system, as a function of the stellar
separation, $d_{\rm sep}$. If the WCR remains optically thin, the
thermal flux scales as $d_{\rm sep}^{-1}$. Moreover, since optically
thin emission scales as $S_{\nu} \propto \nu^{-0.1}$, the flux from
the WCR may dominate the thermal flux from the system, particularly at
cm wavelengths\footnote{Since this emission has a negative spectral
index it can mimic a synchrotron emission component in the system -
see \citet{Pittard:2006} for further details.}.  However, there has
been no investigation to date of the thermal emission from systems
where the WCR is highly radiative. This situation clearly needs to be
addressed.

In this paper we investigate the thermal radio to sub-mm emission from
short period O+O-star binaries computed from the three-dimensional
hydrodynamical models described in \citet[hereafter
Paper~I]{Pittard:2009a}. We aim to illuminate the relative
contribution of the WCR to the total free-free flux, the effect that
this may have on the observed spectral index, and the size and nature
of orbital related variations, for a variety of models where the
behaviour of the plasma in the WCR is either highly radiative or
largely adiabatic.

The layout of this paper is as follows. We briefly describe in
Section~\ref{sec:hydro} the hydrodynamical models and in
Section~\ref{sec:setup} our method of calculating the thermal cm to
sub-mm emission.  In Section~\ref{sec:results} we present our results,
focussing first on calculations of the emission from the wind of a
single star, and then on calculations based on our binary models.  In
Section~\ref{sec:discussion} we compare our findings against
observations, and in Section~\ref{sec:summary} we summarize and
conclude this work.

\begin{table*}
\begin{center}
\caption[]{Assumed binary parameters for the models calculated in
Paper~I. The semi-major axis is $34.26\;\Rsol$ in model cwb1,
$76.3\;\Rsol$ in models cwb2 and cwb3, and $55\;\Rsol$ in model cwb4.
$e$ is the orbital eccentricity and $\eta$ is the (terminal velocity)
momentum ratio of the winds. $v_{\rm orb}$ and $v_{\rm w}$ are the
orbital speeds of the stars and the preshock wind speeds along the
line of centres. $\chi$ is the ratio of the cooling time to the
characteristic flow time of the hot shocked plasma. $\chi \ltsimm 1$
indicates that the shocked gas rapidly cools, while $\chi \gtsimm 1$
indicates that the plasma in the WCR remains hot as it flows out of
the system. Larger values of the ratio $v_{\rm orb}/v_{\rm w}$ produce
a greater aberration angle, $\theta_{\rm ab}$, and tighter downstream
curvature, of the WCR.  The degree of downstream curvature of the WCR
in the orbital plane is given by $\alpha_{\rm coriolis}$, where the
curvature is assumed to trace an Archimedean spiral which in polar
coordinates is described by $r = \alpha_{\rm coriolis}\theta$. The
value of $\alpha_{\rm coriolis}$ corresponds to the approximate downstream
distance (in units of $d_{\rm sep}$) along the WCR for each radian of
arc it sweeps out in the orbital plane. Smaller values indicate
tighter curvature.  The leading and trailing arms of the WCR in model
cwb3 display differing degrees of curvature, so the value quoted for
this model is an average.  The pre-shock orbital and wind speeds in
model cwb3 are also different for each star/wind - the first (second)
value is for the primary (secondary) star/wind. The values of $\chi$,
$v_{\rm orb}/v_{\rm w}$, $\theta_{\rm ab}$ and $\alpha_{\rm coriolis}$
are phase dependent in model cwb4, because of its eccentric orbit -
values at periastron and apastron are quoted. The values for
$\alpha_{\rm coriolis}$ are calculated after comparing the orbital
speeds at periastron and apastron against those in models cwb1 and
cwb2, and represent the ``instantaneous'' curvature at these phases.}
\label{tab:models}
\begin{tabular}{lllllllllll}
\hline
\hline
Model & Stars & Period & $e$ & $\eta$ & $v_{\rm orb}$ & $v_{\rm w}$ & $\chi$ & $v_{\rm orb}/v_{\rm w}$ & $\theta_{\rm ab}$ & $\alpha_{\rm coriolis}$ \\
 & & (d) &  & & ($\kmps$) & ($\kmps$) & & &($^{\circ}$) & ($d_{\rm sep}\,{\rm rad}^{-1}$) \\
\hline
cwb1 & O6V+O6V & 3 & 0.0 & 1 & 290 & 730 & 0.34 & 0.40 & 17 & 3.5\\
cwb2 & O6V+O6V & 10 & 0.0 & 1 & 225 & 1630 & 19 & 0.14 & $3-4$ & 6.5\\
cwb3 & O6V+O8V & 10.74 & 0.0 & 0.4 & 152,208 & 1800,1270 & 28,14 & $0.16-0.084$ & $\sim 2$ & 4.5\\
cwb4 & O6V+O6V & 6.1 & 0.$\overline{36}$ & 1 & $334-156$ & $710-1665$ & $0.34-19$ & $0.47-0.09$ & 21-4 & 3-10 \\
\hline
\end{tabular}
\end{center}
\end{table*}

\section{Summary of the hydrodynamical models}
\label{sec:hydro}
The radio calculations in this paper are based on the
three-dimensional hydrodynamical models described in Paper~I. The
models incorporate the radiative driving of the stellar winds,
gravity, orbital effects, and cooling, and are summarized in
Tables~\ref{tab:models} and~\ref{tab:stellar_params}. The models were
not designed to simulate particular systems, but were aimed at
mimicking much of the interesting phenomena that can occur.

Model cwb1 is of an O6V+O6V system with a circular orbit and a period
of 3 days. The WCR is highly radiative, and significantly distorted by
orbital effects, showing strong aberration and downstream curvature.
Model cwb1 is similar to DH\,Cep \citep[see][and references
therein]{Linder:2007}, HD\,165052 \citep{Arias:2002,Linder:2007}, and
HD\,159176 \citep{DeBecker:2004b,Linder:2007}, all of which have near
identical main-sequence stars of spectral type O6$-$O7, and circular
or near-circular orbits with periods near 3 days. Hence the
hydrodynamics of, and emission from, the WCR in model cwb1 should be a
reasonable approximation to these systems.

Model cwb2 has identical parameters to model cwb1 except for an
increase in the orbital period to 10 days. The winds now have more
room to accelerate before they collide, and the postshock gas remains
largely adiabatic as it flows out of the system. Both the aberration
angle and the downstream curvature of the WCR are lessened relative to
model cwb1. Model cwb2 is similar to HD\,93161A, an O8V + O9V system
with a circular orbit and an orbital period of 8.566 days
\citep{Naze:2005}, albeit with slightly more massive stars and
powerful winds. Another system which is not too dissimilar is Plaskett's star
\citep[HD\,47129,][]{Linder:2006,Linder:2008}, though this object
contains stars which have evolved off the main sequence.

Model cwb3 examines the interaction of unequal winds in a hypothetical
O6V + O8V binary. The stars have the same separation as those in model
cwb2, but a slightly longer orbital period. The stronger wind from the
primary star collides at higher speeds than the slower secondary wind,
resulting in postshock plasma which is hotter and more adiabatic.

Model cwb4 investigates the effect of an eccentric orbit, which takes
the stars through a separation of $34.26-76.3\;\Rsol$ (i.e.  the
separations of the stars in the circular orbits of models cwb1 and
cwb2). The WCR is radiative at periastron and adiabatic at apastron,
and its aberration and downstream curvature are phase dependent. A
surprising finding from Paper~I is that dense cold clumps formed in
the WCR at periastron still persist near the apex of the WCR at
apastron. This is because the clumps have relatively high inertia, and
require a significant fraction of the orbtial period to be accelerated
out of the system. Some well-known O+O binaries with eccentric orbits
include (in order of increasing orbital period) HD\,152248 \citep[$e =
0.127$;][]{Sana:2004}, HD\,93205 \citep[$e=0.46$;][]{Morrell:2001},
HD\,93403 \citep[$e=0.234$;][]{Rauw:2002}, Cyg\,OB2\#8A
\citep[$e=0.24$;][]{DeBecker:2004,DeBecker:2006} and $\iota$\,Orionis
\citep[$e=0.764$;][]{Bagnuolo:2001}.

\begin{table}
\begin{center}
\caption[]{Assumed stellar parameters for the models.}
\label{tab:stellar_params}
\begin{tabular}{lllll}
\hline
\hline
Parameter/Star & O6V & O8V \\
\hline
Mass ($\Msol$) & 30 & 22 \\
Radius ($\Rsol$) & 10 & 8.5 \\
Effective temperature (K) & 38000 & 34000 \\ 
Mass-loss rate ($\Msolpyr$) & $2 \times 10^{-7}$ & $10^{-7}$ \\
Terminal wind speed ($\kmps$) & 2500 & 2000 \\
\hline
\end{tabular}
\end{center}
\end{table}

\begin{table}
\begin{center}
\caption[]{The characteristic radius of emission, $R_{\rm c}$, and
thermal flux, $S_{\nu}$, of the wind of the O6V star as a function of
frequency \citep[see][]{Wright:1975}, assuming that H and He are both
singly ionized, that the wind temperature is 10000\,K, that the wind
is smooth (i.e. no clumping), and that the star is at a distance of
1\,kpc.  The relatively low values of $R_{\rm c}$ and $S_{\nu}$
reflect the adopted mass-loss rates of the stars (which are
nevertheless consistent with recent downward revisions of the {\em
actual} mass-loss rates of massive stars), coupled with our assumption
of smooth winds. Assuming $v(r) = v_{\infty}$ reduces $R_{\rm c}$ and
$S_{\nu}$ below their actual values, particularly at the highest
frequencies which probe the innermost parts of the acceleration region
(in fact, the formal values of $R_{\rm c}$ at 1 and 2\,THz obtained
from Eq.~11 of \citet{Wright:1975} are $< R_{*}$).  For the O8V star,
$R_{\rm c}$ and $S_{\nu}$ are smaller by factors of 0.731 and 0.534,
respectively.}
\label{tab:radio_params}
\begin{tabular}{llll}
\hline
\hline
Frequency & Wavelength & $R_{\rm c}$ & $S_{\nu}$ \\
(GHz) & ($\mu$m) & ($\Rsol$) & (mJy) \\
\hline
5 & 60000 & 386 & 0.0049\\
8.4 & 35714 & 273 & 0.0067\\
15 & 20000 & 185 & 0.0094 \\
22.5 & 13333 & 141 & 0.012 \\
43.3 & 6928 & 92 & 0.017 \\
100 & 3000 & 44.1 & 0.028 \\
150 & 2000 & 32.9 & 0.035 \\
250 & 1200 & 22.7 & 0.046 \\
500 & 600 & 13.6 & 0.067 \\
1000 & 300 & 8.13 & 0.095 \\
2000 & 150 & 4.80 & 0.13 \\
\hline
\end{tabular}
\end{center}
\end{table}

\section{Modelling the radio emission from CWBs}
\label{sec:setup}
We use the model developed by \citet{Dougherty:2003} and
\citet{Pittard:2006} to determine the thermal radio to sub-mm emission
from each of our colliding winds simulations. Detailed notes on our
method can be found in these papers, but the salient points are
summarized below.

To calculate the radio emission we read our hydrodynamical models into
a radiative transfer ray-tracing code, and calculate appropriate
emission and absorption coefficients for each cell. A synthetic image
on the plane of the sky is then generated by solving the radiative
transfer equation along suitable lines of sight through the grid.  
The thermal emission ($\epsilon_{\nu}^{\rm ff}$) and absorption 
($\alpha_{\nu}^{\rm ff}$) coefficients at frequency $\nu$ are given by
\citet{Rybicki:1979} as
\begin{eqnarray}
\varepsilon_{\nu}^{\rm ff}
&=&6.8\times10^{-38}Z^{2}n_{\rm e}n_{\rm i}T^{-1/2}e^{-h\nu/kT}g_{\rm ff},\label{eq:ffem}\\
\alpha_{\nu}^{\rm ff}&=&3.7\times10^{8}T^{-1/2}Z^{2}n_{\rm e}n_{\rm i}\nu^{-3}\label{eq:ffabs}
(1-e^{-h\nu/kT})g_{\rm ff},
\end{eqnarray}
\noindent where $Z$ is the mean ionic charge, $n_{\rm e}$ and $n_{\rm
i}$ are electron and ion number densities, $T$ is the temperature of
the gas, and $g_{\rm ff}$ is a velocity averaged Gaunt factor 
\citep{Hummer:1988}. All quantities are in cgs units. The
densities $n_{\rm e}$ and $n_{\rm i}$ are determined from the cell
density, composition and ionization, where the ionization is specified
as a function of cell temperature.  The appropriate ionization for a
particular temperature regime is determined and the absorption and
emission coefficients are then evaluated. The unshocked winds are
assumed to be isothermal at $T=10000$\,K, and each model is assumed
to be at a distance of 1\,kpc.

In the majority of the models in this work we assume that the winds
are smooth (i.e. not clumped). This enables the characteristic radius
of emission and the $\tau=1$ surface to better fit within our
hydrodynamical grids, at the expense of underestimating the emission
compared to real systems where the winds are often strongly
clumped. Our assumption of smooth winds also removes the need to
specify the radial stratification of the clump volume filling factor,
$f$, which remains poorly known \citep[see e.g.][]{Puls:2008}.
Table~\ref{tab:radio_params} lists the characteristic radius, $R_{\rm
c}$, and flux density, $S_{\nu}$, of free-free emission from our model
O6V star at various frequencies between 5 and 2000\,GHz (corresponding
to wavelengths between 6\,cm and 150\,$\mu$m).  The radius of optical
depth unity, $R_{\tau=1} = 0.623\;R_{\rm c}$.

All of the hydrodynamical grids in our models are large enough to
contain the characteristic radius for free-free emission at $43\;$GHz,
while models cwb2 and cwb3 are also able to contain the characteristic
radius for emission at $15\;$GHz (compare Table~\ref{tab:radio_params}
with Table~3 in Paper~I). This enables us to explore the thermal radio
emission from these systems, in addition to the X-ray emission
explored in a previous paper \citep[][hereafter
Paper~III]{Pittard:2009b}.

Clumping increases the emission and absorption coefficients in
Eqs.~\ref{eq:ffem} and~\ref{eq:ffabs} by a factor of $1/f$, where $f$
is the volume filling factor of the clumps, and increases the
characteristic radius by a factor $f^{-1/3}$, and the flux density by
a factor $f^{-2/3}$. This means that the flux from a model with a
given $\Mdot$ and $f$, if interpreted as arising from a wind with no
clumping, gives (incorrectly) the mass-loss rate $\Mdot/\sqrt{f}$
(note that one can also take an alternative approach: a particular
observed flux can be translated into a mass-loss rate $\Mdot_{\rm
smooth}$ assuming that there is no clumping, or a mass-loss rate
$\Mdot_{\rm clumped} = \Mdot_{\rm smooth} \sqrt{f}$, assuming that
there is clumping).  However, even when the stellar winds are clumpy,
the WCR may be considerably less so. This is expected to be the case
when the postshock gas behaves largely adiabatically. In such
situations the destruction time of the clumps can be significantly
shorter than the flow time of the shocked gas out of the system, and
the WCR, though highly turbulent, is noticeably smoother than the
stellar winds \citep[see][]{Pittard:2007}. Therefore, in previous work
\citep{Dougherty:2003,Pittard:2006,Pittard:2006b} we have assumed that
the winds are clumpy and the WCR is smooth(er).

In this work we also investigate the emission assuming clumpy winds
with $f=0.1$, but a WCR which is smoother to varying degrees i.e.
$0.1 \leq f_{\rm WCR} \leq 1$. This provides more realistic fluxes
from the unshocked winds and insight into how clumping in different
regions of the system affects the observed lightcurves and spectra. In
these models the coefficients in Eqs.~\ref{eq:ffem} and~\ref{eq:ffabs}
are multiplied by a factor of $1/f_{\rm winds}$ ($1/f_{\rm WCR}$) for
cells in the unshocked winds (WCR), respectively.

\begin{figure*}
\psfig{figure=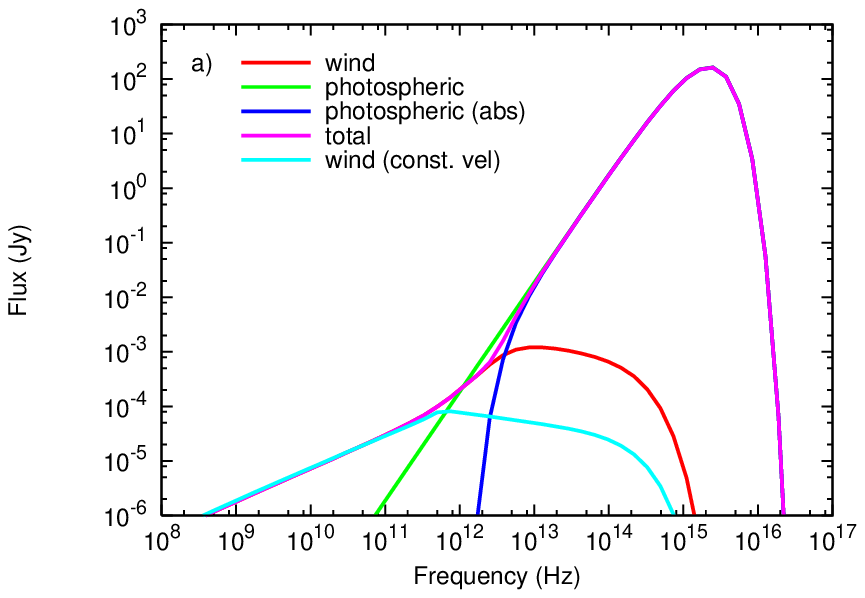,width=5.67cm}
\psfig{figure=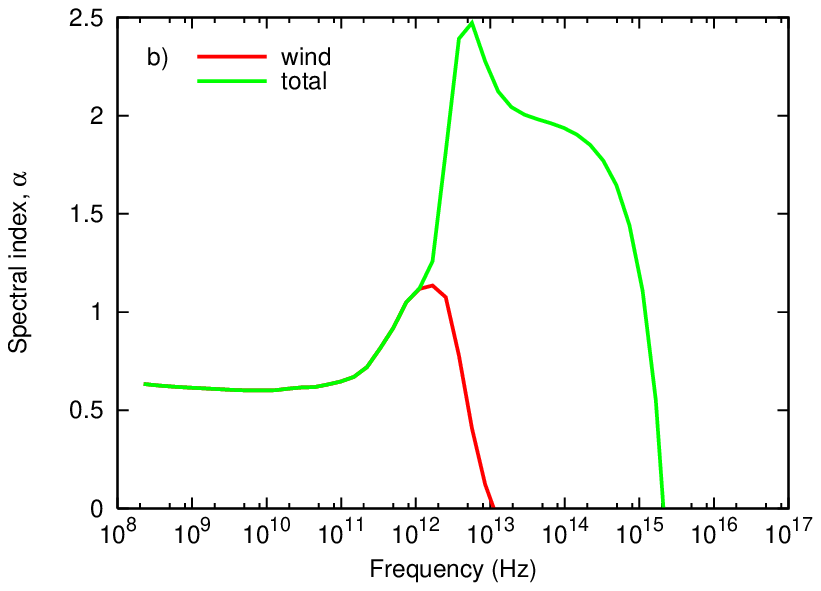,width=5.67cm}
\psfig{figure=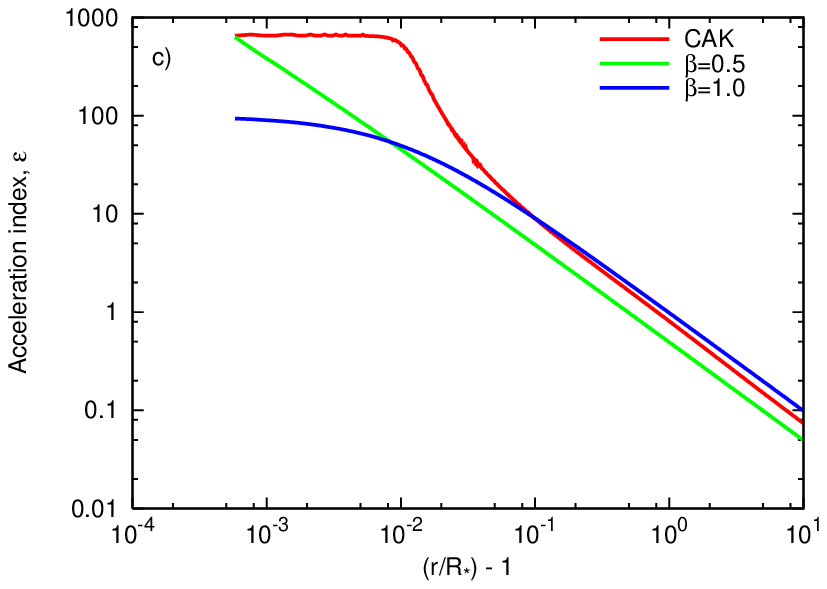,width=5.67cm}
\caption[]{(a) The spectrum from the wind and photosphere of a simple
model of a single O6V star. The photosphere is assumed to emit as a
blackbody at a temperature of $38000$\,K, while the wind is assumed to
be at a constant temperature of $10000$\,K and to have a constant
ionization structure of H$^+$, 50 per cent neutral He and 50 per cent 
He$^+$, and CNO$^+$. The photospheric emission is strongly
absorbed by the wind for $\nu \ltsimm 2 \times 10^{12}$\,Hz.  (b) The spectral
index of the total and wind component of the spectrum.
(c) The variation of the wind acceleration index, $\epsilon$, where $v(r)
\propto r^{\epsilon}$.}
\label{fig:singlestar_spec_tau}
\end{figure*}

\begin{figure*}
\psfig{figure=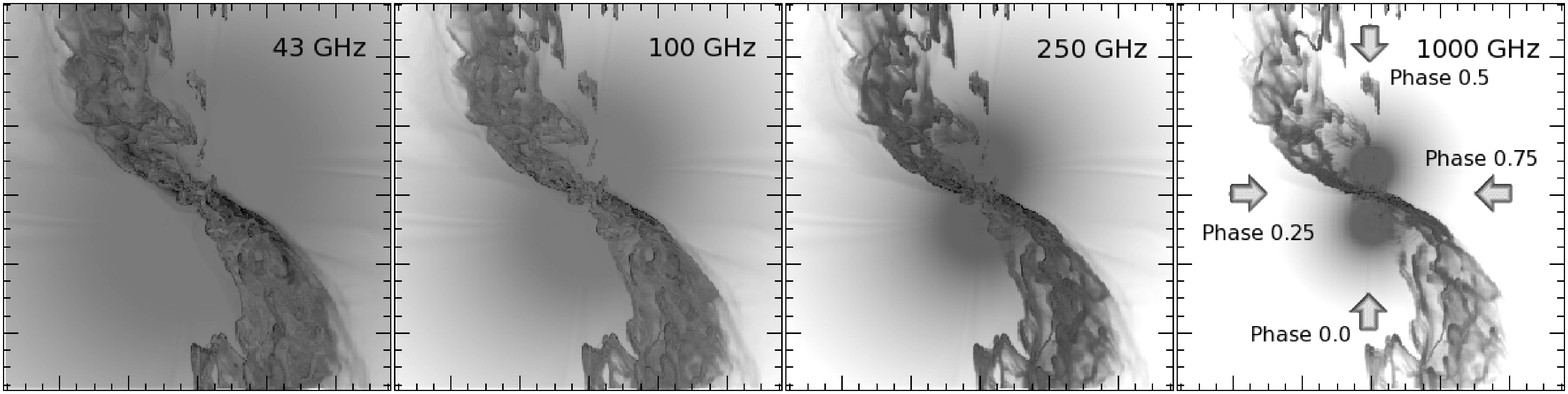,width=13.6cm}
\caption[]{Intensity images from model cwb1 at $i=0^{\circ}$.  
The maximum intensity (black in the images) is
$4.6\times10^{-10}\,{\rm erg\,cm^{-2}\,s^{-1}\,Hz^{-1}\,ster^{-1}}$ at
43\,GHz, $2.7\times10^{-9}\,{\rm erg\,cm^{-2}\,s^{-1}\,Hz^{-1}\,ster^{-1}}$ 
at 100\,GHz, $7.1\times10^{-9}\,{\rm
erg\,cm^{-2}\,s^{-1}\,Hz^{-1}\,ster^{-1}}$ at 250\,GHz, and
$1.1\times10^{-7}\,{\rm erg\,cm^{-2}\,s^{-1}\,Hz^{-1}\,ster^{-1}}$ at
1000\,GHz. The gray scale spans 4 orders of magnitude between maximum
and minimum. The major ticks on each axis mark out 0.2\,mas. Orbital
phases are marked on for an observer in the orbital plane, with
phases 0.0 and 0.5 corresponding to conjunction and phases 0.25 and 0.75
to quadrature. The stars and the WCR rotate anticlockwise. 
We also define an azimuthal viewing angle, $\phi$, which increases 
anticlockwise from the bottom of the image, so that at phase 0.0
$\phi=0^{\circ}$. $\phi=90^{\circ}$ 
corresponds to phase 0.75, and $\phi=180^{\circ}$ and $\phi=270^{\circ}$ 
correspond to phases 0.5 and 0.25, respectively.}
\label{fig:cwb1_radio_image}
\end{figure*}

\begin{figure*}
\psfig{figure=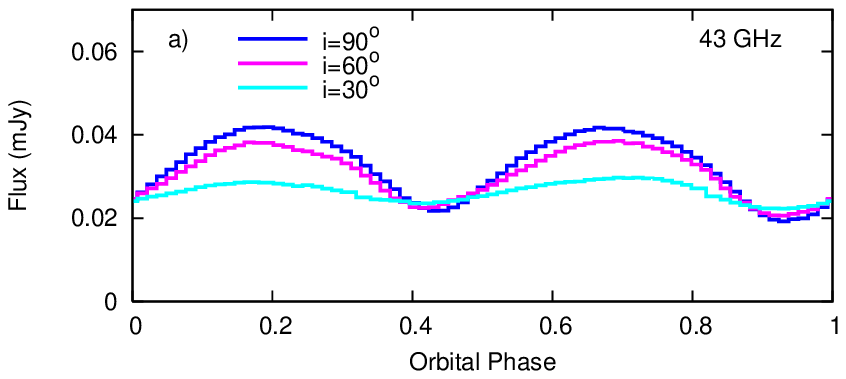,width=5.67cm}
\psfig{figure=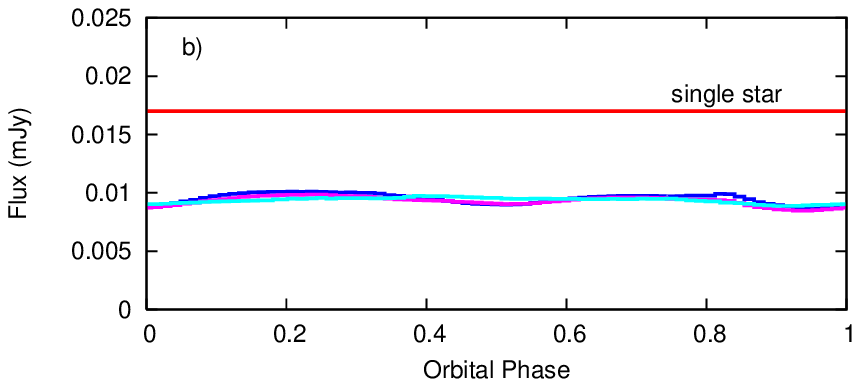,width=5.67cm}
\psfig{figure=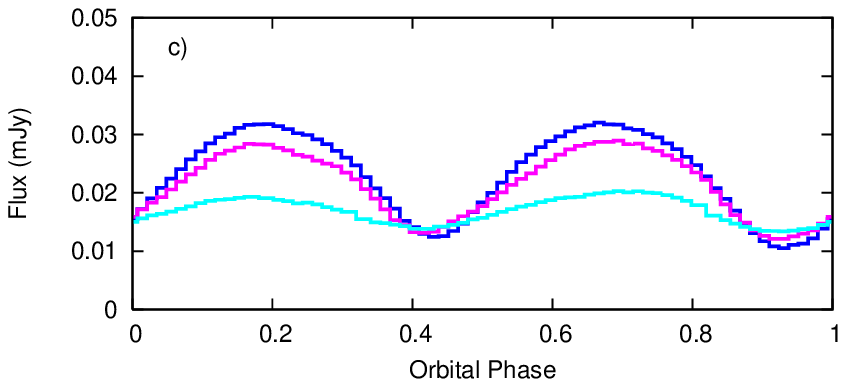,width=5.67cm}
\psfig{figure=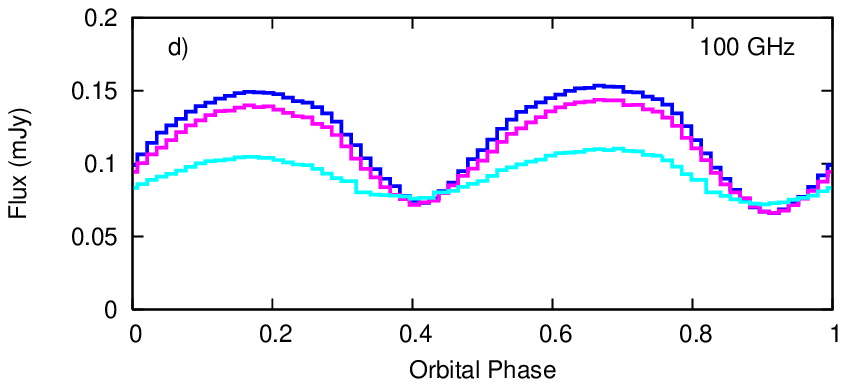,width=5.67cm}
\psfig{figure=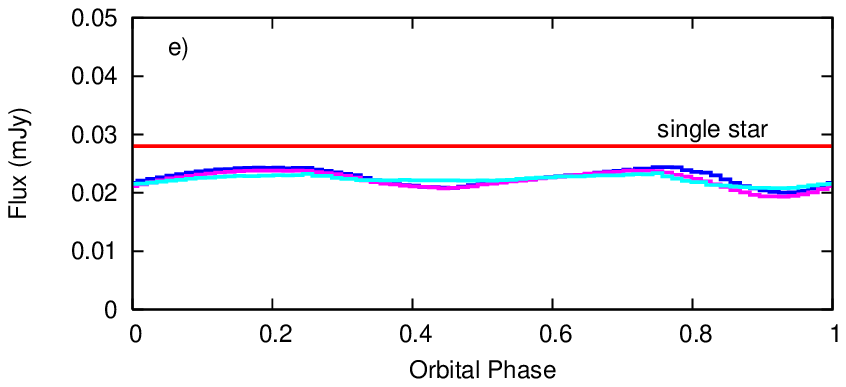,width=5.67cm}
\psfig{figure=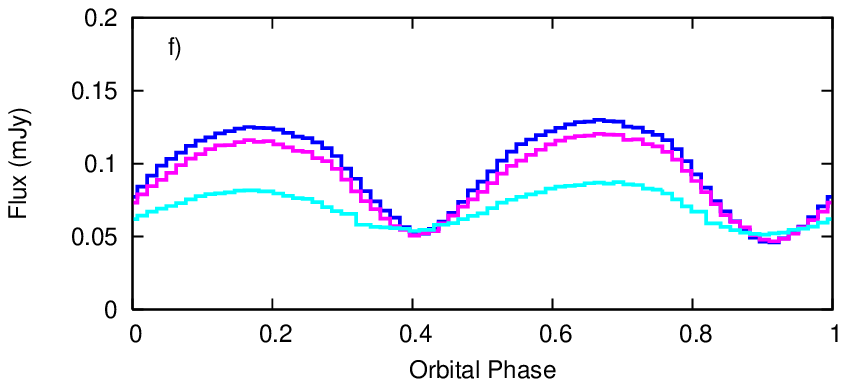,width=5.67cm}
\psfig{figure=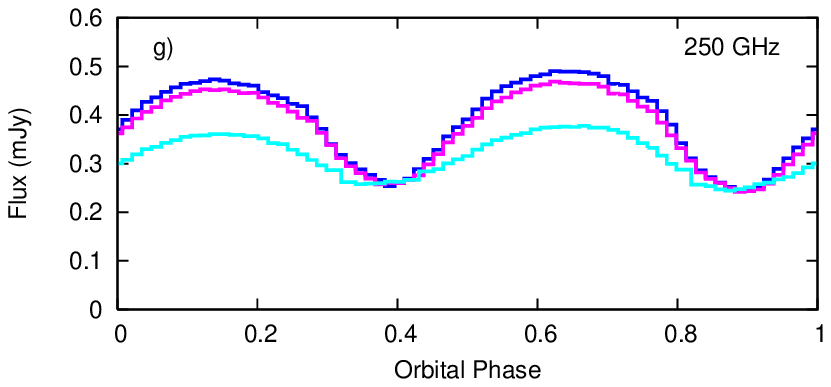,width=5.67cm}
\psfig{figure=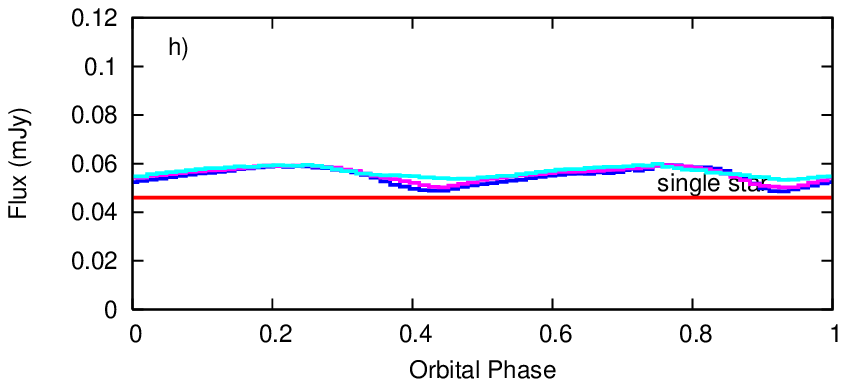,width=5.67cm}
\psfig{figure=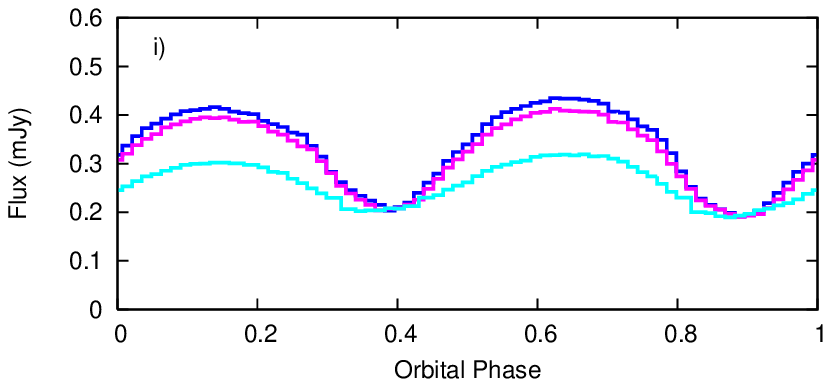,width=5.67cm}
\psfig{figure=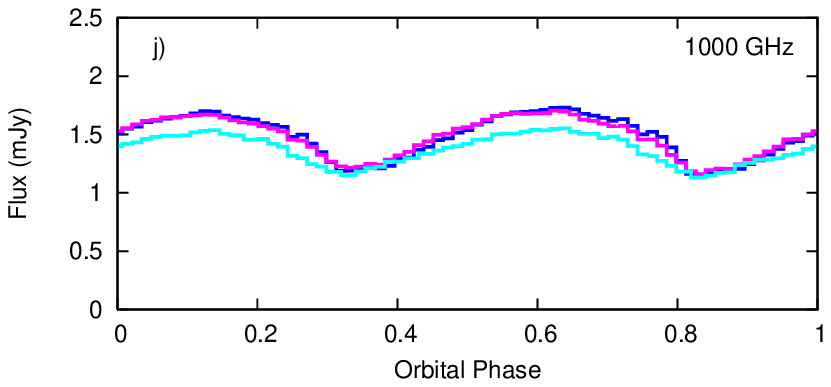,width=5.67cm}
\psfig{figure=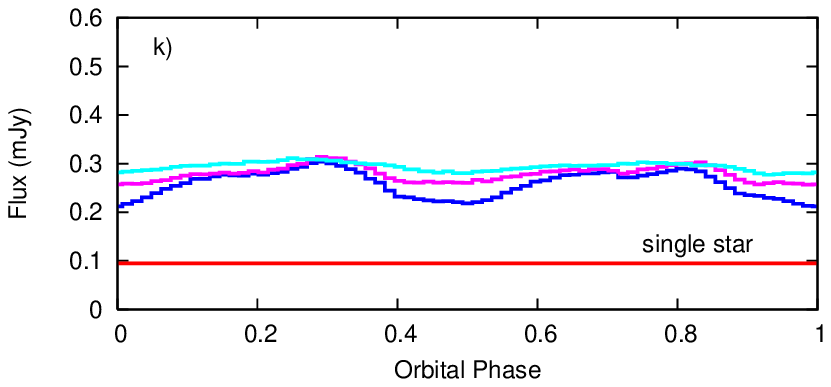,width=5.67cm}
\psfig{figure=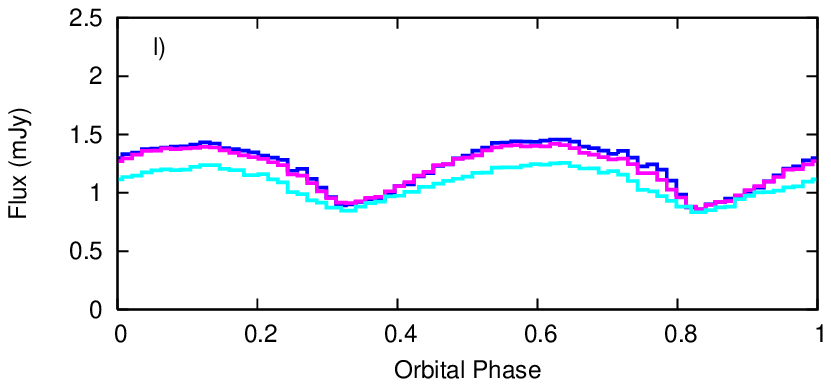,width=5.67cm}
\caption[]{Radio lightcurves of the free-free emission from model cwb1
at 43\,GHz (top), 100\,GHz, 250\,GHz, and 1000\,GHz (bottom) 
for inclination angles $i = 30^{\circ}$, $60^{\circ}$, and $90^{\circ}$.
The total free-free emission is shown in the left panels, and the
contributions from the winds and WCR in the middle and right panels,
respectively. In all cases, the observer is located along a direction vector
specified by $\phi=0^{\circ}$. The stars are at conjunction at phases 0.0 and
0.5, and quadrature at phases 0.25 and 0.75. Note that the y-axes (flux scales)
are different in each panel.}
\label{fig:cwb1_ffradio_lc}
\end{figure*}

\begin{figure*}
\psfig{figure=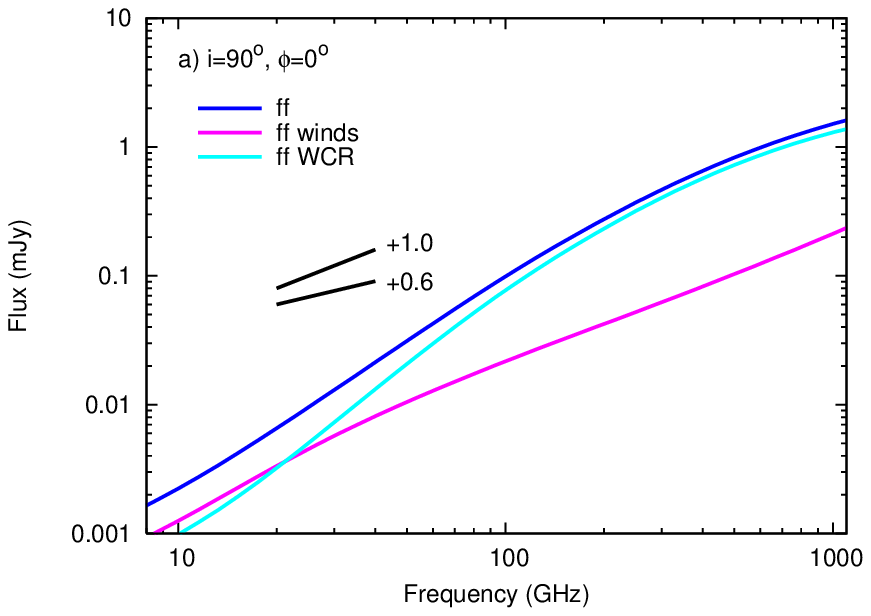,width=5.67cm}
\psfig{figure=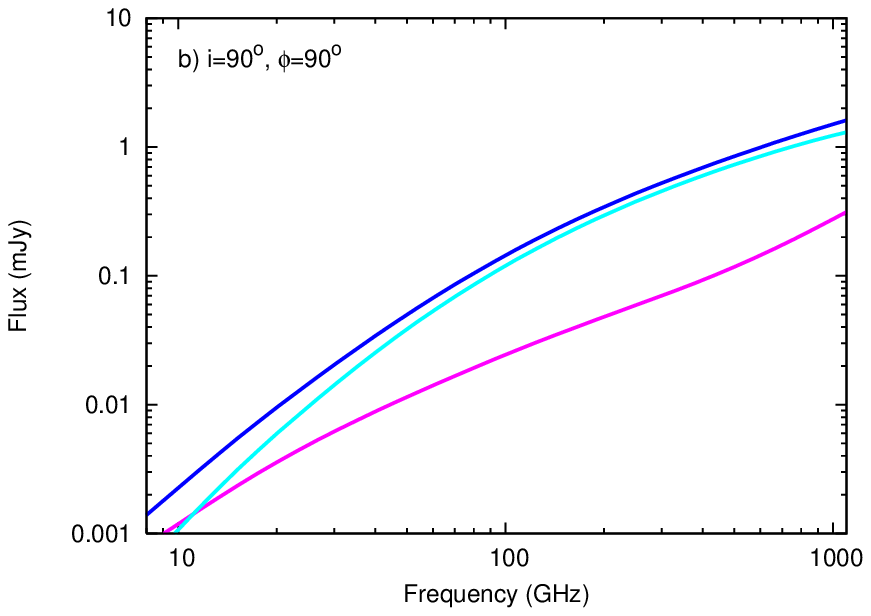,width=5.67cm}
\psfig{figure=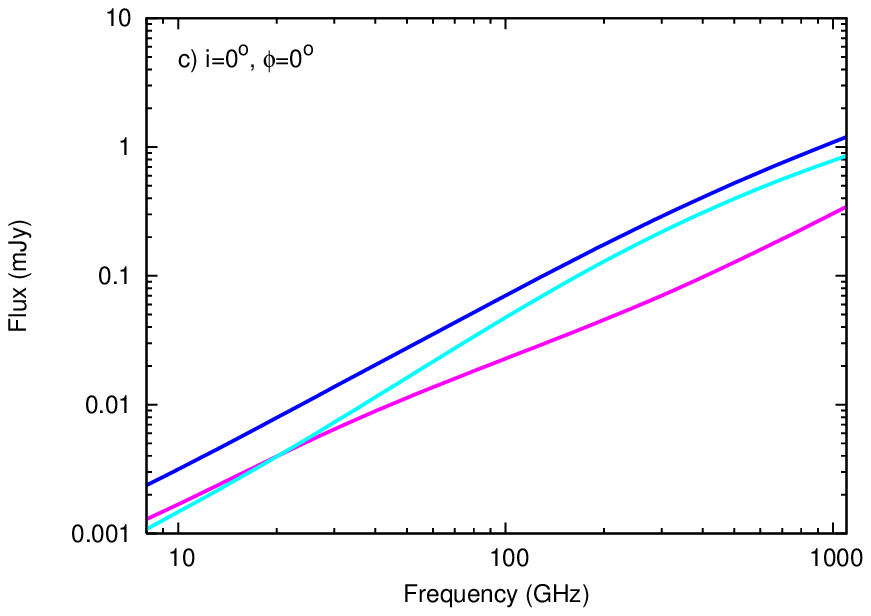,width=5.67cm}
\psfig{figure=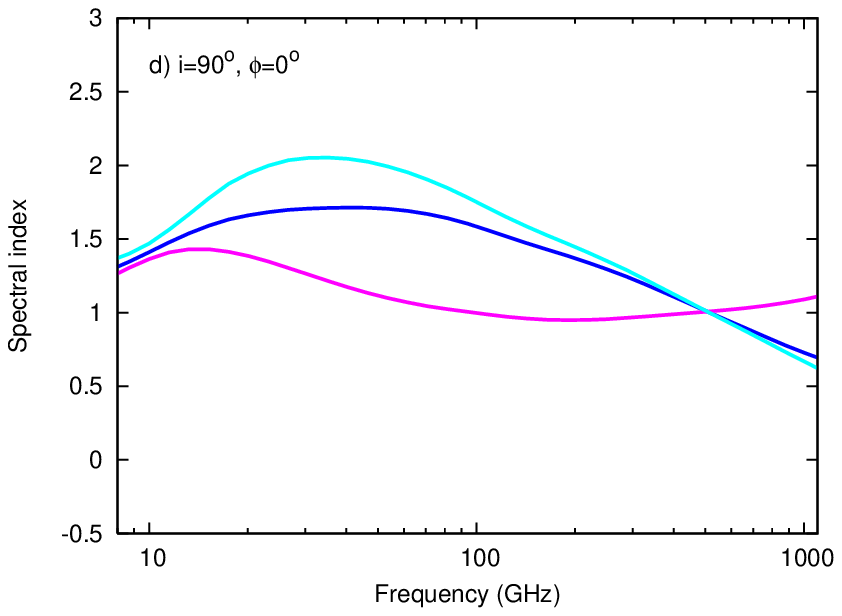,width=5.67cm}
\psfig{figure=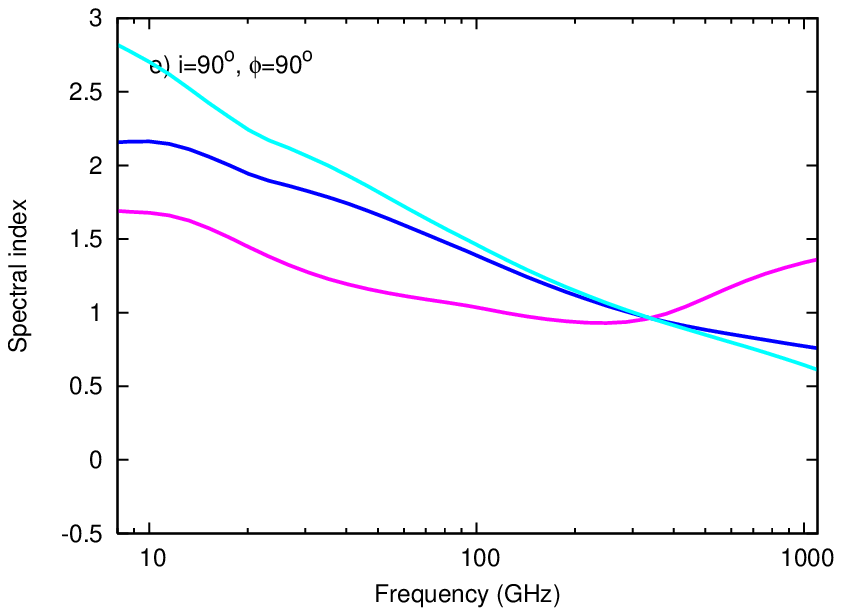,width=5.67cm}
\psfig{figure=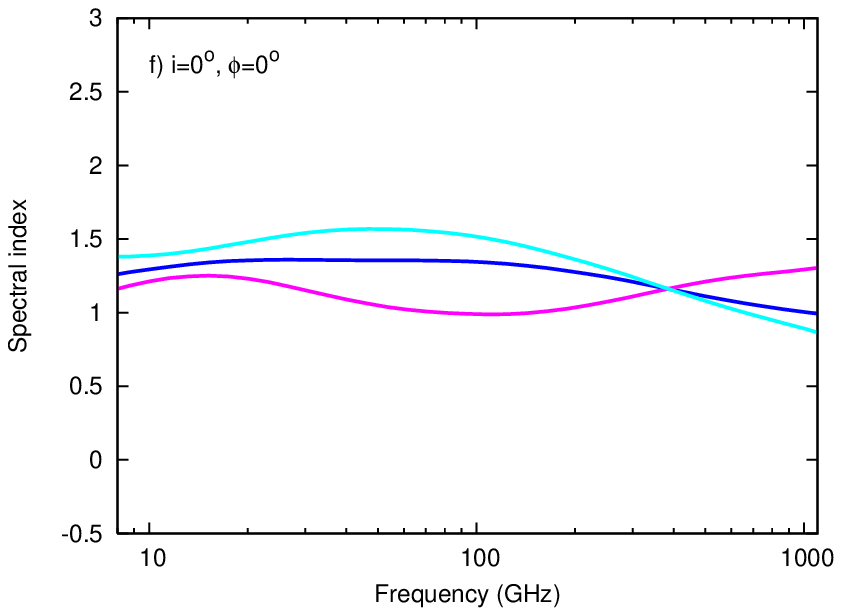,width=5.67cm}
\caption[]{Top: free-free radio spectra from model cwb1 at viewing angles: a)
$i = 90^{\circ}$, $\phi=0^{\circ}$, b) $i = 90^{\circ}$, $\phi=90^{\circ}$,
c) $i = 0^{\circ}$, $\phi=0^{\circ}$. In each case the 
contributions of the free-free emission from the
unshocked winds and the WCR to the total free-free emission are shown.
The stars are at conjunction (phase 0.0) in a), and quadrature (phase 0.75) 
in b). The stars are also at quadrature in c).
There is likely a significant loss of flux at the lowest
frequencies displayed due to the finite size of the numerical grid.
Bottom: the spectral index as a function of frequency from the spectra in
panels a)-c).}
\label{fig:cwb1_radio_spectrum}
\end{figure*}

\begin{figure*}
\psfig{figure=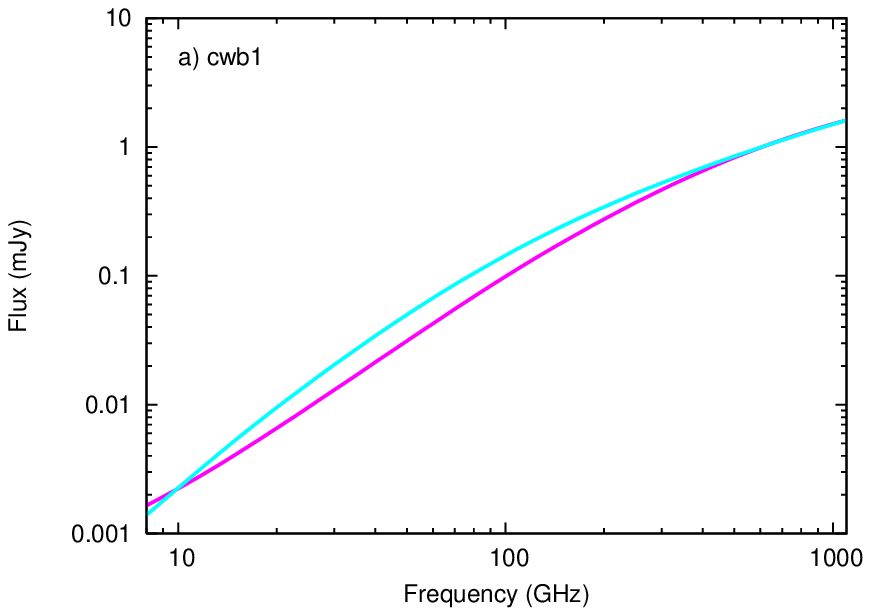,width=5.67cm}
\psfig{figure=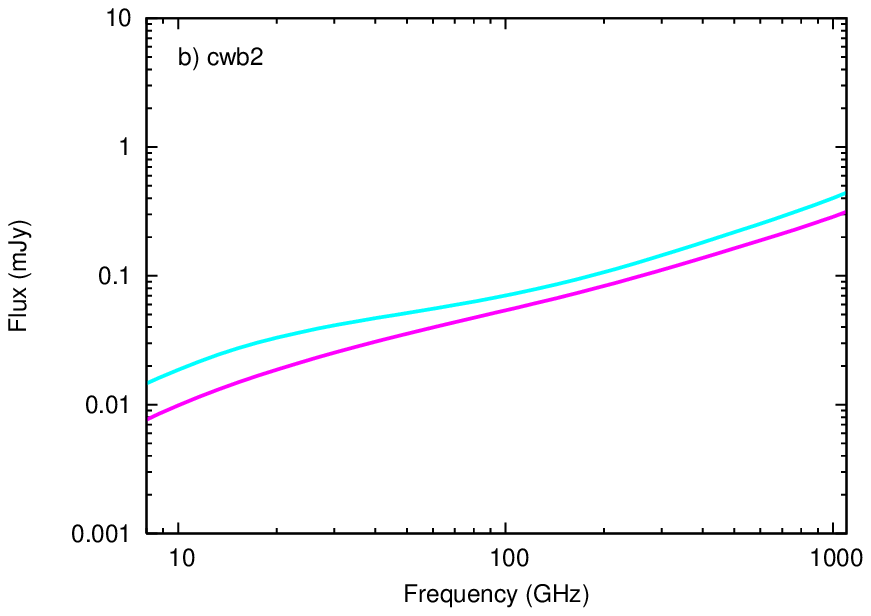,width=5.67cm}
\psfig{figure=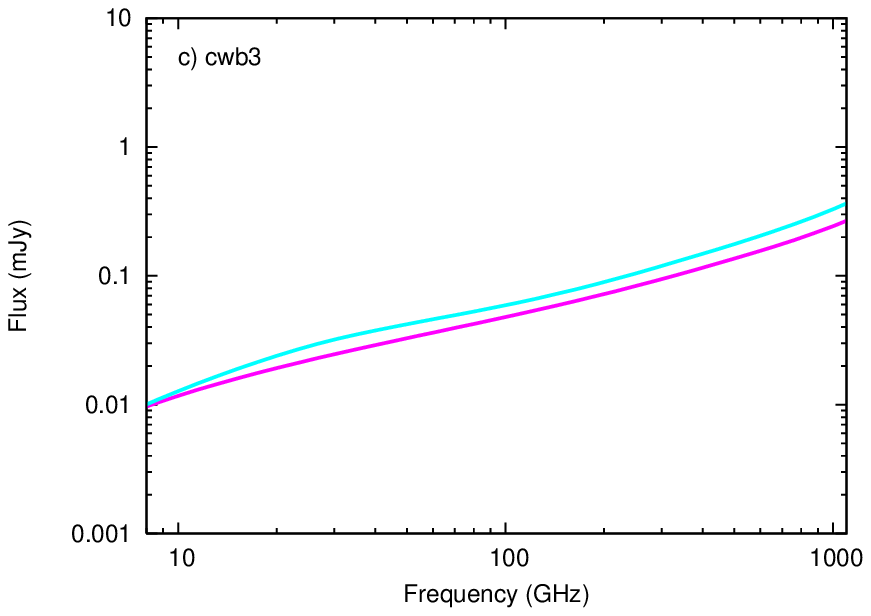,width=5.67cm}
\caption[]{Free-free radio spectra from models cwb1 (a), cwb2 (b), and
cwb3 (c), at viewing angles $i = 90^{\circ}$, $\phi=0^{\circ}$ 
(phase 0.0, pink), and $i = 90^{\circ}$, $\phi=90^{\circ}$ 
(phase 0.75, cyan) - conjunction and quadrature, respectively.
Because of the symmetry in models cwb1 and cwb2, the spectra at phase 0.0
and 0.75 are virtually identical to those at phase 0.5 and 0.25, respectively.
In model cwb5 the O8V star is in front at phase 0.0.}
\label{fig:cwb123_radio_spectrum_var}
\end{figure*}

\section{Results}
\label{sec:results}

\subsection{Emission from the wind of a single star}
\label{sec:singlestar}
We first investigate the cm to infrared free-free emission from the wind
of a single star, namely from the O6V star used in our binary
simulations. The wind density as a function of radial distance is
obtained from our solution to the modified CAK equations
\citep{Castor:1975,Pauldrach:1986}.  At high frequencies
the emission arises from within the acceleration zone of the wind, and
it is no longer possible to use, e.g., Eq.~8 in \citet{Wright:1975} to
evaluate the flux. We therefore numerically integrate the flux as a
function of impact parameter, following the method of
\citet{Wright:1975}.  The wind is assumed to begin at the point
where the outflow velocity is equal to the isothermal sound speed,
which is $\approx 25\,\kmps$ in our models. 
As in the binary models which follow we assume
that the wind is smooth, isothermal and maintains a constant
ionization: these assumptions allow us to isolate the effect of the
wind acceleration on the resulting emission. We also include the
photospheric flux, which is modelled as a black body for
simplicity. The photospheric flux is attenuated at wavelengths
where the wind is optically thick.

Fig.~\ref{fig:singlestar_spec_tau}(a) shows the resulting spectrum
obtained from the wind and photosphere of our model O6V star.  At GHz
frequencies (cm wavelengths) the total emission is dominated by the
wind and arises from large distances where the wind has very nearly
reached its terminal speed. The spectral index
(Fig.~\ref{fig:singlestar_spec_tau}b) is near $+0.6$, as expected. The
emission starts to probe the acceleration region of the wind at $\nu
\gtsimm 100$\,GHz, where a steepening of the spectrum becomes
apparent. It is from this point that the flux increasingly
exceeds that from a calculation where the wind is forced to emanate
from the star at its terminal speed (i.e. instantaneous
acceleration). As shown in Fig.~\ref{fig:singlestar_spec_tau}(a), 
such a terminal speed wind becomes optically thin at
frequencies above 700\,GHz \citep[Eq.~12 of][]{Wright:1975}, with
declining flux thereafter \citep[see also][]{Bertout:1985}. In
contrast, the sharp increase in the wind density of the CAK model as
the photosphere is approached results in a steep rise in the thermal
flux. At frequencies around $2 \times 10^{12}$\,Hz ($\lambda =
150\,{\rm \mu m}$), the spectral index of the emission from the
{\em wind} steepens to reach a maximum of
$\approx +1.1$. This steepening is due purely to the observed emission
occuring from regions where the wind is rapidly accelerating.

The wind also starts to become optically thin at $\nu \sim 2 \times
10^{12}$\,Hz ($\lambda = 150\,{\rm \mu m}$), and there is a steep rise
at higher frequencies in the spectral index of the total emission as
the photospheric emission becomes dominant at $\nu \gtsimm 4 \times
10^{12}$\,Hz ($\lambda = 75\,{\rm \mu m}$; for stars with denser
winds, the emission from the wind may dominate the photospheric
emission until $\mu$m wavelengths).
The maximum spectral index for thermal free-free emission is $+2.0$,
the value for a blackbody source where the flux $S_{\nu} \propto
\nu^2$, as given by the Rayleight-Jeans approximation to the Planck
function. This is indeed the case in a calculation where the wind and
photospheric temperatures are equal. However, when there is a sharp
change between the wind and photospheric temperatures, such as in our
simple model, spectral indices greater than +2.0 can be obtained. This
is apparent in Fig.~\ref{fig:singlestar_spec_tau}(b), where the
spectral index actually reaches $+2.5$ at $\nu \approx
5\times10^{12}$\,Hz, due to the steeply rising photospheric flux as
the absorption of the overlying wind declines.\footnote{In reality,
there will not be such a sharp division in temperature between the
wind and photosphere, and one would expect that the maximum value of
the spectral index determined from observations would be closer to
+2.0. Spectral indices greater than +2.0 have in fact
been previously reported from Be stars \citep*{Dougherty:1991},
but we note that these were based on 3 upper limits to the flux, 
and so are meaningless.}  The
photospheric emission from our model O6V star peaks at $\nu \approx
3\times10^{15}$\,Hz ($\lambda = 100\,{\rm nm}$), and subsequently
declines.

For a smooth, spherical, isothermal wind of constant ionization,
the spectral index, $\alpha$, is related to the
acceleration index, $\epsilon$ (where $v(r) \propto r^{\epsilon}$), by
\citep[e.g.][]{Leitherer:1991}
\begin{equation}
\label{eq:accnindex}
\alpha = \frac{4\epsilon + 1.8}{2\epsilon + 3}.
\end{equation}
Note that $v(r) \propto r^{\epsilon}$ is a {\em local}
approximation (assumed valid over the formation region
corresponding to a particular wavelength). The value of $\epsilon$
is {\em not} valid over the whole distance range, in the same
way that the value of $\beta$ in the ``$\beta$-law'' (see below) is.

To obtain $\alpha = +1.1$ requires that $\epsilon = 0.83$. This is easily
achieved in accelerating winds.
Fig.~\ref{fig:singlestar_spec_tau}(c) shows $\epsilon$ 
as a function of radius for the CAK wind profile and
standard ``$\beta$-law'' parameterizations (where
\begin{equation}
\label{eq:betalaw}
v(r) = v_{0} + (v_{\infty} - v_{0})(1 - R_{*}/r)^{\beta},
\end{equation}  
and $v_{0}$ is the initial wind velocity at the stellar radius,
assumed here to be $25 \,\kmps$).
The CAK velocity profile is reasonably well
approximated by a $\beta$ velocity law with $\beta=0.8$.  Winds with
slower acceleration have larger values of $\beta$, and vice-versa. Winds
with large $\beta$ have smaller values of $\epsilon$ close to the
star, but larger values further from the star where their winds 
continue to accelerate. Values of $\epsilon$ greater than 100 are achieved
in the innermost regions of strongly accelerating winds. 
In the CAK model $\epsilon = 0.83$ at $r/{\rm R_{*}} - 1 = 0.97$.

The maximum frequency which can be explored in the 3D hydrodynamical
models is set by the necessity of sufficiently resolving the
acceleration region where the characteristic radius of emission
occurs. Models cwb1 and cwb4 have a numerical resolution of
$0.5\;\Rsol$.  In models cwb2 and cwb3 the numerical resolution is
$1.25\;\Rsol$. Therefore, it is only possible to examine frequencies
up to $\sim 1$\,THz, since the optical depth for an observer at
infinity to $r=1.1\,{\rm R_{*}}$ is less than unity at frequencies
greater than this. In the following we restrict our investigation to
$\nu \leq 1$\,THz ($\lambda \geq 300$\,$\mu$m).

\subsection{Emission from binary systems}
\label{sec:results_binary}
\subsubsection{Model cwb1}
The hydrodynamical grid used to compute model cwb1 contains within its
volume the characteristic radius of emission from the O6V wind at
$\nu \geq 43$\,GHz. The optical depth unity surface,
which is slightly smaller than $R_{\rm c}$, fits completely inside the
grid for $\nu > 15$\,GHz. Therefore, we limit our study of the
free-free intensity images and lightcurves from model cwb1 to $\nu
\geq 43 $\,GHz, but show spectra down to 8\,GHz for ease of comparison
with our other models.

Fig.~\ref{fig:cwb1_radio_image} shows intensity images of the
free-free emission from model cwb1 at 43, 100, 250, and 1000\,GHz, for
an observer located directly above the orbital plane ($i =
0^{\circ}$).  It is immediately apparent that the WCR dominates the
emission, even at a frequency as high as 1\,THz. The morphology of the
emission from the WCR reflects its underlying structure and
clumpiness, caused by its high susceptibility to dynamical
instabilities (see Paper~I for a detailed discussion of the
hydrodynamics). Near the apex of the WCR the projected emission from
different inhomogeneities merges together, but further downstream
individual clumps and bowshocks can be identified.  It is clear that a
great deal of potential emission is not captured due to the limited
physical extent of the hydrodynamical grid used in the model. For
obvious reasons it is impossible to know just how much flux is
``lost'' in this way as dense clumps in the WCR flow out through the
grid boundaries, but it is clear that this problem is worse at lower
frequencies where structure in the WCR remains optically thick for
larger downstream distances\footnote{One can estimate a rough upper
limit for the amount of flux lost by taking the flux at 1000\,GHz, and
extrapolating to lower frequencies with a spectral index of $+0.6$ -
this reveals that the actual flux from the WCR at 43\,GHz may be up to
an order of magnitude higher.}.  Nevertheless, these images, and the
lightcurves and spectra which follow, provide important new insight
into the emission at cm and sub-mm wavelengths from CWBs. 

The spatial scale of the images is 1.12 by 1.12\,mas, which is too
small to be resolved with the EVLA (angular resolution of 4\,mas at
50\,GHz) or ALMA (highest angular resolution expected to be about
10\,mas). In contrast, the VLBA has the necessary spatial
resolution, but even with all current upgrade plans may struggle with
the necessary sensitivity to obtain images of this system within a
reasonable integration time (e.g. 10\,hrs) to avoid smearing of the
essential features due to orbital motion.  In 10\,hrs, with
256\,Mbit/s bandwidth at 43\,GHz, the VLBA + GBT + phased-VLA obtains
a sensitivity of 0.029\,mJy/beam. Bandwidth improvements to 4\,Gbit/s
will increase the sensitivity by a factor of 4. A signal to noise
ratio (SNR) of 5 (the minimum needed to provide imaging constraints)
in 10\,hrs thus requires a flux of 0.0725\,mJy/beam. The spatial
resolution at 43\,GHz is $\approx 0.3$\,mas, so each arm of the WCR
falls within a beam (see Fig.~\ref{fig:cwb1_radio_image}), which is
roughly 50 per cent of the total flux, or 0.02\,mJy. Hence a 10\,hr
exposure is insufficient to obtain a SNR of 5 - instead $\approx
130$\,hrs is required. Having said this, it has already been noted
that there is a significant loss of flux at such frequencies from
model cwb1 due to the finite size of the hydrodynamical grid, so the
future detection of such a system with the VLBA is not totally beyond
the realm of possibility (e.g. if the flux from model cwb1 at 43\,GHz
was actually $3.6\times$ higher).  Systems with stronger winds (higher
mass-loss rates) than assumed for model cwb1 may further improve the
possibility of such imaging.

Fig.~\ref{fig:cwb1_ffradio_lc} shows the free-free emission
lightcurves at 43\,GHz (top panels) to 1000\,GHz (bottom panels), as a
function of the inclination angle, $i$, of the system. The left panels
show the combined emission from the unshocked winds and the WCR, while
the middle and right panels display the unshocked and shocked
components, respectively.  The stars are aligned along the
lines-of-sight at phases 0.0 and 0.5, and are at quadrature at phases
0.25 and 0.75. The variations in flux are entirely due to changes in
the occultation and circumstellar absorption, as the intrinsic
emission is constant. Note that the presence of the companion star
breaks the spherical symmetry of the unshocked winds. For example,
because of the twin radiation fields, the winds accelerate
more slowly towards each other, and more quickly away from each
other. There are also regions of lower velocity wind in the
``shadows'' behind each star (see Paper~I for further details).

The lightcurves of the WCR emission (right panels in
Fig.~\ref{fig:cwb1_ffradio_lc}) would display dual symmetry about
phases corresponding to both quadrature (0.25, 0.75) and conjunction
(0, 0.5) if there were no orbital effects on the WCR \citep[i.e. no
aberration or coriolis induced curvature - cf.][]{Pittard:1997}, because of
the equal winds and constant stellar separation.  Orbital effects
break the symmetry about quadrature, so that in the case of identical
winds a double periodicity in the lightcurve is expected. However,
dynamical instabilities which form in the WCR can break this symmetry
too, as revealed by close examination of Fig.~\ref{fig:cwb1_ffradio_lc}.

At the lower frequencies the flux from the unshocked winds is
typically only a small fraction of the expected combined flux from the
winds (see the horizontal red lines across the plots in the middle
column of Fig.~\ref{fig:cwb1_ffradio_lc}). For instance, at 43\,GHz
the thermal flux from a single smooth (unclumped) wind of the same
parameters as used in model cwb1 is 0.017\,mJy, while the combined
unshocked winds contribution in Fig.~\ref{fig:cwb1_ffradio_lc}(b) is
$\approx 0.01\,$mJy.  This disparity reduces at higher frequencies -
at 250\,GHz, the two unshocked winds contribute $\approx 0.055\,$mJy,
whereas the expected single star value is 0.046\,mJy - and is
eliminated by 1000\,GHz.

The lack of observed flux from the unshocked winds as shown in
Figs.~\ref{fig:cwb1_ffradio_lc}(b) and~(e) is likely due to the
removal of a large fraction of the unshocked material from both of the
winds by the presence of the WCR.  In contrast, at the highest
frequencies considered (250 and 1000\,GHz), the characteristic radius
of emission moves so close to the surface of the stars (particularly
at 1000\,GHz) that the emission from this region is not strongly
disturbed by the presence of the WCR at larger distances. A small
additional effect at the lowest frequencies arises from the limited
extent of the computational grid compared to the characteristic radius
of emission, which causes an underestimate of the flux. This can be
seen in the spectra in Fig.~\ref{fig:cwb1_radio_spectrum}, where a
distinct downturn in the unshocked (``ff winds'') flux is visible at
the lower frequencies investigated.  However, at higher frequencies
between $43-250$\,GHz the removal of unshocked wind material by the
WCR is the main cause of the ``missing'' flux.

At 1000\,GHz, the combined flux from the unshocked winds is $\approx
0.25\,$mJy, which actually slightly exceeds twice the theoretical
value of 0.095\,mJy for our single star. This reflects the fact that
the characteristic radius of emission is now deep within the
acceleration zones of the winds, while the theoretical flux is
estimated assuming that the entire wind is moving at its terminal
speed. This excess over the ``theoretical'' value was also 
seen in the single-star calculations presented in the previous section.

The thermal emission from the unshocked winds is generally pretty
steady with orbital phase. Slight maxima occur near to quadrature,
with minima near to conjunction, as expected. The 1000\,GHz lightcurve
is not so smooth in appearance - its more ``noisy'' behaviour is a
consequence of lines of sight through the WCR becoming optically thin
in some directions, but not in others. The density of the thin layer
of cooled postshock gas is $\approx 6\times 10^{-14}\,{\rm
g\,cm^{-3}}$ near the apex of the WCR. The linear absorption
coefficient of this gas at 1000\,GHz is $\approx
1.5\times10^{-11}\,{\rm cm^{-1}}$, while the thickness of this layer
is typically $1\,\Rsol$, giving an optical depth of approximately
unity. Hence we see the emitting regions of the WCR and the unshocked
winds near the stellar photospheres through a semi-opaque, semi-porous
WCR.

The free-free emission from the WCR, on the other hand, displays much
greater phase variability. It has already been noted that the 
flux from the WCR dominates the emission from the unshocked winds.
Fig.~\ref{fig:cwb1_ffradio_lc} shows that the WCR flux is 
around $5\times$ greater at $250-1000$\,GHz, though the true value is
likely to be somewhat higher.  Future calculations with a larger grid
are necessary to accurately determine this ratio.
The lightcurves of the emission from the WCR typically display two
minima per orbital cycle, with a phase-lead with respect to conjunction
of the stars which appears to increase with frequency. Whereas the
X-ray lightcurves of model cwb1 in Paper~III produce reasonably sharp
maxima, the lightcurves in the left column of
Fig.~\ref{fig:cwb1_ffradio_lc} display much broader maxima. 

An explanation for the behaviour of the lightcurve of the emission
from the WCR now follows.  Figs.~\ref{fig:cwb1_radio_image}
and~\ref{fig:cwb1_ffradio_lc} show that at radio to sub-mm
frequencies, the most intense emission from model cwb1 occurs from the
cold, dense sheet of post-shock material of cooled gas within the
WCR. The intrinsic emissivity of this region is very high, since the
free-free emission coefficient $\varepsilon_{\rm ff} \propto
\rho^{2}T^{-1/2}$, and the density in the WCR is orders of magnitude
greater than in the unshocked winds.  The WCR is also optically thick,
and, because of the downstream curvature induced by the coriolis
effect, it displays a larger projected surface to the observer, and
hence a larger flux, when the stars are at quadrature than at
conjunction (see Fig.~\ref{fig:cwb1_radio_image}). The amplitude of
the variability decreases slightly with frequency as the optical depth
through the WCR declines.

Fig.~\ref{fig:cwb1_radio_spectrum} shows radio-to-sub-mm spectra as a
function of orbital phase for an observer in the orbital plane (panels
a and b), and directly above/below it (panel c).  In all cases the
optically thick emission from the WCR dominates the observed flux
(remember that there is likely a significant loss of flux at the
lowest frequencies displayed). However, the emission from the WCR is
less dominant over the unshocked winds for an observer above or below
the orbital plane (compare panel c to panels a or b), since the
orientation of the layer of cold dense gas to the observer is less
favourable (which reduces the flux observed from it), while that of
the winds is more favourable (both winds are seen side-to-side on the
sky).

The spectrum of the total thermal flux shows considerable curvature
for an observer in the orbital plane (panels a and b), but
significantly less curvature for an observer with $i = 0^{\circ}$.
Fig.~\ref{fig:cwb1_radio_spectrum}(d)-(f) shows the frequency
dependence of the spectral index from these spectra. Between 100 and
250\,GHz the spectral index of the unshocked winds, $\alpha_{\rm
winds}$, is $+0.95$, $+0.95$, and $+1.00$, for viewing angles of
($i=90^{\circ}$, $\phi=0^{\circ}$), ($i=90^{\circ}$,
$\phi=90^{\circ}$), and ($i=0^{\circ}$), respectively. These are
steeper than the canonical $+0.6$ for optically thick, isothermal,
terminal speed winds, reflecting the fact that the winds in our model
are accelerating and asymmetrical. $\alpha_{\rm winds}$ increases
towards 1000\,GHz as the emission probes deeper into the acceleration
region of the winds. At 1000\,GHz, $\alpha_{\rm winds} = +1.09$,
$+1.34$, and $+1.29$ for the same orientations. The variation of
$\alpha_{\rm winds}$ with viewing angle clearly shows that the
presence of the companion star has a non-negligible impact on this
parameter, with large variations in $\alpha_{\rm winds}$ between
conjunction and quadrature.

The reduction in the spectal index at the highest frequencies of the
emission from the WCR, $\alpha_{\rm WCR}$, is an indication that parts
of the WCR are becoming optically thin. Between 100 and 250\,GHz,
$\alpha_{\rm WCR} = +1.51$, $+1.22$, and $+1.39$ for the three viewing
angles noted above. At 1000\,GHz, $\alpha_{\rm WCR} = +0.66$, $+0.64$, and
$+0.89$, respectively. For the total free-free emission between 100 and
250\,GHz, $\alpha = +1.29$, $+1.00$, and $+1.39$, while at 1000\,GHz,
$\alpha = +0.73$, $+0.77$, and $+1.00$. The less dominant
contribution of the WCR at $i=0^{\circ}$ results in the least
variation in $\alpha$ with frequency (see 
Fig.~\ref{fig:cwb1_radio_spectrum}(d)-(f)).

Fig.~\ref{fig:cwb123_radio_spectrum_var}(a) shows the variation of the
total thermal emission for an observer in the orbital plane between
conjunction and quadrature. The emission at quadrature is dominant
over the emission at conjunction up to $\nu \approx 500$\,GHz.
Between 500 and 1000\,GHz the emission at these phases is almost
identical. Note that in this frequency range the maxima and minima in the
lightcurves no longer occur at quadrature and conjunction.  It is not
possible to state whether this behaviour continues beyond 1000\,GHz,
for the reasons mentioned earlier in Sec.~\ref{sec:singlestar}.

\begin{figure*}
\psfig{figure=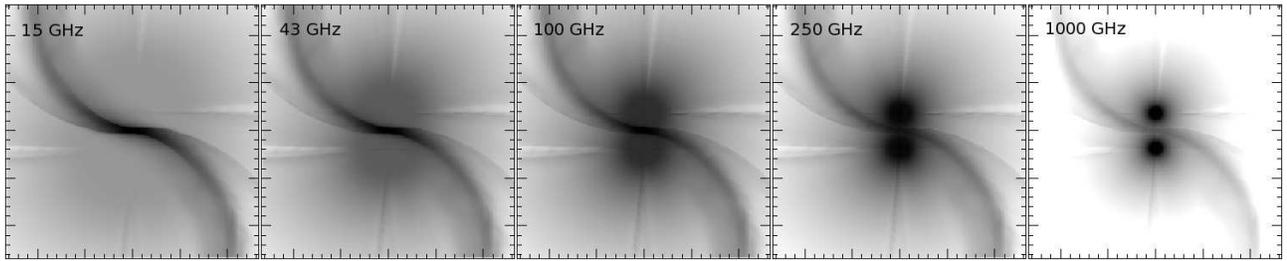,width=17cm}
\caption[]{Intensity images from model cwb2 at $i=0^{\circ}$. The
stars and the WCR again rotate anticlockwise with time. In this and
Figs.~\ref{fig:cwb2_radio_image2} and~\ref{fig:cwb2_radio_image3}, the
maximum intensity (black in the images) is $1.6\times10^{-10}\,{\rm
erg\,cm^{-2}\,s^{-1}\,Hz^{-1}\,ster^{-1}}$ at 15, 43 and 100\,GHz,
$3\times10^{-10}\,{\rm erg\,cm^{-2}\,s^{-1}\,Hz^{-1}\,ster^{-1}}$ at
250\,GHz, and $3.5\times10^{-9}\,{\rm
erg\,cm^{-2}\,s^{-1}\,Hz^{-1}\,ster^{-1}}$ at 1000\,GHz. The gray
scale spans 4 dex between maximum and minimum. The major ticks on each
axis mark out 0.5\,mas. Some numerical artefacts from the
hydrodynamics are visible, but have negligible impact on the flux.}
\label{fig:cwb2_radio_image}
\end{figure*}

\begin{figure*}
\psfig{figure=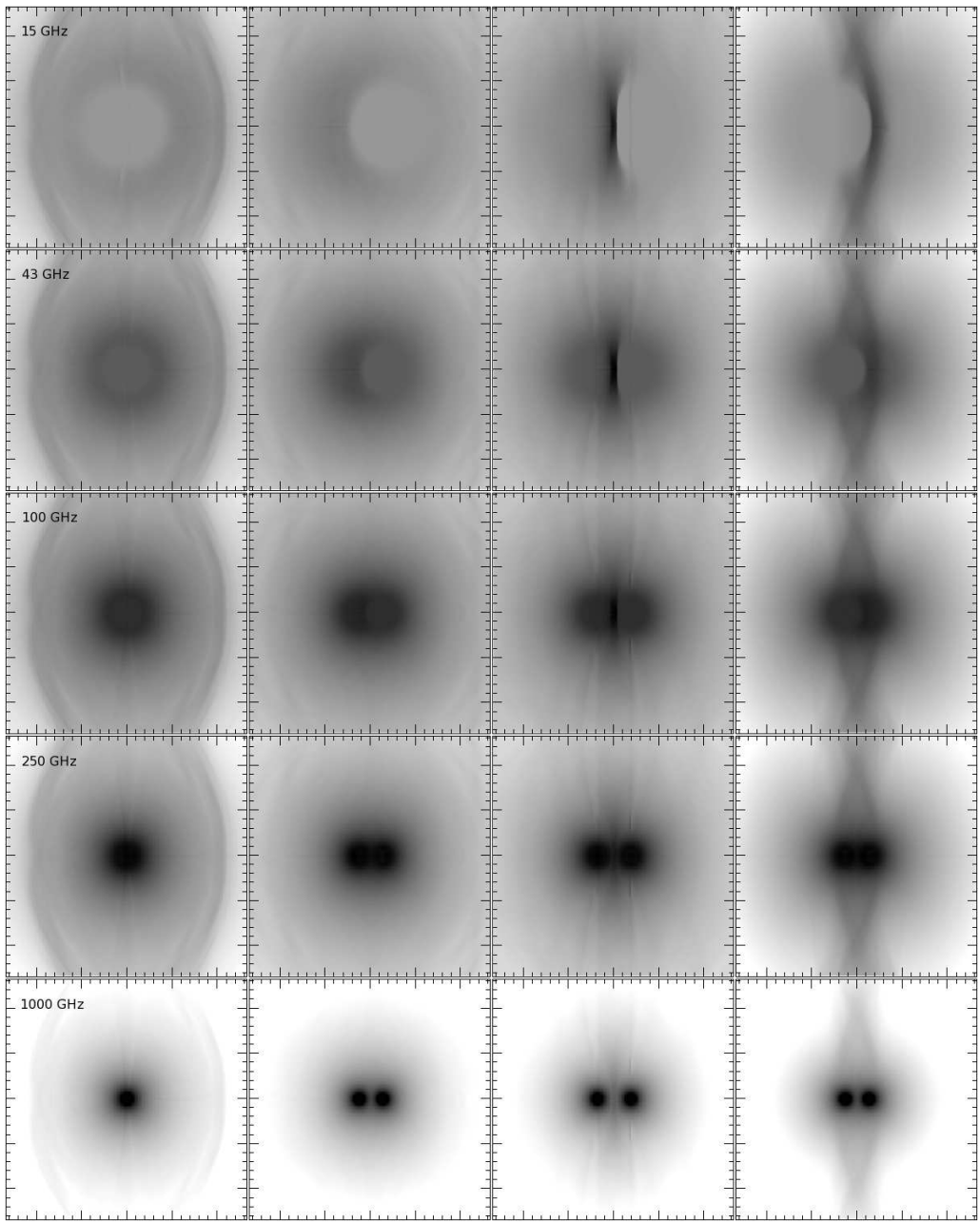,width=13.6cm}
\caption[]{Intensity images from model cwb2 at $i=90^{\circ}$. From
top to bottom the images are at a frequency of 15, 43, 100, 250, and
1000\,GHz. From left to right the phase of the observation increases
from 0.0 (conjunction, $\phi = 0^{\circ}$), to 0.125 ($\phi = 315^{\circ}$),
to 0.25 (quadrature, $\phi = 270^{\circ}$), to 0.375 ($\phi = 225^{\circ}$).
The grey scale is the same as in Fig.~\ref{fig:cwb2_radio_image}.}
\label{fig:cwb2_radio_image2}
\end{figure*}

\begin{figure*}
\psfig{figure=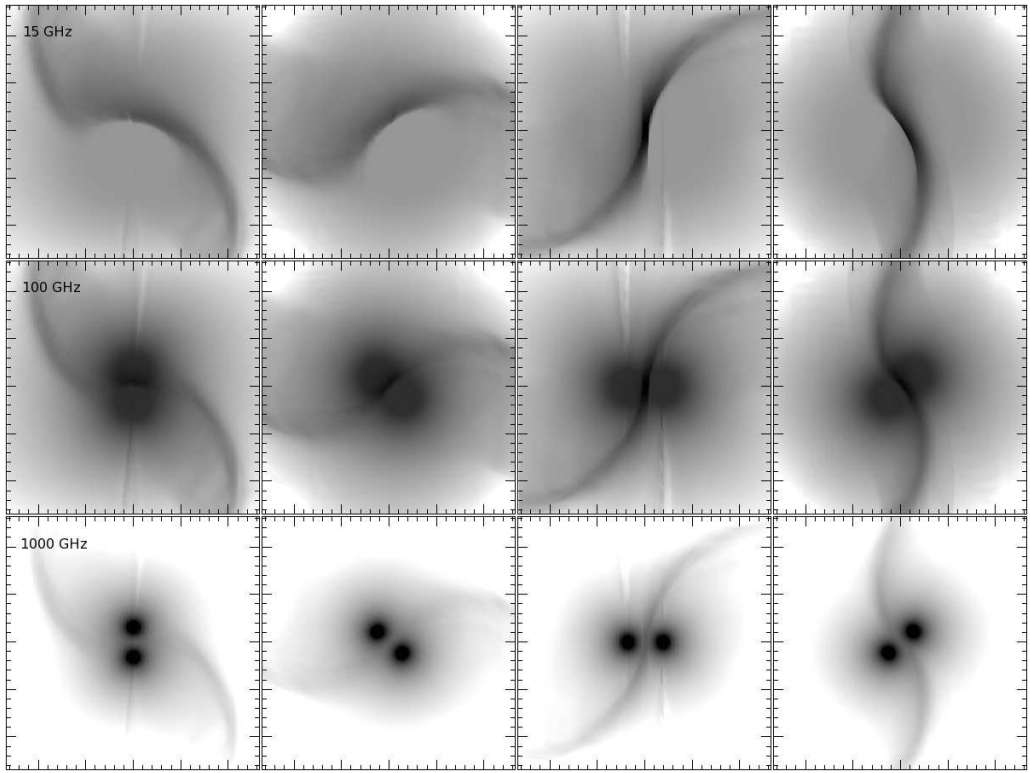,width=13.6cm}
\caption[]{As Fig.~\ref{fig:cwb2_radio_image2} but for an inclination angle
$i=30^{\circ}$. From top to bottom the images are at 15, 100, and 1000\,GHz,
and from left to right the orbital phases are 0.0, 0.125, 0.25, and 0.375. 
The stars and the WCR again rotate anticlockwise, while the grey 
scale is the same as in Fig.~\ref{fig:cwb2_radio_image}.}
\label{fig:cwb2_radio_image3}
\end{figure*}

\begin{figure*}
\psfig{figure=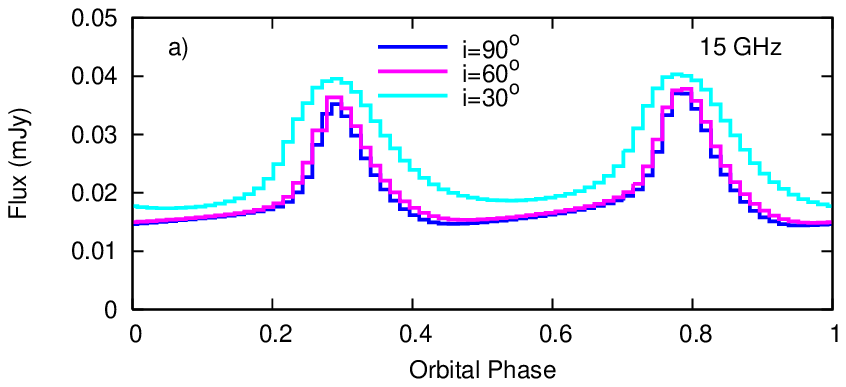,width=5.67cm}
\psfig{figure=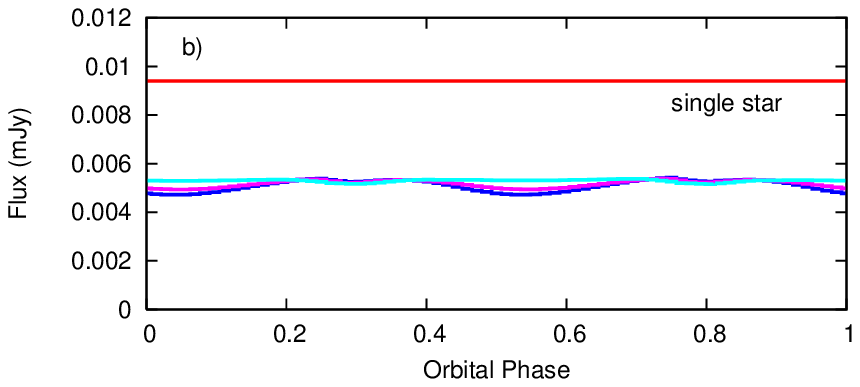,width=5.67cm}
\psfig{figure=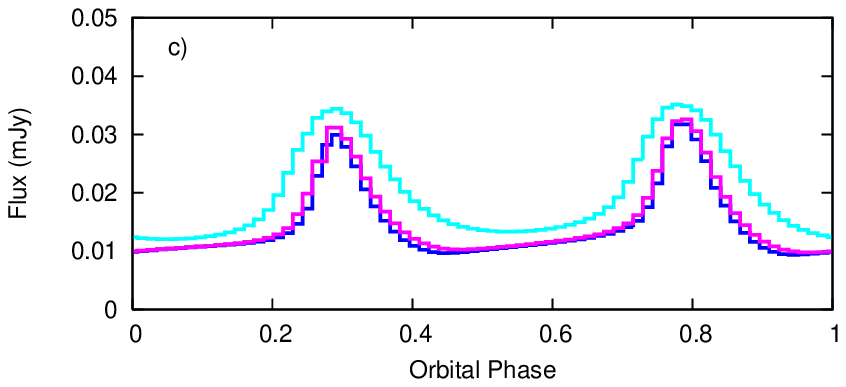,width=5.67cm}
\psfig{figure=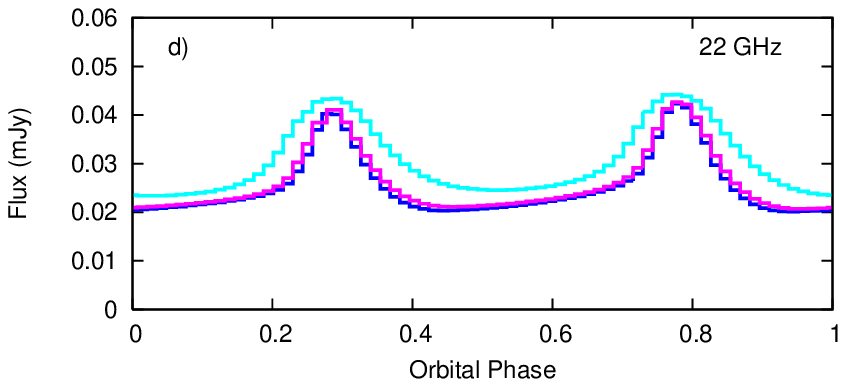,width=5.67cm}
\psfig{figure=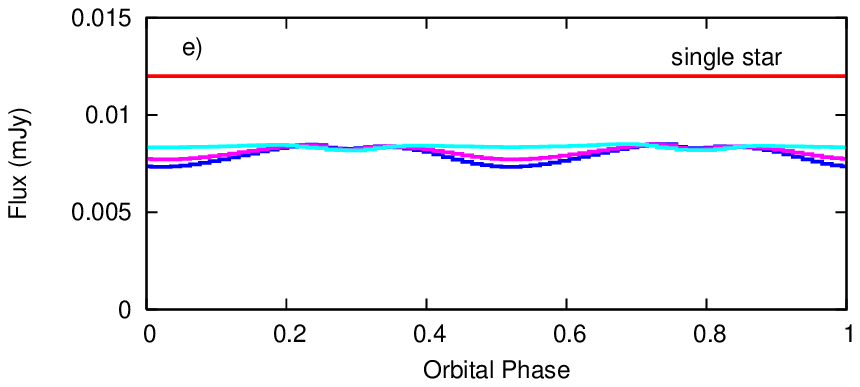,width=5.67cm}
\psfig{figure=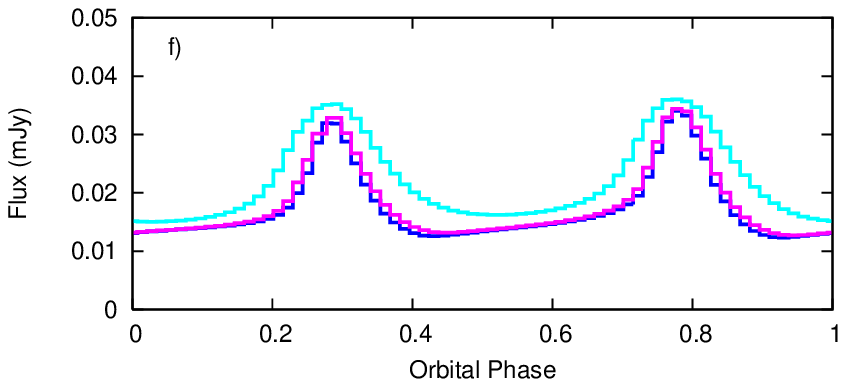,width=5.67cm}
\psfig{figure=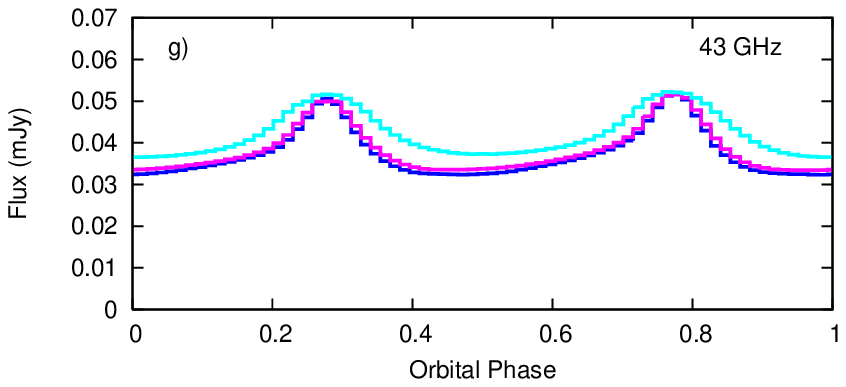,width=5.67cm}
\psfig{figure=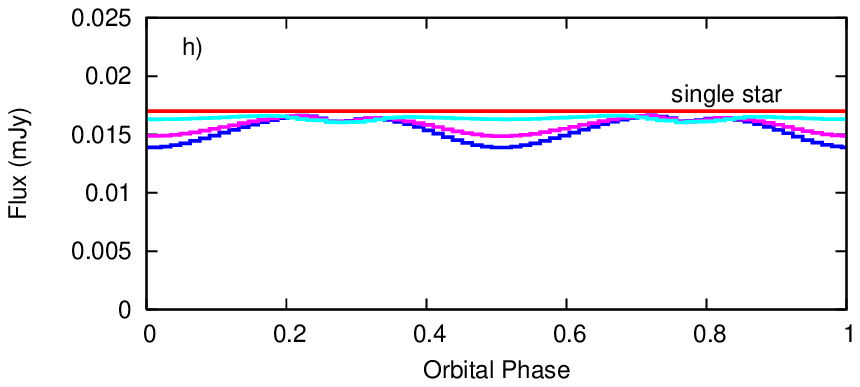,width=5.67cm}
\psfig{figure=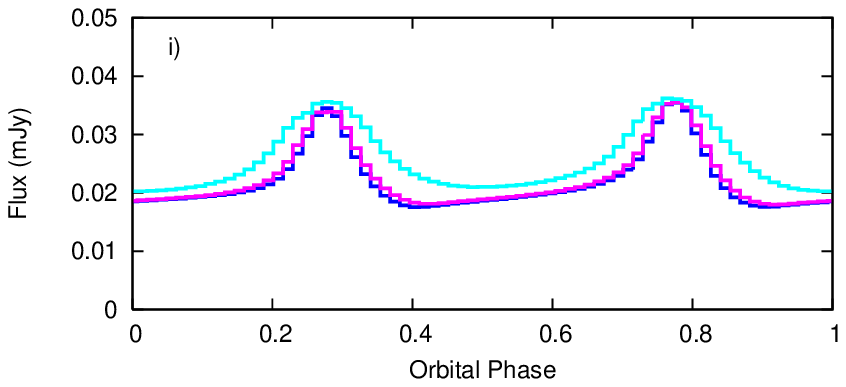,width=5.67cm}
\psfig{figure=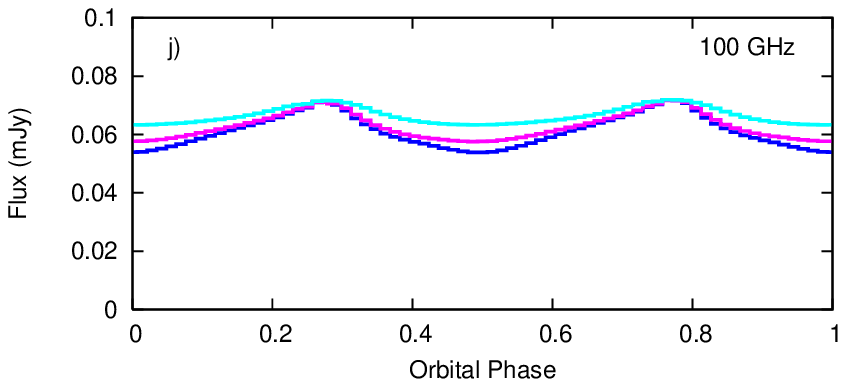,width=5.67cm}
\psfig{figure=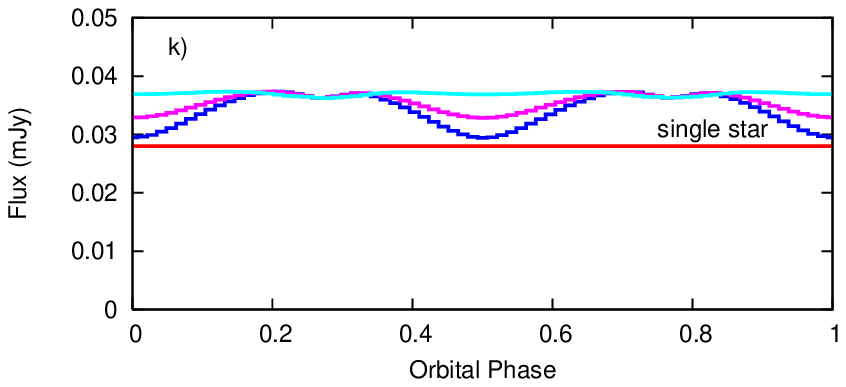,width=5.67cm}
\psfig{figure=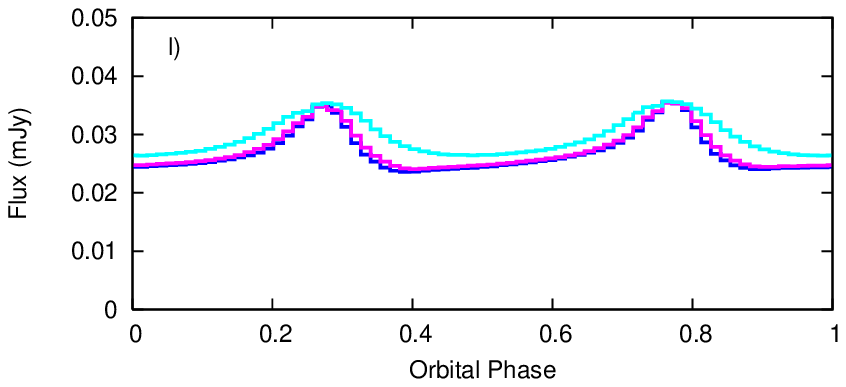,width=5.67cm}
\psfig{figure=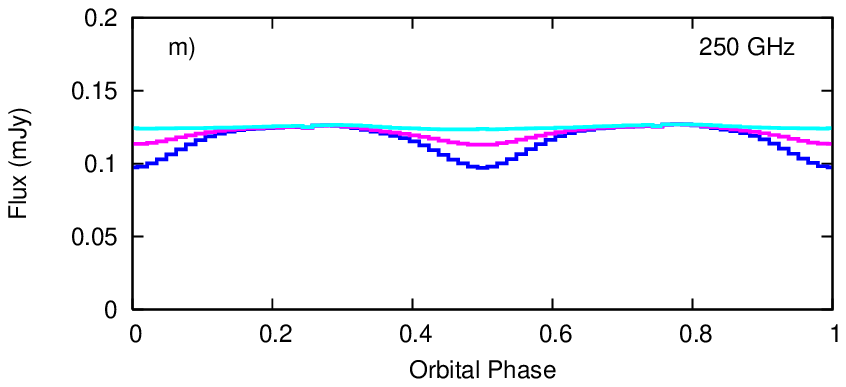,width=5.67cm}
\psfig{figure=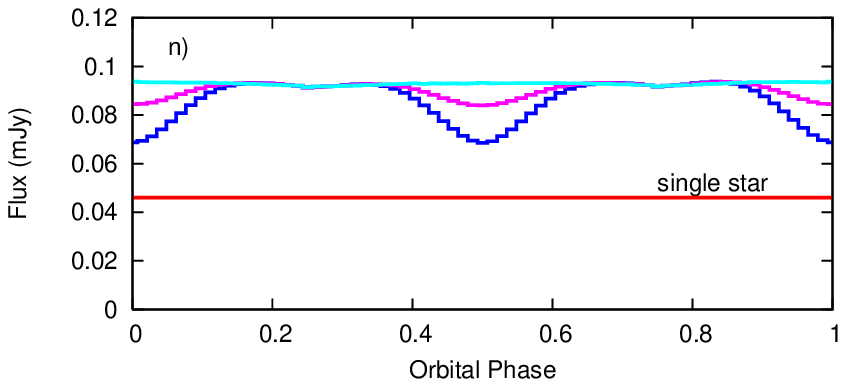,width=5.67cm}
\psfig{figure=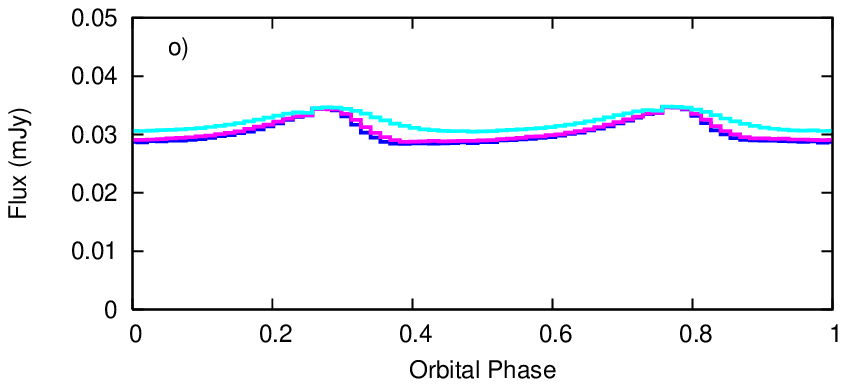,width=5.67cm}
\psfig{figure=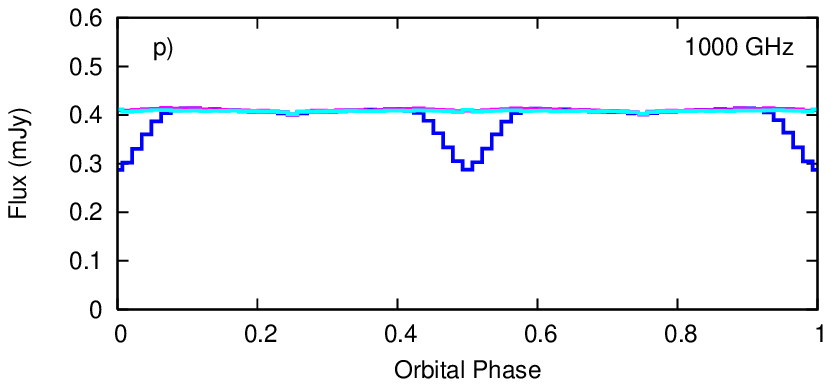,width=5.67cm}
\psfig{figure=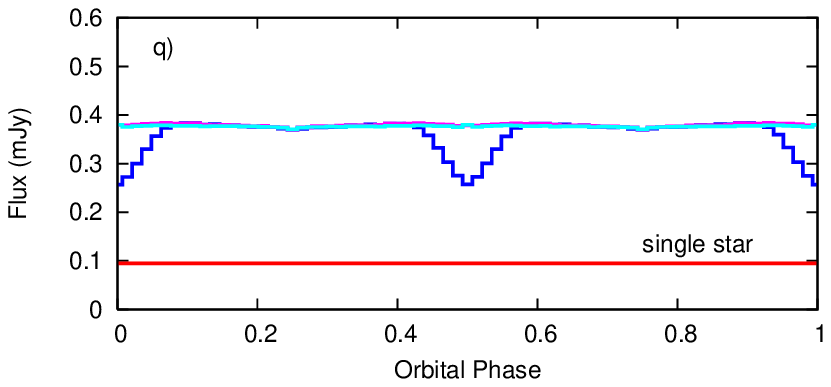,width=5.67cm}
\psfig{figure=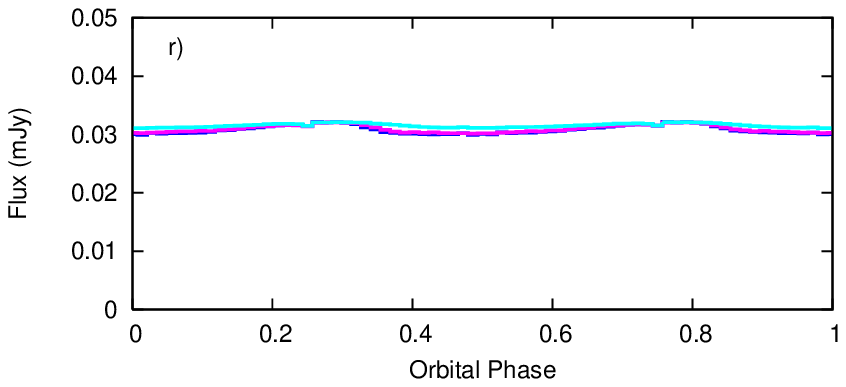,width=5.67cm}
\caption[]{Radio lightcurves of the free-free emission from model cwb2
at 15\,GHz (top), 22\,GHz, 43\,GHz, 100\,GHz, 250\,GHz, and 
1000\,GHz (bottom) for
inclination angles $i = 30^{\circ}$, $60^{\circ}$, and $90^{\circ}$.
The total free-free emission is shown in the left panels, and the
contributions from the winds and WCR in the middle and right panels,
respectively. In all cases, the observer is located along a direction vector
specified by $\phi=0^{\circ}$. The stars are at conjunction at phases 0.0 and
0.5, and quadrature at phases 0.25 and 0.75. Note that the y-axes (flux scales)
are different in each panel.}
\label{fig:cwb2_ffradio_lc}
\end{figure*}

\begin{figure*}
\psfig{figure=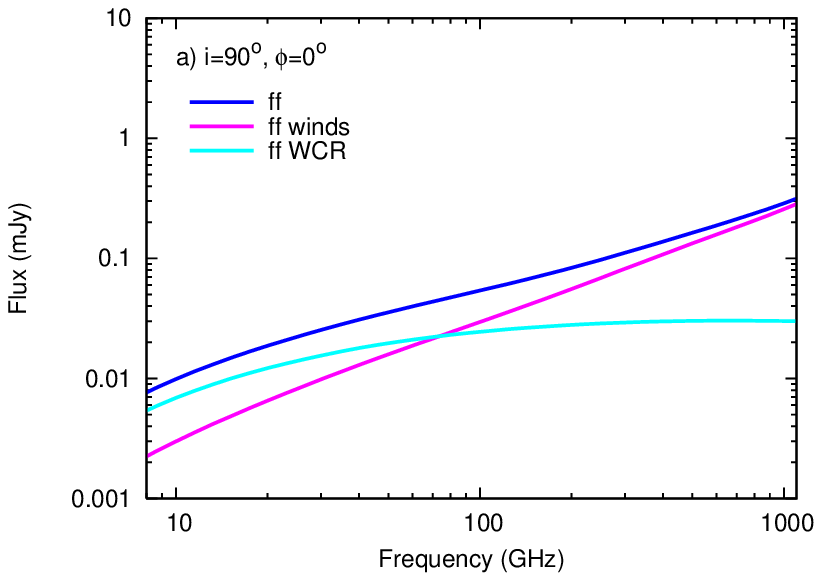,width=5.67cm}
\psfig{figure=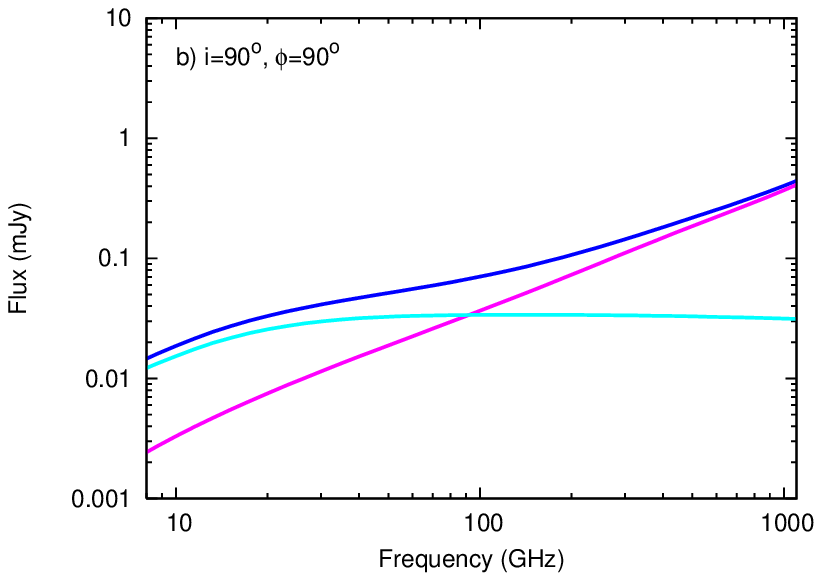,width=5.67cm}
\psfig{figure=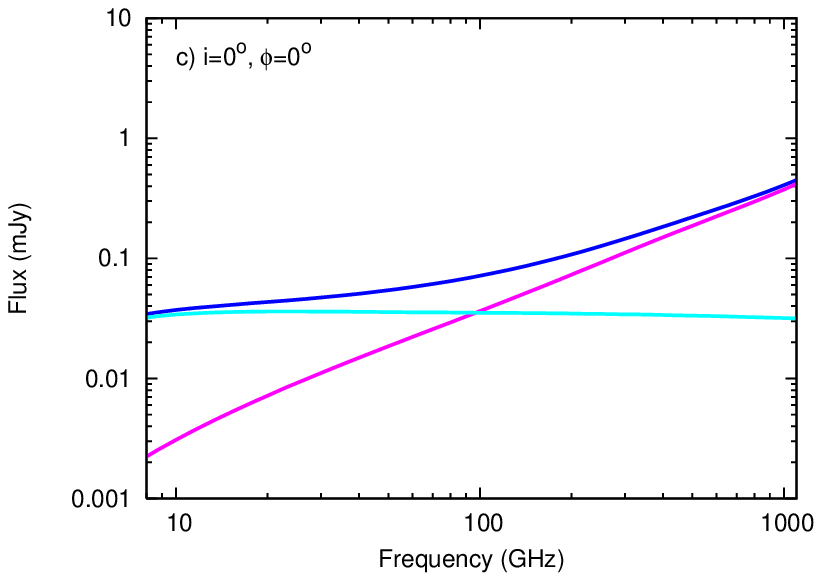,width=5.67cm}
\psfig{figure=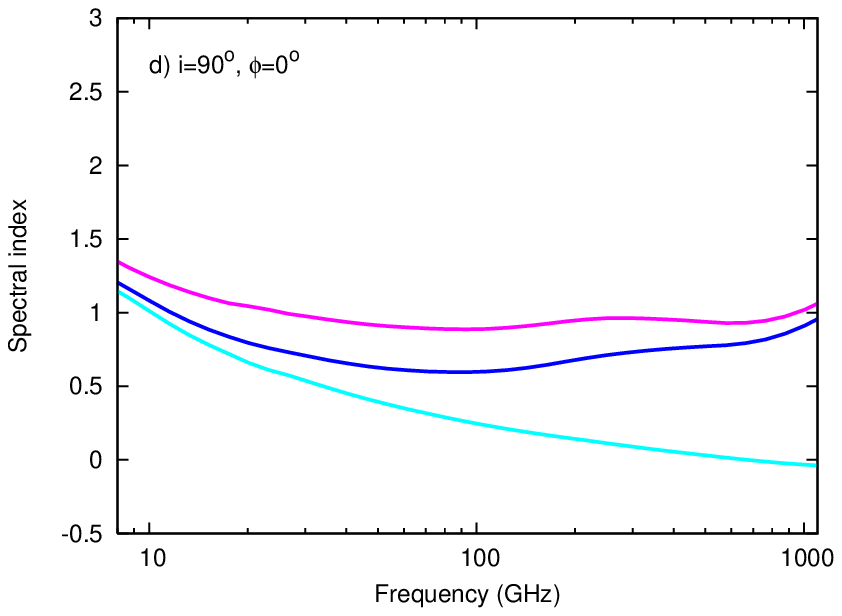,width=5.67cm}
\psfig{figure=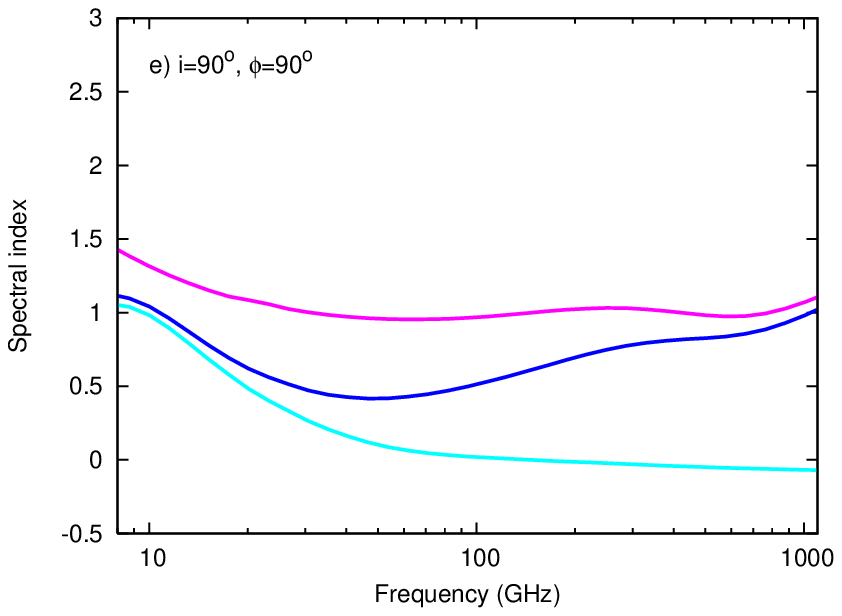,width=5.67cm}
\psfig{figure=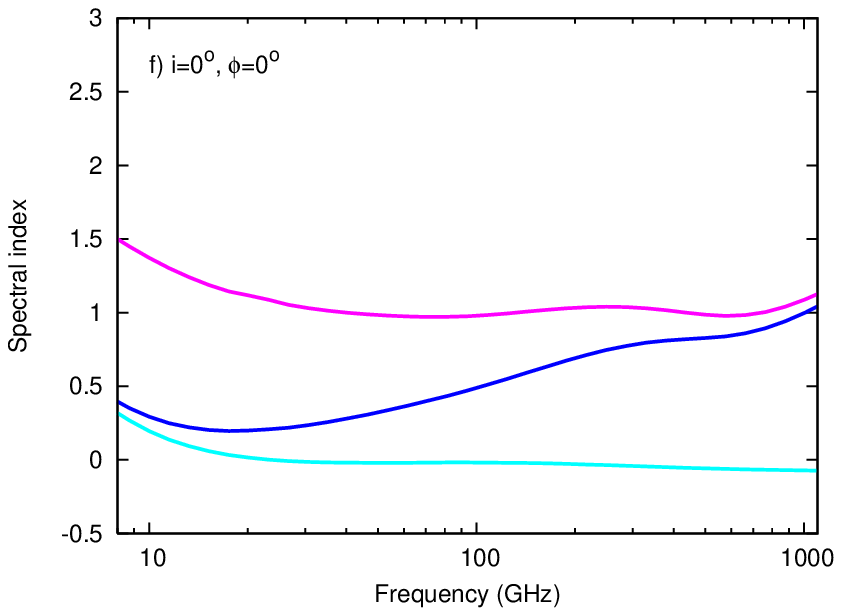,width=5.67cm}
\caption[]{As Fig.~\ref{fig:cwb1_radio_spectrum} but for model cwb2. 
The free-free emission from the WCR becomes optically thin at 
$\nu \gtsimm 5$\,GHz (i.e. the free-free opacity of the shocked
plasma has an impact on the spectrum below $\sim 5$\,GHz - see also the
discussion in \citet{Dougherty:2003}).}
\label{fig:cwb2_radio_spectrum}
\end{figure*}

\subsubsection{Model cwb2}
Fig.~\ref{fig:cwb2_radio_image} shows intensity images of the
free-free emission from model cwb2 at 15, 43, 100, 250, and 1000\,GHz,
for an observer located directly above the orbital plane ($i =
0^{\circ}$).  The curved shape of the WCR is clearly seen. The WCR
dominates the emission at the lower frequencies ($\nu \ltsimm
100$\,GHz), while the unshocked winds dominate the emission at
higher frequencies. At 15\,GHz the WCR is very bright in comparison to
the winds. The brightest part of the WCR is its apex, between the
stars. Further downstream the gas at the trailing edge of the WCR is
brighter than at the leading edge, reflecting the higher densities
there (see Paper~I). The angular scale of all of the images in
Figs.~\ref{fig:cwb2_radio_image}-\ref{fig:cwb2_radio_image3} is 2.7 by
2.7\,mas, which is again too small to be resolved with EVLA/ALMA
but may be resolved with the VLBA.

Intensity maps from model cwb2 for an observer in the orbital plane
are shown in Fig.~\ref{fig:cwb2_radio_image2}. The features in these
images depend on the relative orientation and the degree of
emission/absorption of the winds and the WCR, and display significant
changes as a function of frequency and orbital phase. At conjunction
only the foreground wind is seen, since this blocks the emission from
the wind of the more distant star. The surface brightness of the
foreground wind is less than that of the WCR at 15 GHz, so there is a
``hole'' in the emission from the WCR at conjunction. In contrast, the
surface brightness of the unshocked winds is greater than that of the
WCR when $\nu \gtsimm 100$\,GHz.  The sharp, curved edges to the
emission from the WCR near the borders of the images is due to
limb-brightening. If the hydrodynamical grid were large enough to
contain several spirals of the WCR around the stars one would expect
to see an expanding series of progressively fainter, concentric,
edges. The unshocked winds emission at 1000\,GHz arises from a region
very close to the surface of the stars.

Some very interesting features arise in the synthetic images as the 
phase of the orbit advances. From the observer's point-of-view, the
foreground star slides to the right, which at $15-100$\,GHz exposes 
more of the brighter WCR to the left by phase 0.125.  
At higher frequencies the winds are brighter than the WCR. At phase 0.25
(quadrature) the apex of the WCR is seen side on, with the unshocked
winds to either side. The brightest part of the WCR is between the
stars as expected, and is visible in these images even at 1\,THz. At
low frequencies stronger absorption of the WCR occurs on the
right-hand-side of the image due to the greater depth of intervening
wind material. The right-hand column of Fig.~\ref{fig:cwb2_radio_image2}
shows the system viewed at phase 0.375, halfway between quadrature 
and conjunction. The image now displays a prominent double-helix-like
structure formed through the limb-brightening of the WCR. The vertical
curvature of the WCR induced by the increasing phase-lag of shocked gas 
with height/depth from the orbital plane is strikingly displayed (see also
the warped shock surfaces of the WCR shown in Paper~I). At low frequencies,
this double-helix-like structure is partially obscured by the 
foreground wind (now to the left of the image).

Fig.~\ref{fig:cwb2_radio_image3} displays images obtained for an
observer with an inclination angle $i=30^{\circ}$. The brightest parts
of the WCR are again those which are limb-brightened, with the overall
morphology of the WCR being shaped like an ``S''. In some of the
images displayed (e.g. the middle-left image with $\nu=100$\,GHz and
$\phi = 0^{\circ}$) the arms of the WCR do not trace back to a central
point, due to the additional emission from the winds. This figure also
demonstrates that the observed shape of the WCR, particularly at the
lower frequencies, owes much to the location of any foreground
absorber (i.e. a wind), and is not necessarily a good indicator of the
actual shape and position of the WCR \citep[a point made already
in][]{Dougherty:2003}.

Lightcurves of the thermal emission from model cwb2 are shown in
Fig.~\ref{fig:cwb2_ffradio_lc}. Because of the larger size of the
computational volume, lightcurves at 15 and 22\,GHz are also
displayed. The lightcurves of the total observed emission differ
distinctly in their shape compared to those from model cwb1 (left
panels of Fig.~\ref{fig:cwb1_ffradio_lc}) - those from model cwb2
display maxima which are less broad. This reflects the differences in
the optical depth through the WCR in the two models.
Fig.~\ref{fig:cwb2_radio_spectrum} illustrates that the emission from
the WCR of model cwb2 becomes optically thin at relatively low
frequencies, in contrast to that of model cwb1 where it remains
optically thick to $\nu \gtsimm 1000$\,GHz.

The free-free emission from the unshocked winds at 43\,GHz is still
substantially below the combined emission from the winds of two single
stars (panel h in Fig.~\ref{fig:cwb2_ffradio_lc}), but compared to
model cwb1 (see Fig.~\ref{fig:cwb1_ffradio_lc}b) the flux from model
cwb2 is roughly twice as high.  This reflects both the larger
computational domain in model cwb2, which means more of the emission
at 43\,GHz is captured, plus the wider separation of the stars i.e.
the WCR removes a smaller segment of the 43\,GHz emitting region
from the unshocked winds.  Once again, it is only by 250\,GHz that the
flux from the unshocked winds is comparable to twice the flux from a
single star.  And, as was also the case for model cwb1, by 1000\,GHz
the unshocked wind flux exceeds twice the single star value (due to
the emission probing deep into the acceleration region of the winds).
However, the 1000\,GHz lightcurve of unshocked wind emission from
model cwb2 (Fig.~\ref{fig:cwb2_ffradio_lc}q) is almost completely
constant (the main exception being a triangular shaped minimum
centered on each conjunction), in stark contrast to the greater
variability seen from model cwb1 (Fig.~\ref{fig:cwb1_ffradio_lc}k).
The minimum in Fig.~\ref{fig:cwb2_ffradio_lc}(q) fails to approach 50
per cent of the unshocked wind flux. This is expected, since
two perfectly symmetric winds around stars in conjunction will
not completely occult, because wind emission from the background star
will not be completely absorbed for impact parameters where the winds start to
get optically thin. Therefore, a perfect occultation by two stars with
stellar winds is not possible, and hence a dip of 50 per cent is never
reached. An additional (smaller) part may be played by the fact that the
winds are no longer spherically symmetric, since their acceleration
is inhibited towards each other, but is boosted in other directions.  A
slight lag in the phase of the minima relative to conjunction
is increasingly apparent at lower frequencies.

\begin{figure}
\begin{center}
\psfig{figure=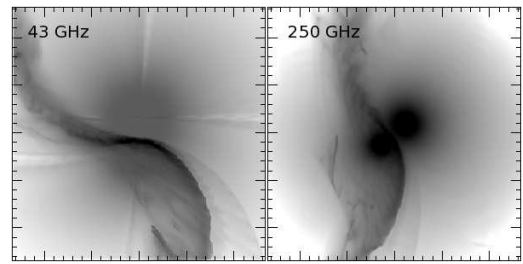,width=6.8cm}
\end{center}
\caption[]{Intensity images from model cwb3. The left panel shows the
emission at 43\,GHz with $i=0^{\circ}$. The right panel displays the
emission at 250\,GHz with $i=30^{\circ}$ and $\phi = 45^{\circ}$
(phase 0.875). The maximum intensity is
$1.6\times10^{-10}\,{\rm erg\,cm^{-2}\,s^{-1}\,Hz^{-1}\,ster^{-1}}$ and
the major ticks on each axis mark out 0.5\,mas. $\phi=0^{\circ}$
corresponds to the O8V star in front.}
\label{fig:cwb3_radio_image}
\end{figure}

Now consider the observed emission arising from the WCR (see the right
column of Fig.~\ref{fig:cwb2_ffradio_lc}). A maximum in the emission
again occurs near quadrature, just as in model cwb1, though this time
for a totally different reason. In model cwb2, maxima occur when the
line of sight of the observer to the apex of the WCR (where the
intrinsic emissivity is greatest) is through one of the low-opacity
arms of the WCR. This contrasts with model cwb1, where maxima occur
when the optically thick WCR is oriented almost face on.

Although the free-free emissivity per unit volume in the WCR is much
smaller in model cwb2 than in cwb1 due to the reduced densities and
higher temperatures, the optical depth within this material is also
reduced.  The net result is that the observable flux from the WCR at
$\nu \ltsimm 50$\,GHz is much higher in model cwb2 than in model
cwb1. But above $50$\,GHz, when the opacity through the circumstellar
environment is low, the reduced densities in the WCR limit the
intrinsic emission compared to model cwb1, so that the observed flux
from the WCR in model cwb1 exceeds that from model cwb2 (compare
Figs.~\ref{fig:cwb1_radio_spectrum} and~\ref{fig:cwb2_radio_spectrum}).

At 43\,GHz the free-free flux from the WCR varies by a factor of 2 for
observers with a $90^{\circ}$ inclination
(Fig.~\ref{fig:cwb2_ffradio_lc}i).  At lower frequencies the variation
is even more pronounced, as the relative contribution of free-free
emission from the WCR becomes increasingly dominant (the free-free
flux from the winds scales as $\nu^{+0.6}$, whereas it scales as
$\nu^{-0.1}$ from the optically thin WCR).

Interestingly, for $\nu \leq 43$\,GHz, the cwb2 WCR lightcurves
display a remarkable resemblance to the $0.1-0.5$\,keV X-ray
lightcurves from this model shown in Paper~III, including the
asymmetrical slope of the minimum and the faster rise to maximum than
the fall from it.  Moreover, the 100 and 250\,GHz WCR lightcurves
(panels l and o in Fig.~\ref{fig:cwb2_ffradio_lc}) are similar to the
$0.5-2.5$\,keV X-ray lightcurve. On the other hand, the $2.5-10$\,keV
X-ray lightcurve is most similar to the 43 and 100\,GHz lightcurves
from the unshocked wind emission (see panels h and k in 
Fig.~\ref{fig:cwb2_ffradio_lc}).

Radio-to-sub-mm spectra for a variety of viewing orientations are
shown in Fig.~\ref{fig:cwb2_radio_spectrum}. The emission from the WCR
produces a ``bump'' on top of the unshocked winds spectrum in a manner
which is not too dissimilar to a weak synchrotron source. Note,
however, that this work {\em only} considers the {\em free-free}
emission from the WCR.  Such behaviour was previously noted in
\citet{Pittard:2006}.  It is interesting that \citet{Contreras:1996}
purposefully observed a sample of O and WR stars at 43\,GHz to
minimize the effects of non-thermal emission on determinations of
mass-loss rates. In this respect Fig.~\ref{fig:cwb2_radio_spectrum}
shows that even at 43\,GHz the contribution to the total flux due to
{\em thermal} emission from the WCR may not be negligble, complicating
this goal.

Nevertheless, at high frequencies the thermal emission from the
unshocked winds always dominates the emission. This occurs at $\nu
\gtsimm 100$\,GHz in model cwb2.  Naively, therefore, one might expect
that sub-mm observations may be preferred to radio observations in
respect of determining the mass-loss rates of the winds. However, 
there is then the additional complication that the acceleration regions of
the winds may be being probed. Another problem is that the effects
of clumping in the winds may be more severe at mm wavelengths than
in the radio \citep[see e.g.][]{Runacres:1996,Blomme:2002}. Moreover,
the flux from sub-mm observations is typically not as well calibrated as 
the flux from radio observations.

The lower panels of Fig.~\ref{fig:cwb2_radio_spectrum} show the
spectral indices as a function of frequency. $\alpha_{\rm winds}$ (as
shown by the pink line) is consistently around +1.0, irrespective of
the orientation of the system to the observer. At lower frequencies
$\alpha_{\rm winds}$ is slightly larger, though this may simply
reflect some of the flux loss which occurs from the model due to the
finite grid size.  At high frequencies the emission from the WCR has a
spectral index $\alpha_{\rm WCR} \approx -0.1$. $\alpha_{\rm WCR}$
increases at lower frequencies as the circumstellar environment
absorbs some of the intrinsic free-free emission from the WCR. The
spectral index of the total free-free emission, $\alpha$, lies between
$\alpha_{\rm winds}$ and $\alpha_{\rm WCR}$. Between 43 and 1000\,GHz,
$\alpha = +0.69$, +0.67, and +0.65 in
Figs.~\ref{fig:cwb2_radio_spectrum}(d)-(f).  These are comparable to
the values observed from some stars
\citep[e.g.][]{Williams:1990,Leitherer:1991,Altenhoff:1994,Nugis:1998}.
However, the spectral index from the unshocked winds over this
frequency range is steeper ($\alpha_{\rm winds} = +0.93$, +0.99, and
+1.00), being offset by the flatter spectra from the WCR emission
($\alpha_{\rm WCR} = +0.16$, +0.00, and $-0.04$, in
Figs.~\ref{fig:cwb2_radio_spectrum}(d)-(f), respectively).

The variation of the total thermal emission for an observer in the
orbital plane between conjunction and quadrature is shown in
Fig.~\ref{fig:cwb123_radio_spectrum_var}(b). 
The emission at quadrature is dominant over the emission at
conjunction over the entire frequency range considered.

\begin{figure*}
\psfig{figure=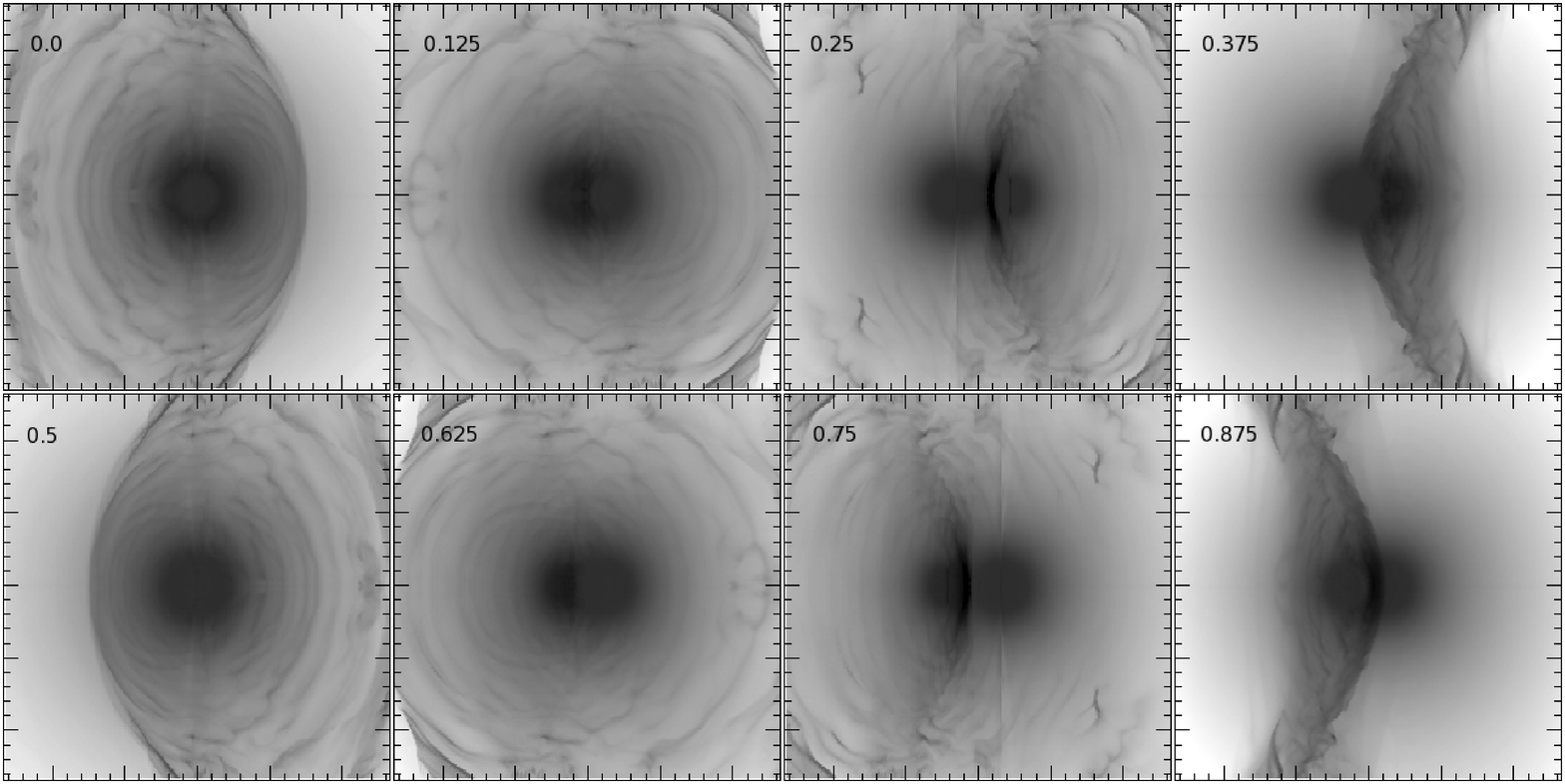,width=13.6cm}
\caption[]{Intensity images from model cwb3 at 100\,GHz and
$i=90^{\circ}$. From top to bottom and left to right the orbital phase
is 0.0 ($\phi = 0^{\circ}$, conjunction, O8V star in front), 0.125, 
0.25 ($\phi=270^{\circ}$, quadrature), 0.375, 0.5 
($\phi=180^{\circ}$, conjunction, O6V star in front), 0.625, 0.75
($\phi = 90^{\circ}$, quadrature), and 0.875. The maximum intensity is
$1.6\times10^{-10}\,{\rm erg\,cm^{-2}\,s^{-1}\,Hz^{-1}\,ster^{-1}}$ and
the major ticks on each axis mark out 0.5\,mas.}
\label{fig:cwb3_radio_image2}
\end{figure*}

\begin{figure*}
\psfig{figure=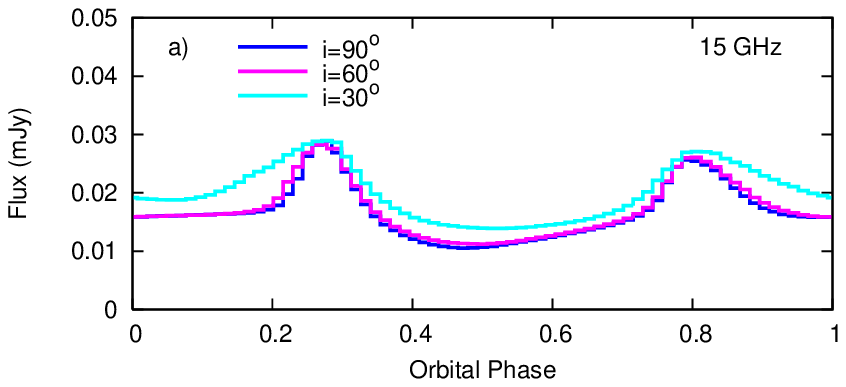,width=5.67cm}
\psfig{figure=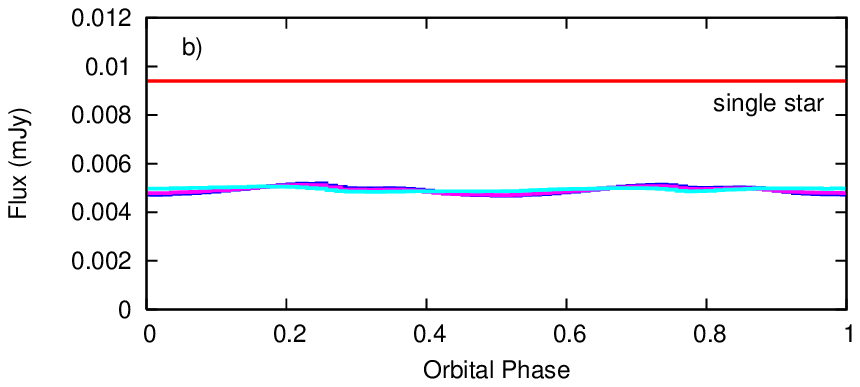,width=5.67cm}
\psfig{figure=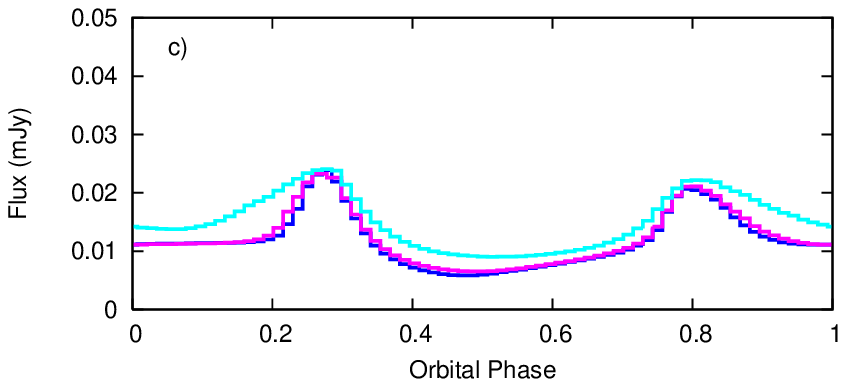,width=5.67cm}
\psfig{figure=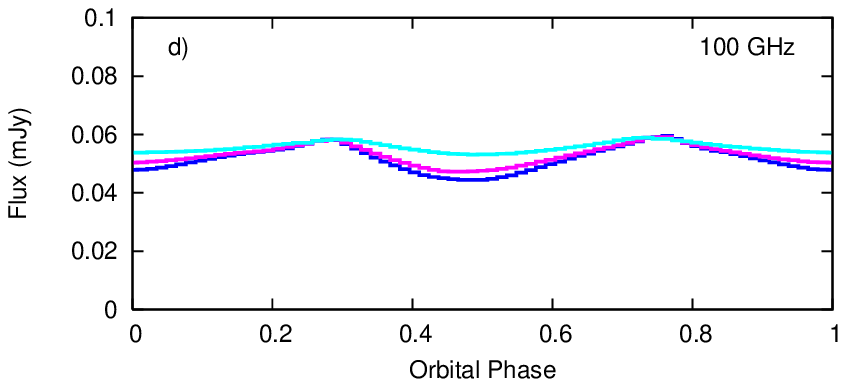,width=5.67cm}
\psfig{figure=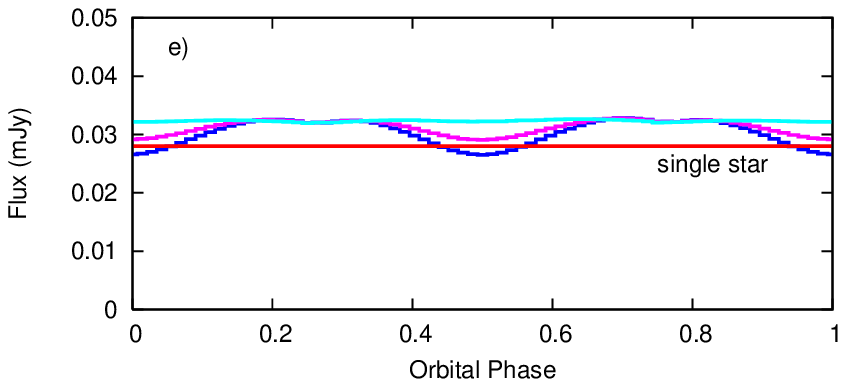,width=5.67cm}
\psfig{figure=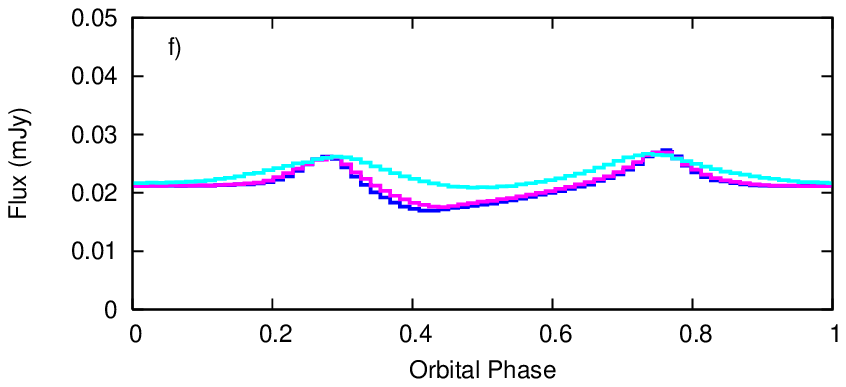,width=5.67cm}
\psfig{figure=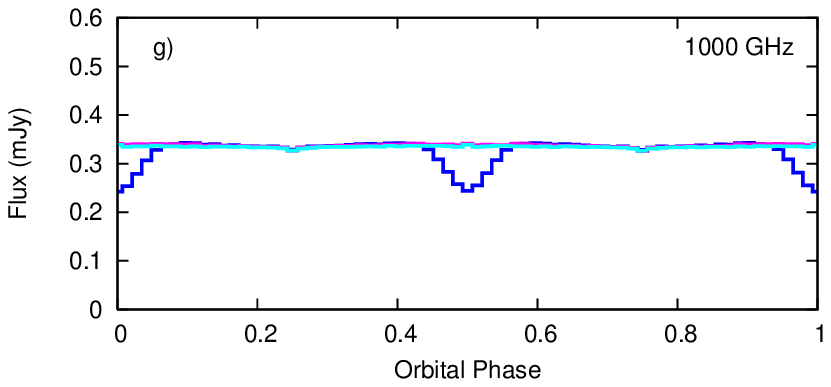,width=5.67cm}
\psfig{figure=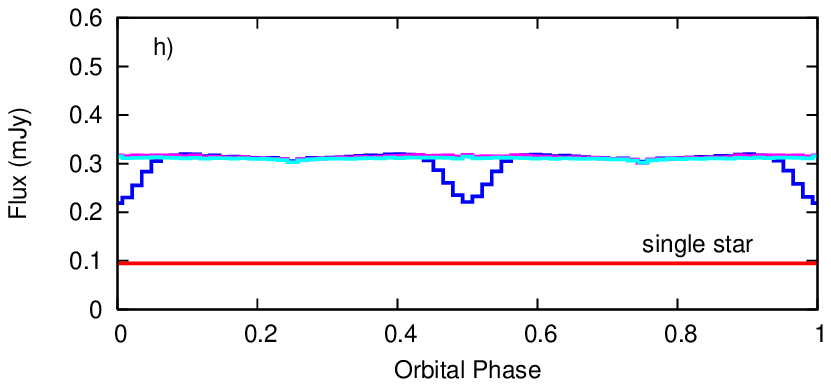,width=5.67cm}
\psfig{figure=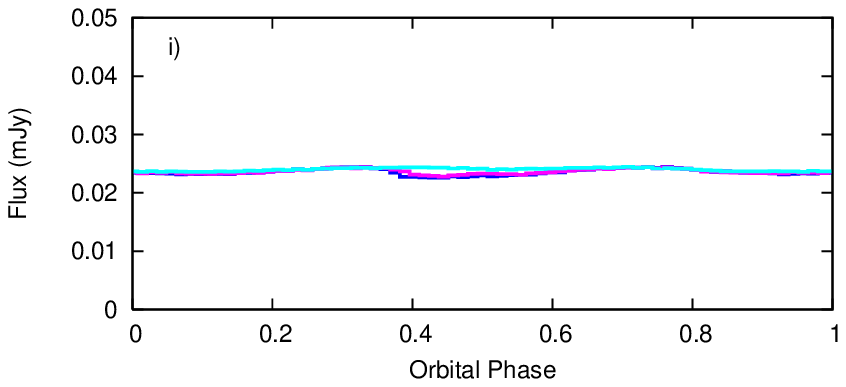,width=5.67cm}
\caption[]{As Fig.~\ref{fig:cwb2_ffradio_lc} but for model cwb3, at
$\nu = 15$, 100, and 1000\,GHz. The theoretical flux from a single,
terminal velocity, wind from an O6V star is again displayed: the
flux from the wind of the O8V companion is 73 per cent of this value.
Note that the y-axes (flux scales) are different in each panel.}
\label{fig:cwb3_ffradio_lc}
\end{figure*}

\begin{figure*}
\psfig{figure=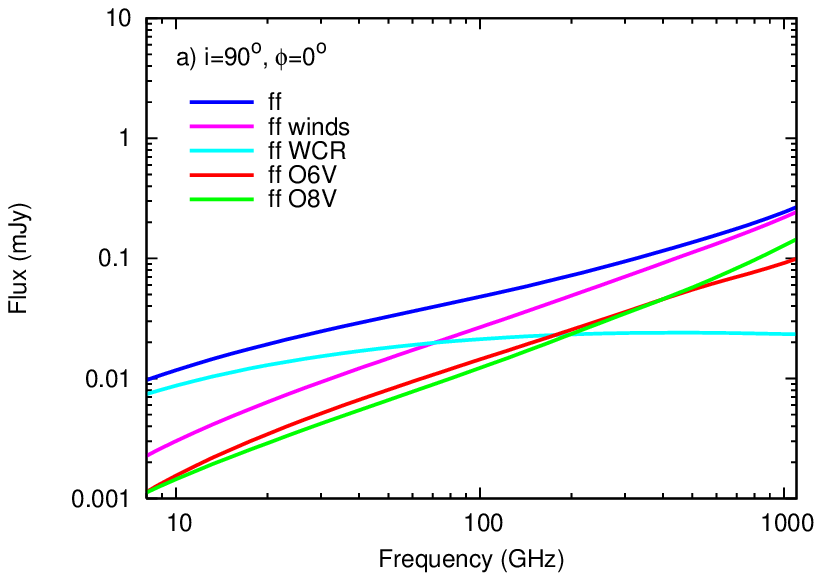,width=5.67cm}
\psfig{figure=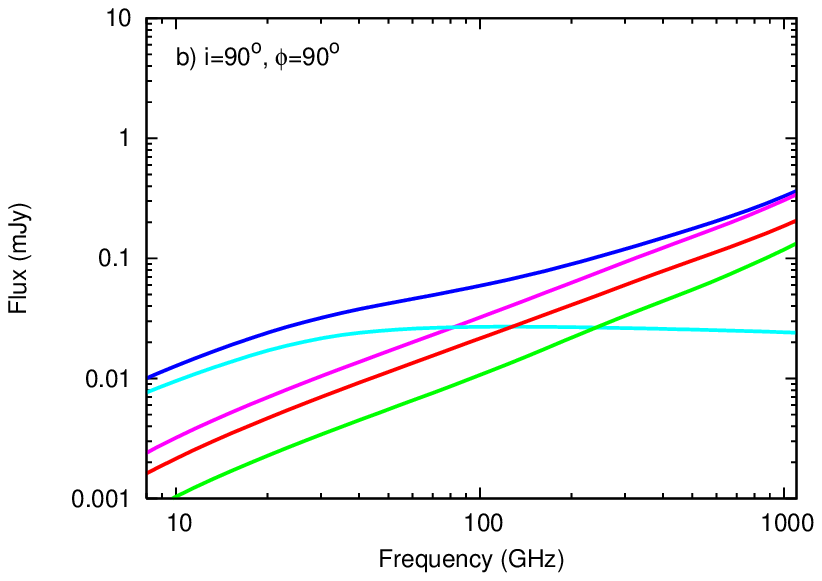,width=5.67cm}
\psfig{figure=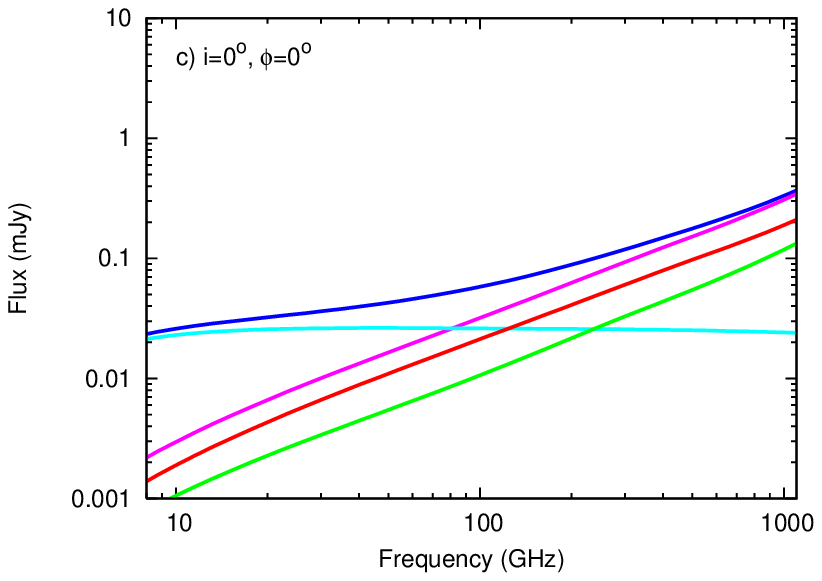,width=5.67cm}
\psfig{figure=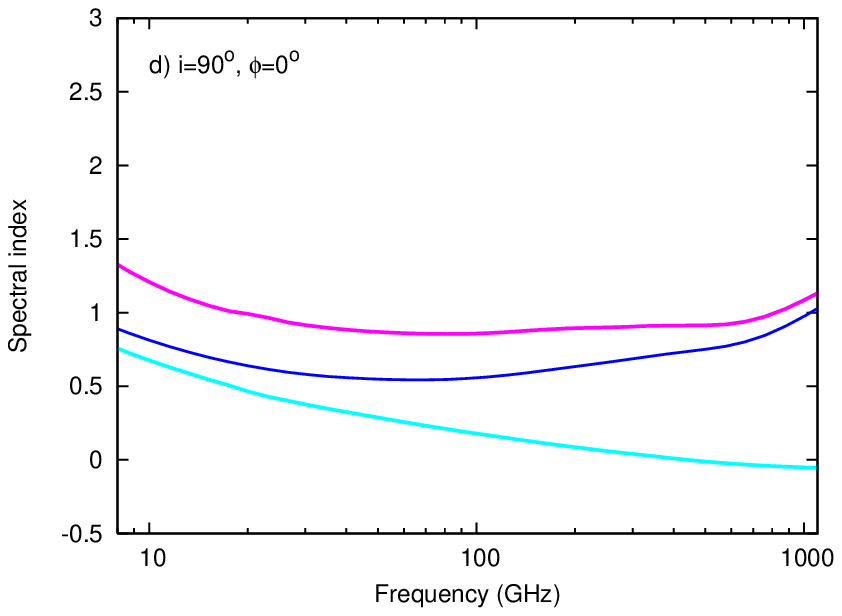,width=5.67cm}
\psfig{figure=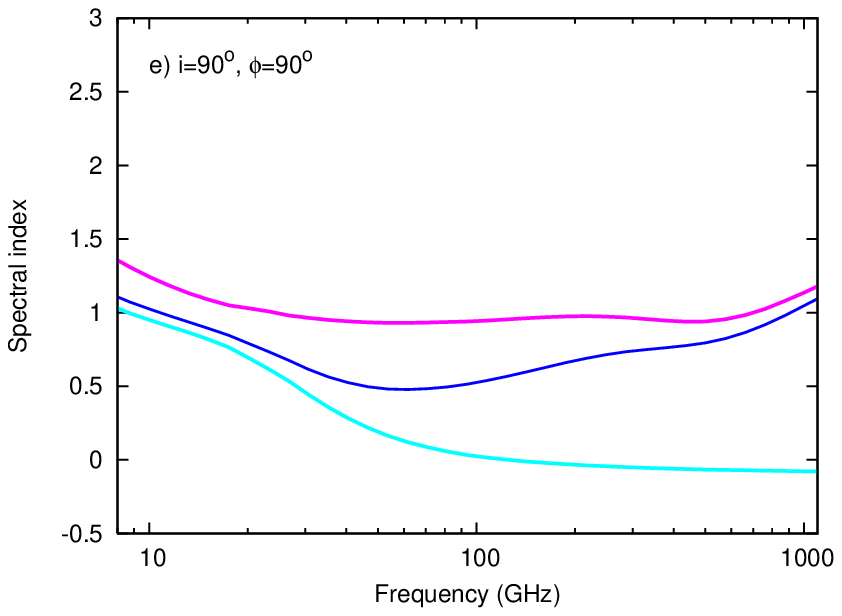,width=5.67cm}
\psfig{figure=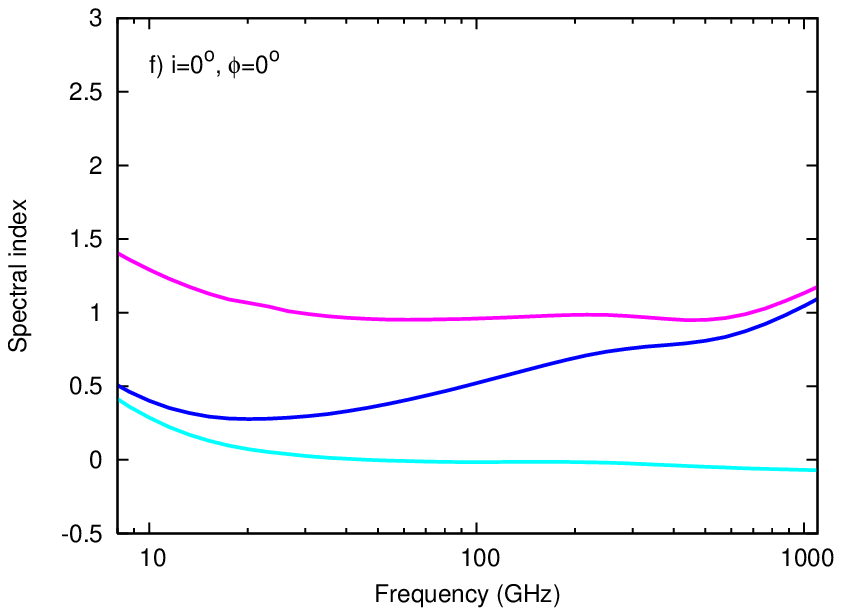,width=5.67cm}
\caption[]{As Fig.~\ref{fig:cwb1_radio_spectrum} but for model
cwb3. Since the stars differ in this model, the individual flux
contributions of the winds from the O6V and O8V stars are plotted in
the top panels. In a) and d) the O8V star is in front. Because of the
asymmetry of the WCR the spectra at phases 0.25 and 0.75 are
different, unlike the case in Fig.~\ref{fig:cwb1_radio_spectrum}.}
\label{fig:cwb3_radio_spectrum}
\end{figure*}

\subsubsection{Model cwb3}
The left panel of Fig.~\ref{fig:cwb3_radio_image} shows intensity
images of the free-free emission from model cwb3 (an unequal winds
simulation) at 43\,GHz for an observer directly above the orbital
plane. A comparison against the 43\,GHz synthetic image in
Fig.~\ref{fig:cwb2_radio_image} reveals several differences. Firstly,
the position of maximum brightness (at the apex of the WCR) is
different in the two images: in Fig.~\ref{fig:cwb3_radio_image}, the
apex of the WCR is pushed closer to the O8V star (which is the
southern star in this image). Secondly, the emission from the
downstream arms of the WCR is also in different locations, reflecting
the changed positions of ram-pressure balance between the two
winds. Finally, there is a greater emission contrast between the
leading and trailing edges of the leading arm of the WCR, which
reflects the changes in the underlying density distribution (see
Paper~I). Conversely, there is also a reduced emission contrast
between the leading and trailing edges of the trailing arm, compared
to that from model cwb2 (again see Paper~I for an explanation of the
density distributions within the WCRs of these models).

The right panel in Fig.~\ref{fig:cwb3_radio_image} shows a synthetic
image at 250\,GHz for an observer with $i=30^{\circ}$ and
$\phi=45^{\circ}$.  The greater emission from the O6V wind is clearly
visible, along with the ``S''-shaped emission from the WCR.

Fig.~\ref{fig:cwb3_radio_image2} shows intensity images of the
free-free emission from model cwb3 at 100\,GHz as a function of phase
for an observer located in the orbital plane ($i =
90^{\circ}$). Again, significant differences are apparent compared to
the 100\,GHz images from model cwb2 (see
Fig.~\ref{fig:cwb2_radio_image2}). At conjunction with the weaker wind
from the O8V star in front ($\phi=0^{\circ}$), there are two main
differences. First, the limb brightened edge on the leading arm (to
the right side of the image) is projected closer to the centre of the
image and shows greater curvature, while the corresponding region of
limb brightening on the trailing arm is located off the side of the
image and is not visible. Second, the reduced radius of the optical
depth unity surface in the O8V wind creates a smaller silhouette
against the brighter WCR than occurs in model cwb2.

Differences also occur at other viewing angles. At phase 0.125
($\phi=315^{\circ}$), the WCR is seen mostly face on, with the
foreground O8V wind silhouetted against it (slightly to the right of
centre). The vertical curvature of the WCR is again clear at phase
0.25 (quadrature), reflecting the wrapping of the WCR around the O8V
star (to the right of centre) due to the stronger wind from the O6V
star (to the left of centre). The apex of the WCR, which occurs closer
to the O8V star (reflecting the position of the ram pressure balance
of the winds), is seen sideways on. The greater emission from the O6V
wind, relative to that from the O8V wind, is also clearly visible. The
double-helix-like structure seen from model cwb2 at phase 0.375
disappears in model cwb3. Instead, the brightest regions of the
projected emission from the WCR primarily traces the limb-brightened,
dense, O8V gas on the trailing edge of the leading arm (see Paper~I).

The images from the second half of the orbit ($0.5 \leq$\,phase\,$\leq
1.0$) resemble those in the upper row of
Fig.~\ref{fig:cwb3_radio_image2} which show the first half of the
orbit.  However, some small differences are seen, which lead also to
differences in the lightcurves. For example, the conjunction at phase
0.5 now has the denser O6V wind in the foreground, whereas the weaker
O8V wind was in front of the WCR at phase 0.0.  This changes the
nature of the silhouette.

Fig.~\ref{fig:cwb3_ffradio_lc} shows lightcurves at a range of
frequencies from radio to sub-mm of the free-free emission from model
cwb3. The lightcurves are reasonably similar to those shown in
Fig.~\ref{fig:cwb2_ffradio_lc} from model cwb2, so we show a smaller
sample of frequencies. The slight differences in the total flux and
the morphology of the variation have their origin in the slighter
weaker secondary wind in model cwb3.  The maxima now show asymmetrical
peaks due to the different wind strengths. The highest peak occurs
when viewing down the trailing arm. This is straighter than the
leading arm, as the curvature of the WCR opposes the coriolis-induced
curvature, and allows more flux to escape through the low opacity
plasma within the WCR.

Fig.~\ref{fig:cwb3_radio_spectrum} shows radio-to-sub-mm spectra from
model cwb3 for three different viewing orientations. Strong similarities
to the results from model cwb2 are again seen. Since model cwb3
consists of O6V and O8V stars, rather than two identical O6V stars as
in the other models, we also plot in Fig.~\ref{fig:cwb3_radio_spectrum} the
individual contributions of the unshocked winds to the total thermal
flux. As expected, the flux from the denser O6V wind usually dominates that
from the O8V wind (see Fig.~\ref{fig:cwb3_radio_spectrum}b and c).
However, for favourable orientations (e.g. when the O8V star is directly
in front of the O6V star), its contribution to the total thermal
flux can be comparable to, and at some frequencies exceed, 
the contribution from the O6V wind (Fig.~\ref{fig:cwb3_radio_spectrum}a).

The spectral index as a function of frequency is shown in
Fig.~\ref{fig:cwb3_radio_spectrum}(d)-(f). The values of $\alpha$ are
very similar to those in Fig.~\ref{fig:cwb2_radio_spectrum}(d)-(f) from
model cwb2. Note that the high
frequency spectral index from the O8V wind exceeds that from the O6V wind.
This is because the emission probes deeper into
the acceleration region of the O8V wind at a given (high) frequency.

Fig.~\ref{fig:cwb123_radio_spectrum_var}(c) shows the variation of the
total thermal emission for an observer in the orbital plane between
conjunction (phase 0.0, O8V star in front) and quadrature (phase
0.75). Again, the behaviour is similar to that from model cwb2, though
the variation between phases is slightly weaker.  The flux, of course,
is also reduced.

\begin{figure*}
\psfig{figure=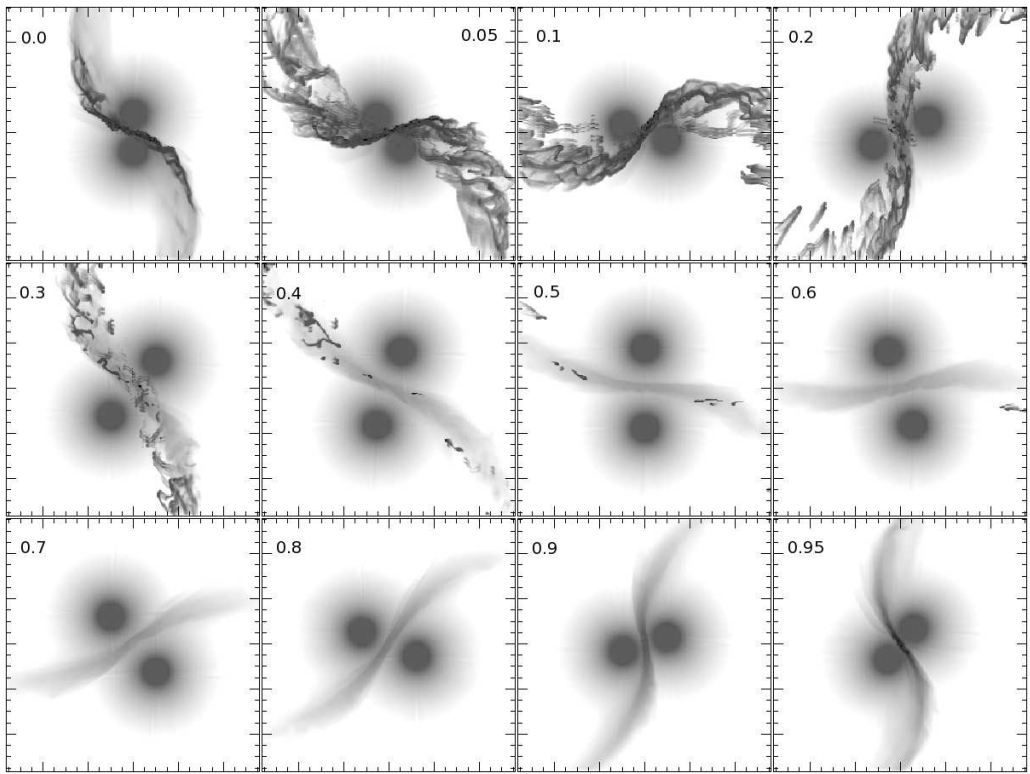,width=13.6cm}
\caption[]{Intensity images from model cwb4 at 1000\,GHz for an observer
with $i=0^{\circ}$. The orbital phase is noted on each panel.
The maximum intensity (black in the images) is
$8.8\times10^{-8}\,{\rm erg\,cm^{-2}\,s^{-1}\,Hz^{-1}\,ster^{-1}}$. 
The major ticks on each axis mark out 0.2\,mas.}
\label{fig:cwb4_radio_image}
\end{figure*}

\begin{figure*}
\psfig{figure=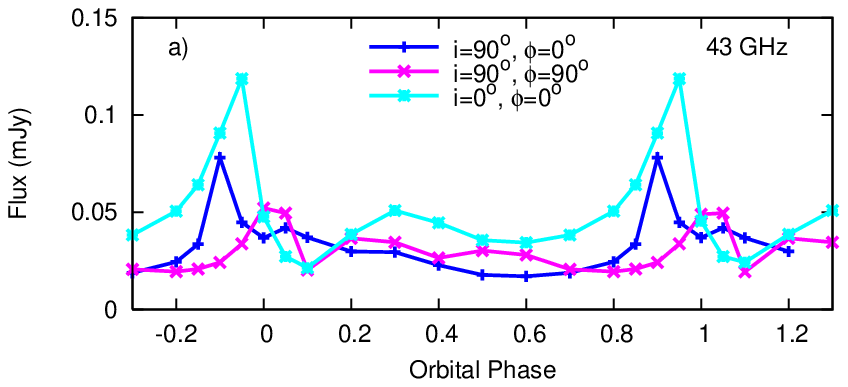,width=5.67cm}
\psfig{figure=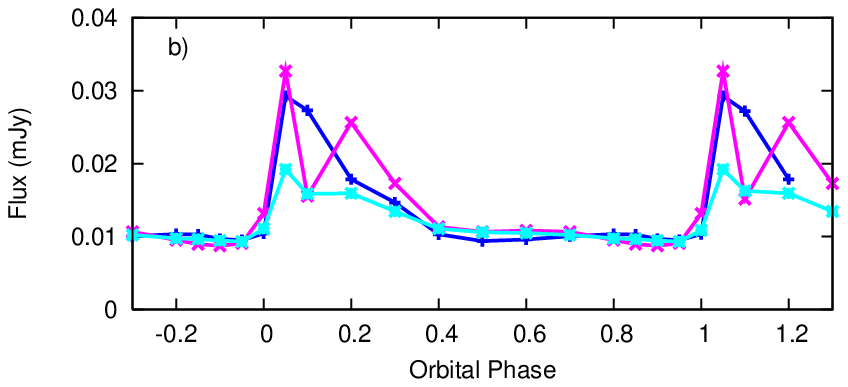,width=5.67cm}
\psfig{figure=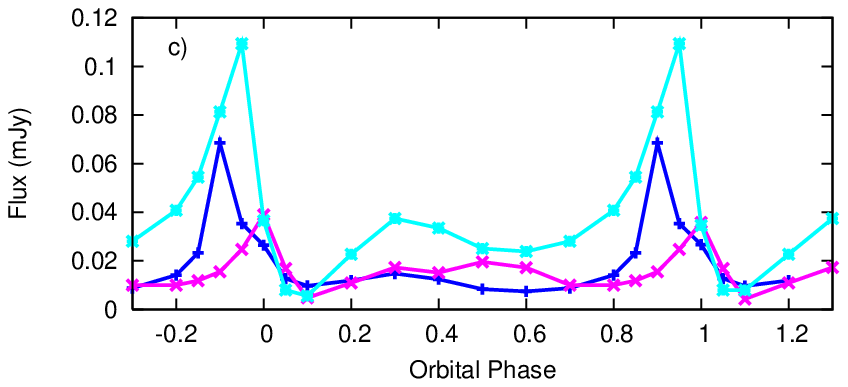,width=5.67cm}
\psfig{figure=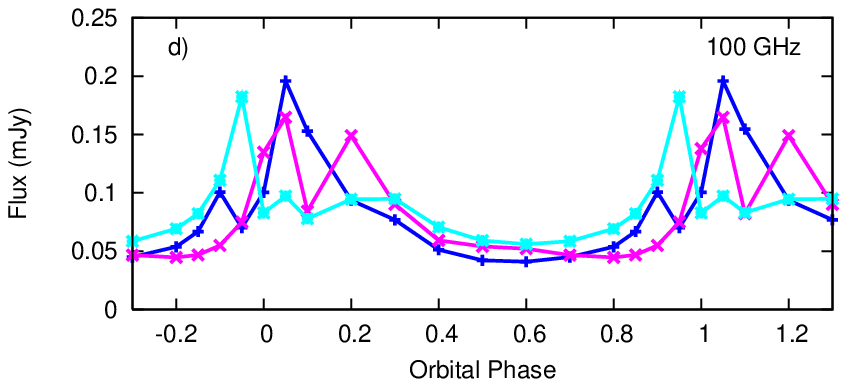,width=5.67cm}
\psfig{figure=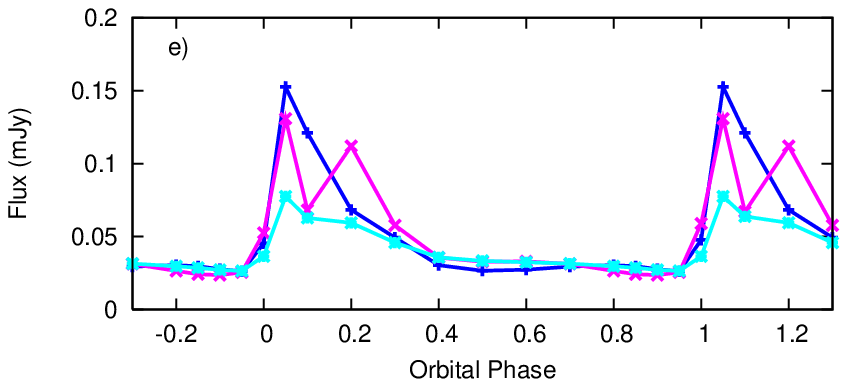,width=5.67cm}
\psfig{figure=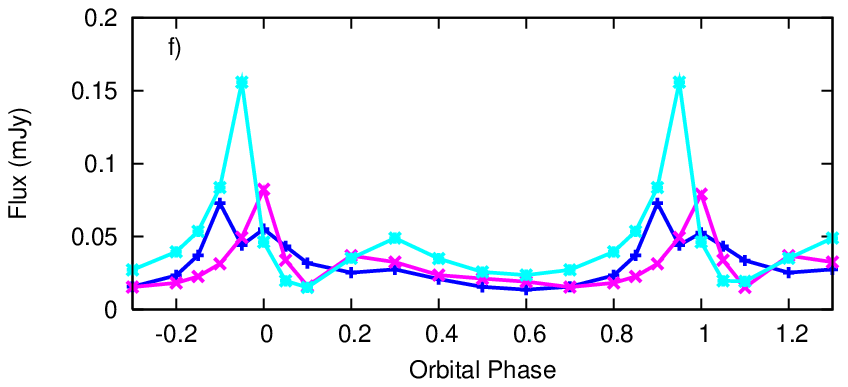,width=5.67cm}
\psfig{figure=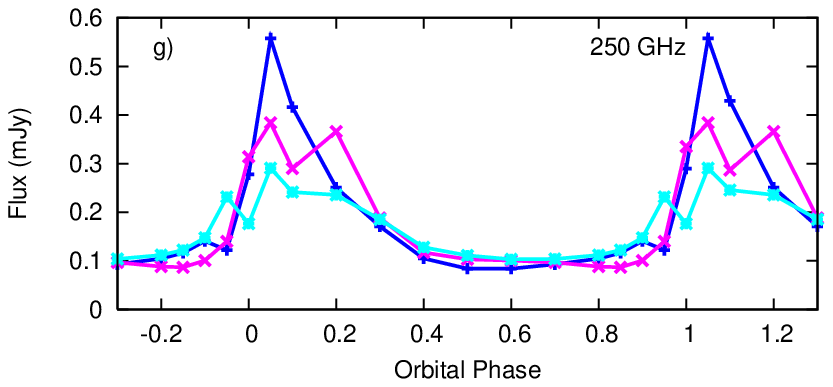,width=5.67cm}
\psfig{figure=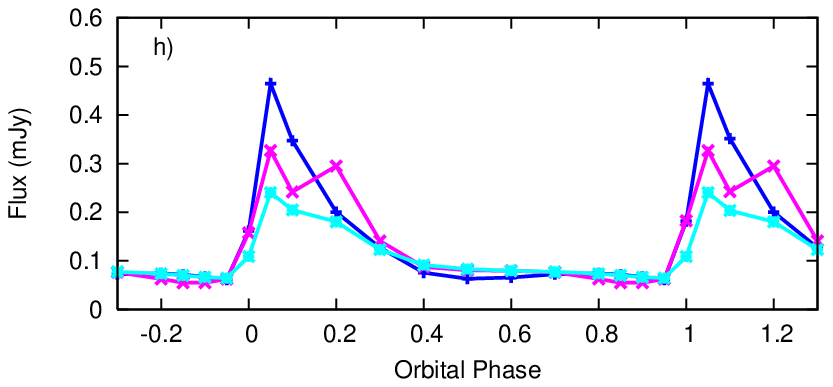,width=5.67cm}
\psfig{figure=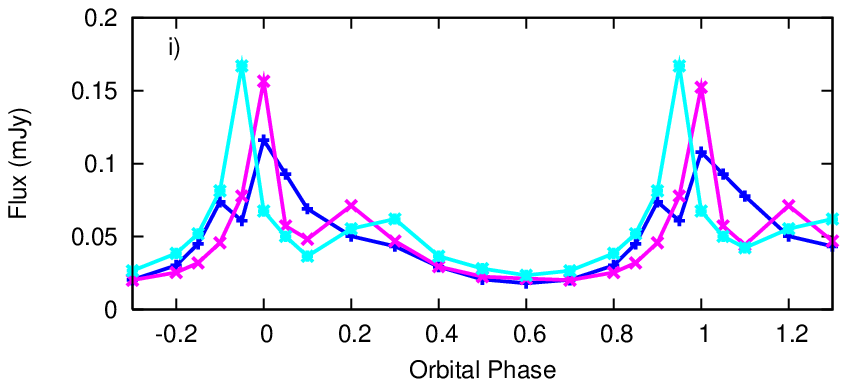,width=5.67cm}
\psfig{figure=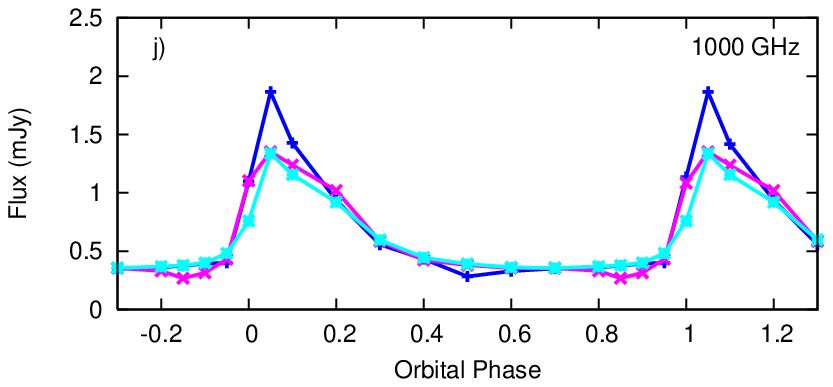,width=5.67cm}
\psfig{figure=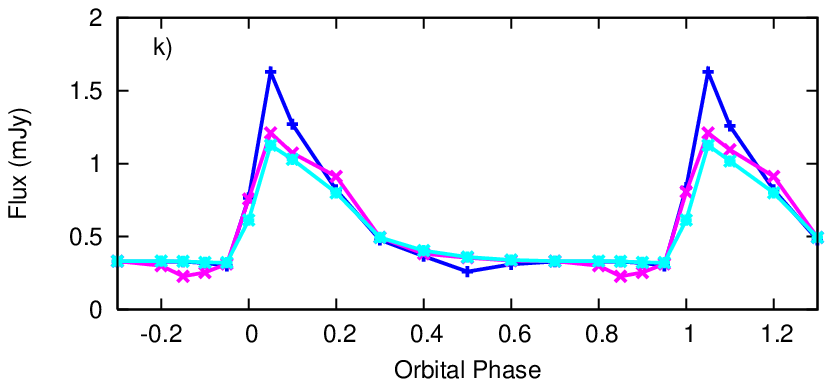,width=5.67cm}
\psfig{figure=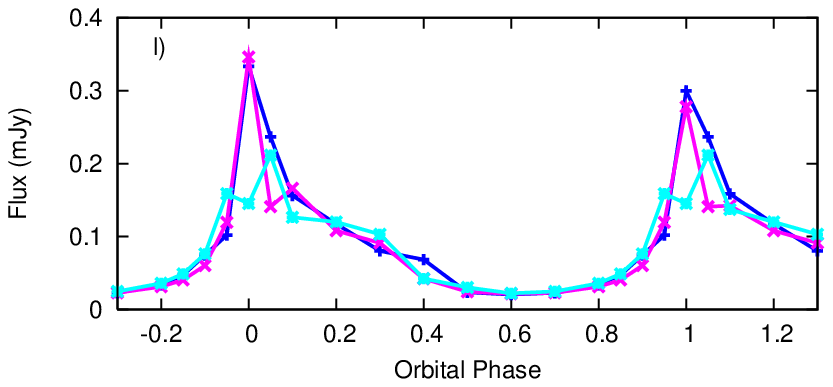,width=5.67cm}
\caption[]{Radio lightcurves of the free-free emission from model cwb4
at 43\,GHz (top), 100\,GHz, 250\,GHz, and 1000\,GHz (bottom). The
emission for three different viewing orientations is presented: in the
orbital plane ($i = 90^{\circ}$) with $\phi=0^{\circ}$ (blue) and
$\phi=90^{\circ}$ (pink), and directly above/below the orbital plane
($i = 0^{\circ}$, cyan).  For the blue curves the stars are at
conjunction at phases 0.0 and 0.5, and quadrature at phases 0.14 and
0.86. For the pink curves the stars are at conjunction at phases 0.14
and 0.86, and quadrature at phases 0.0 and 0.5.  The total free-free
emission is shown in the left panels, and the contributions from the
{\em cold} and {\em hot} gas in the middle and right panels,
respectively. Note that what is shown in the middle and right
panels differs from previous similar figures (see text for details). 
Note also that the y-axes (flux scales) are different in each panel.}
\label{fig:cwb4_ffradio_lc}
\end{figure*}

\begin{figure*}
\psfig{figure=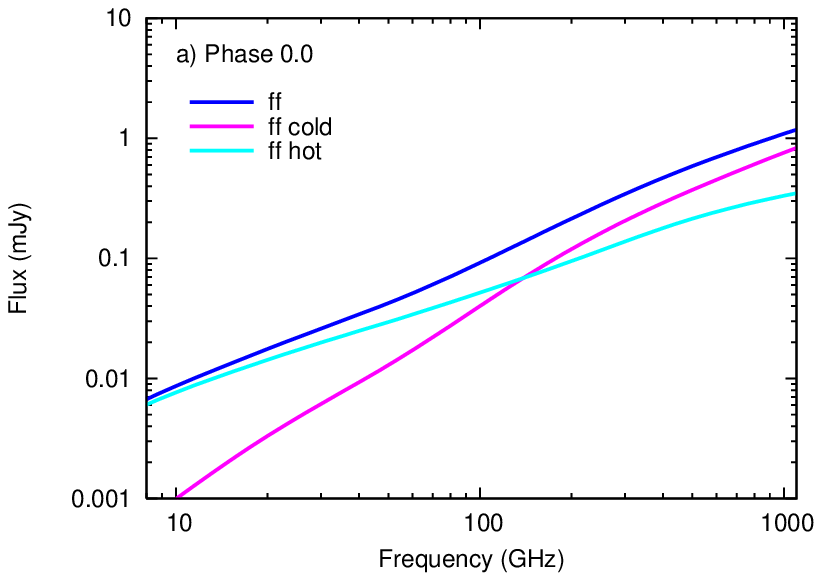,width=5.67cm}
\psfig{figure=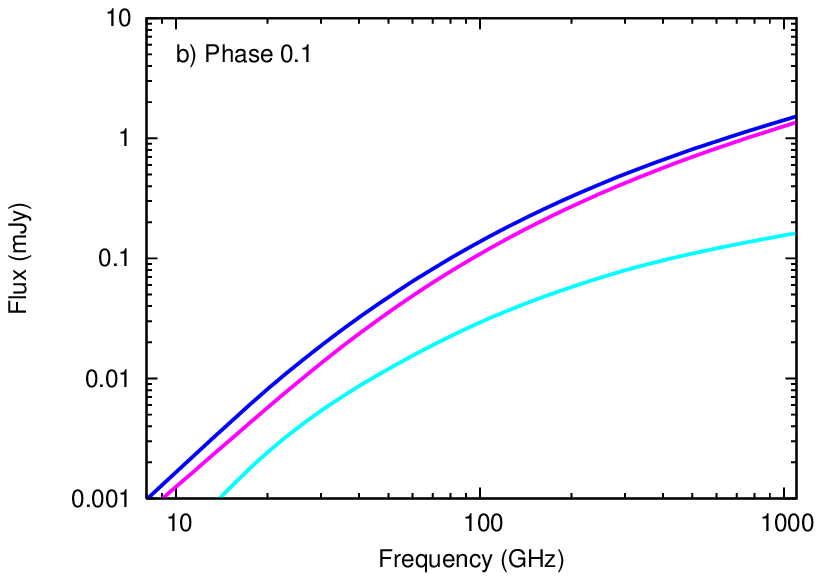,width=5.67cm}
\psfig{figure=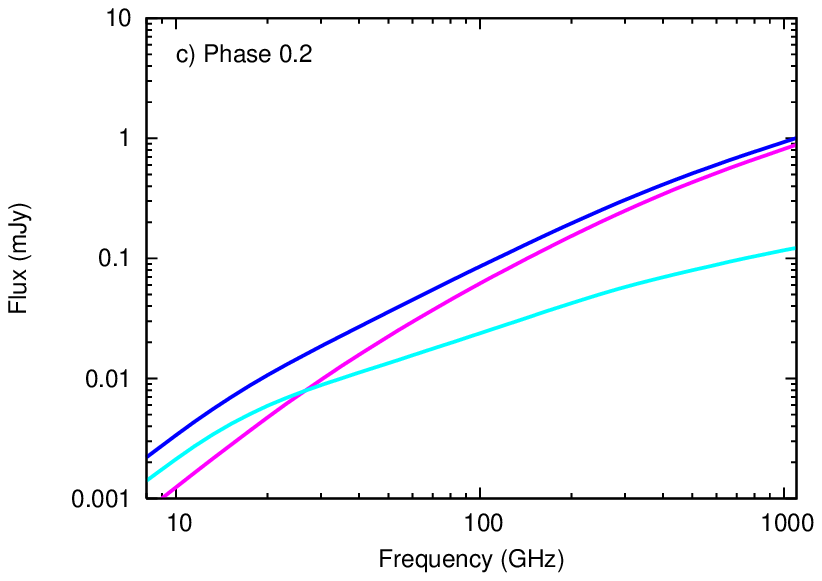,width=5.67cm}
\psfig{figure=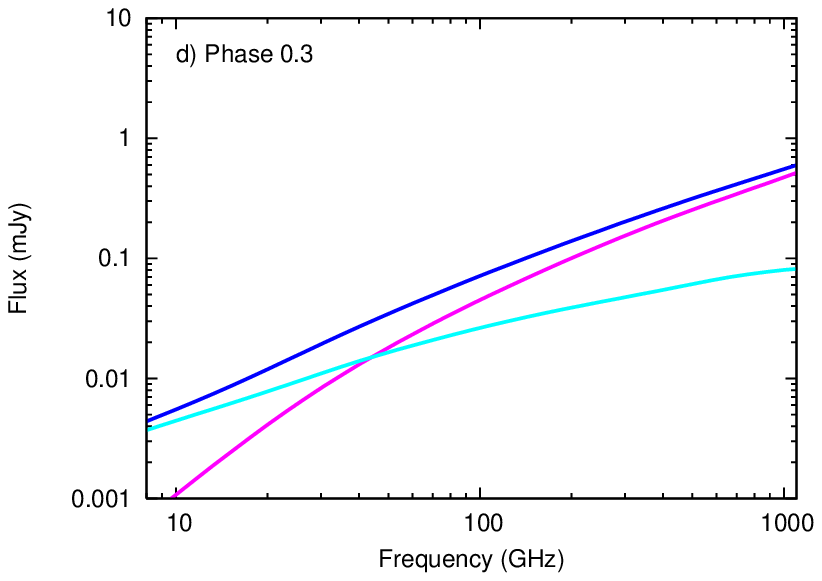,width=5.67cm}
\psfig{figure=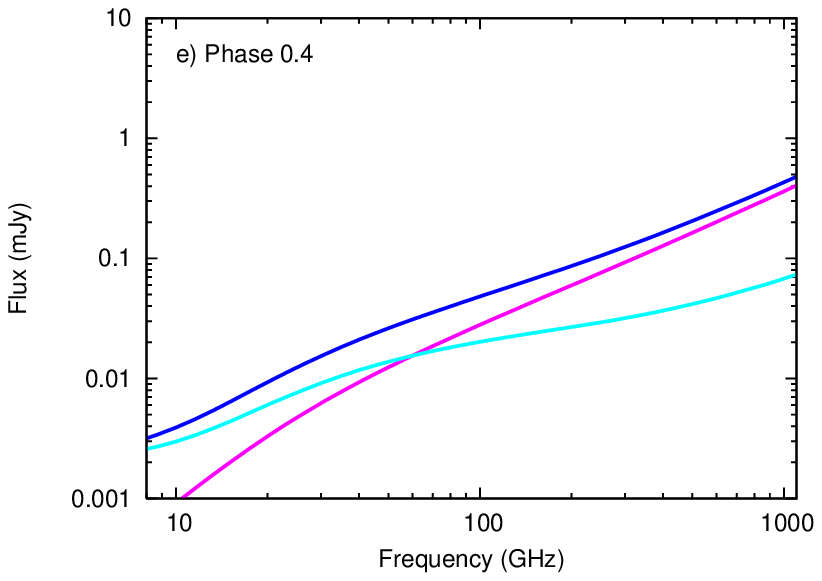,width=5.67cm}
\psfig{figure=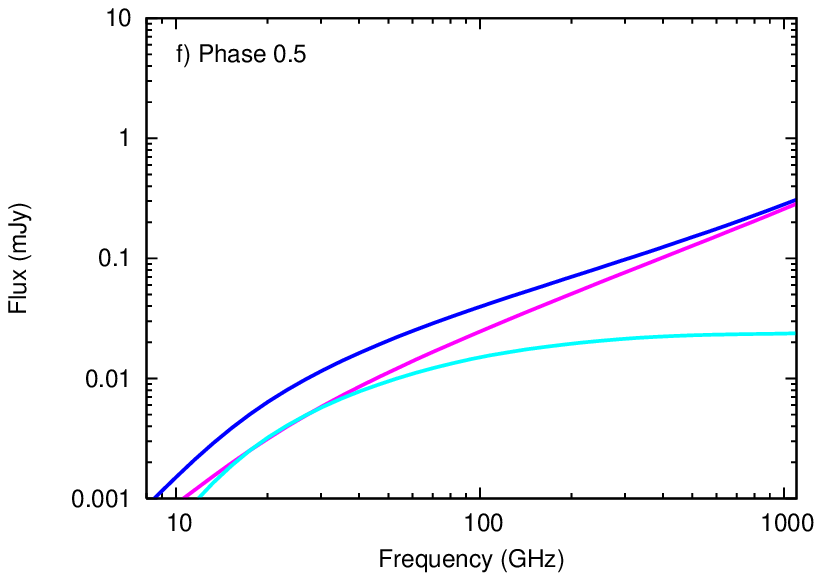,width=5.67cm}
\psfig{figure=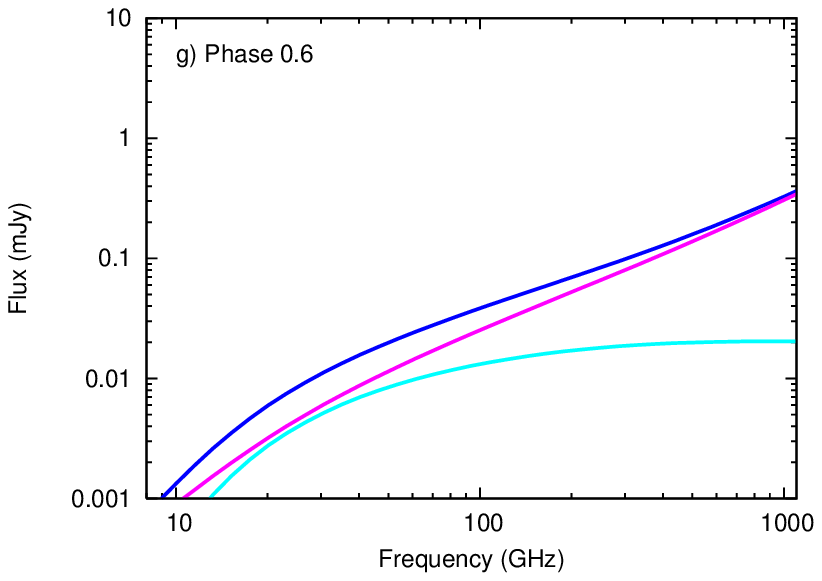,width=5.67cm}
\psfig{figure=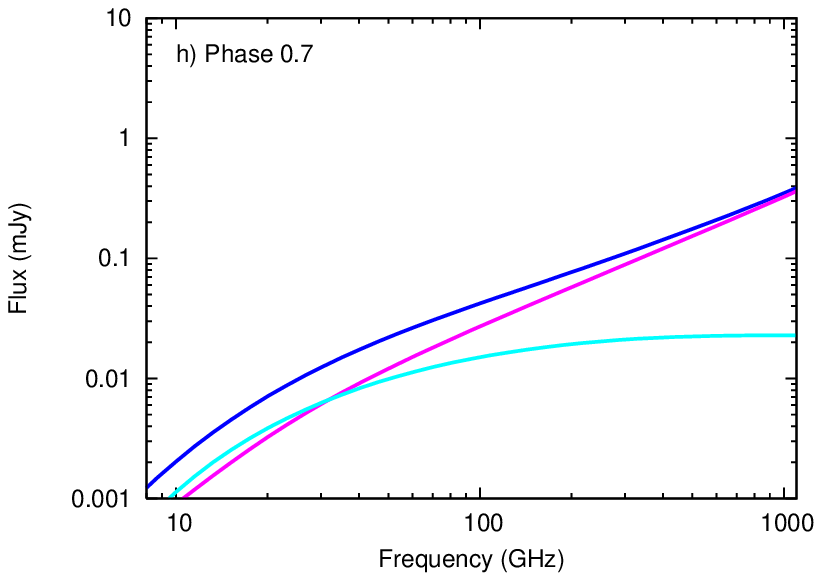,width=5.67cm}
\psfig{figure=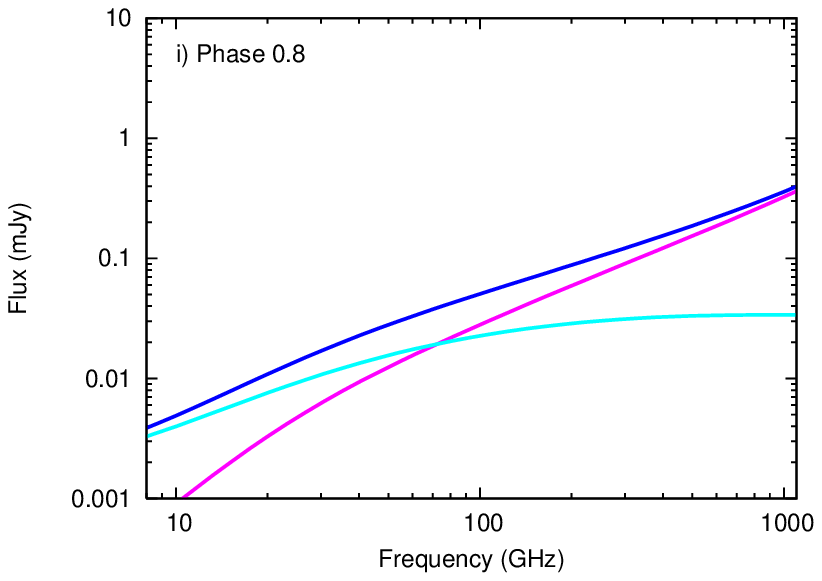,width=5.67cm}
\psfig{figure=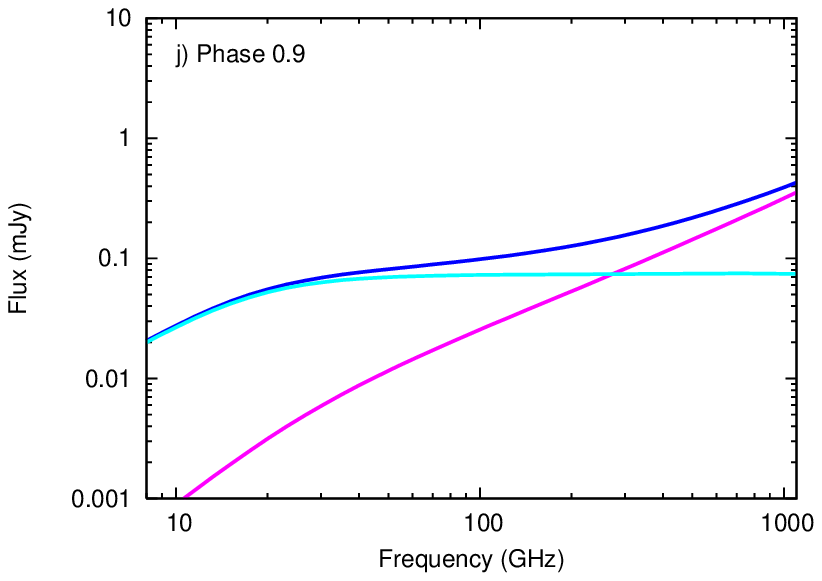,width=5.67cm}
\psfig{figure=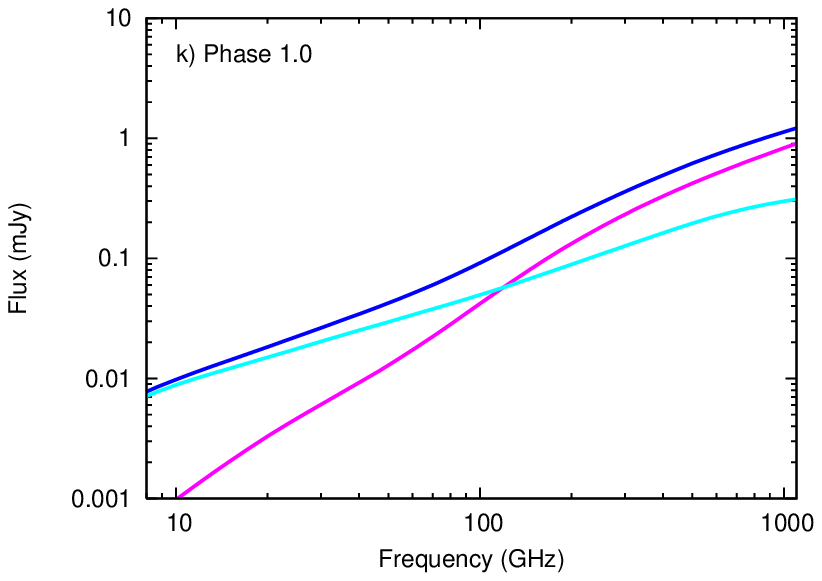,width=5.67cm}
\psfig{figure=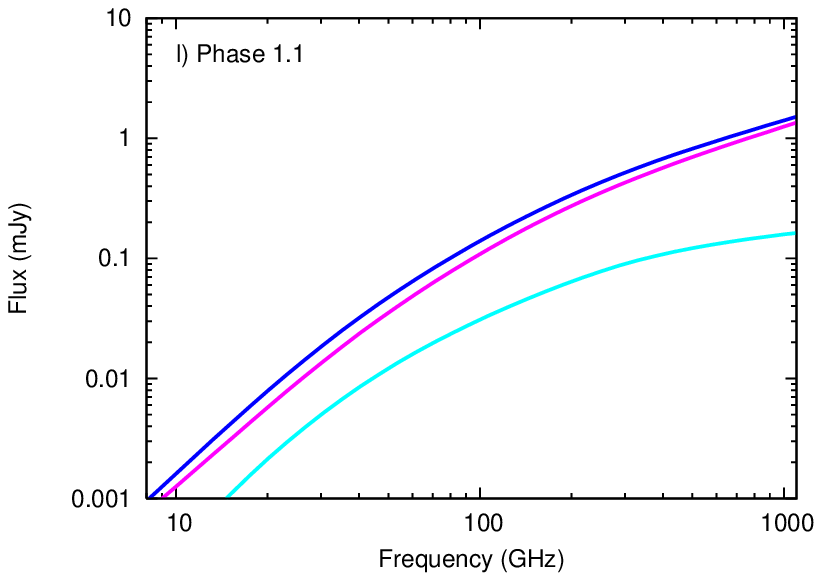,width=5.67cm}
\caption[]{Free-free radio spectra from model cwb4 as a function of
orbital phase for a viewing angle $i = 90^{\circ}$,
$\phi=0^{\circ}$.  The stars are at conjunction at phases 0.0 and
0.5.}
\label{fig:cwb4_radio_spectrum}
\end{figure*}

\begin{figure*}
\psfig{figure=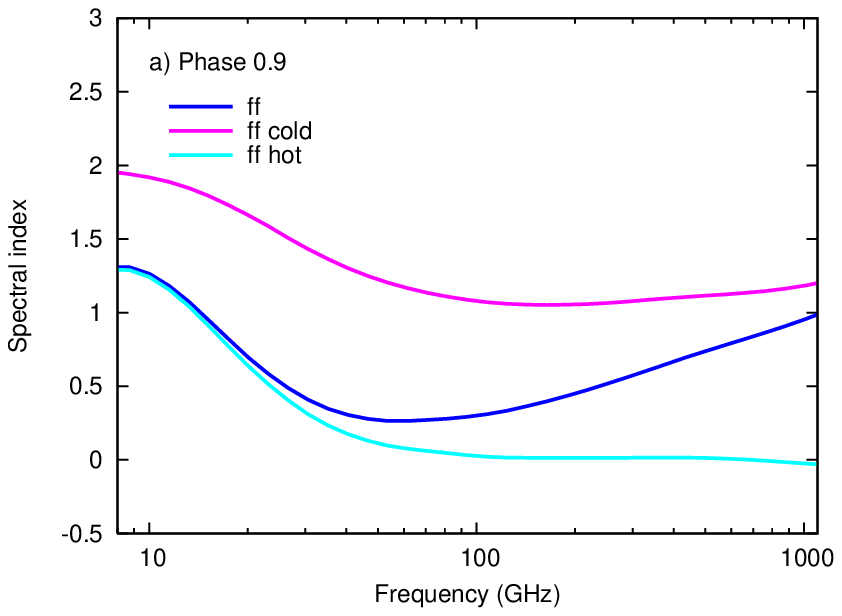,width=5.67cm}
\psfig{figure=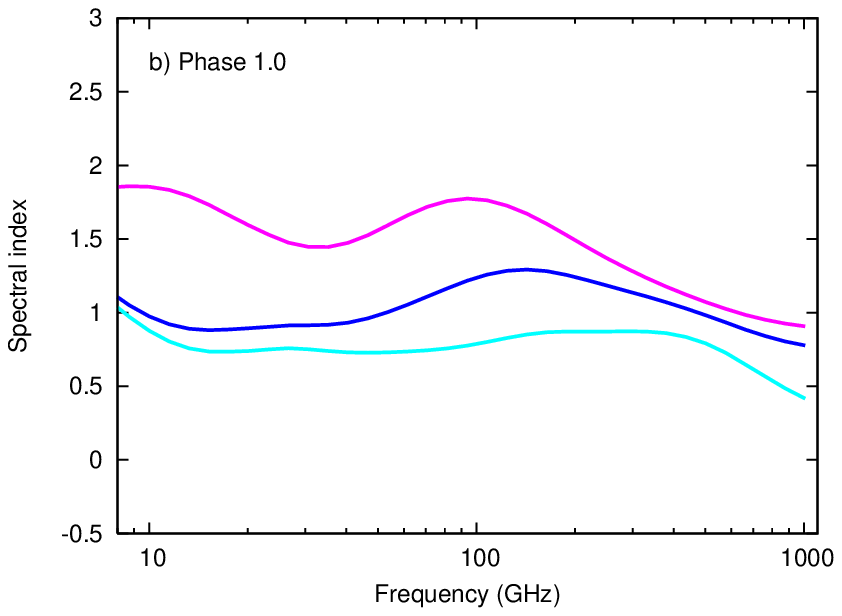,width=5.67cm}
\psfig{figure=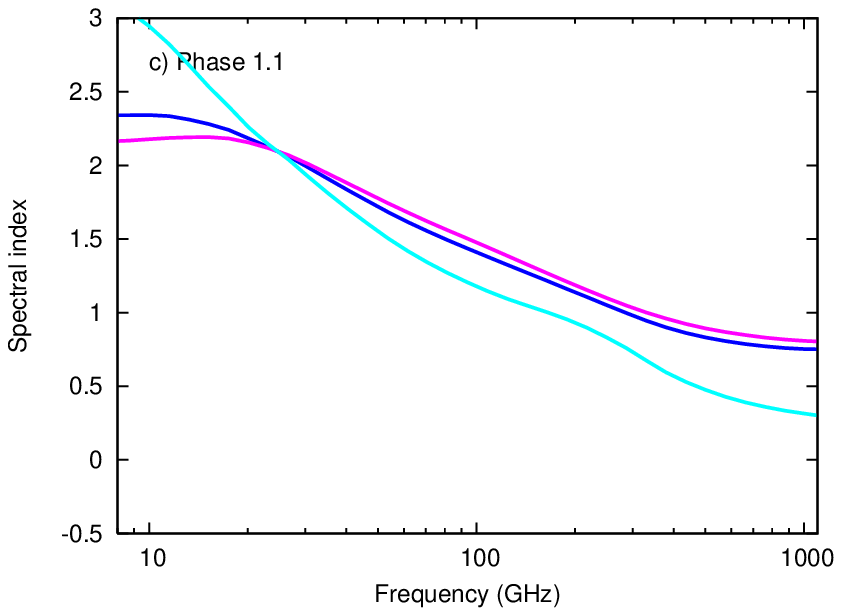,width=5.67cm}
\caption[]{The spectral index as a function of frequency from the spectra
in Fig.~\ref{fig:cwb4_radio_spectrum} at orbital phase 0.9, 1.0, and 1.1 
(which encompasses the greatest spectral changes). The observer has a
viewing angle $i = 90^{\circ}$ and $\phi=0^{\circ}$.}
\label{fig:cwb4_radio_spectrum_index}
\end{figure*}

\begin{figure*}
\psfig{figure=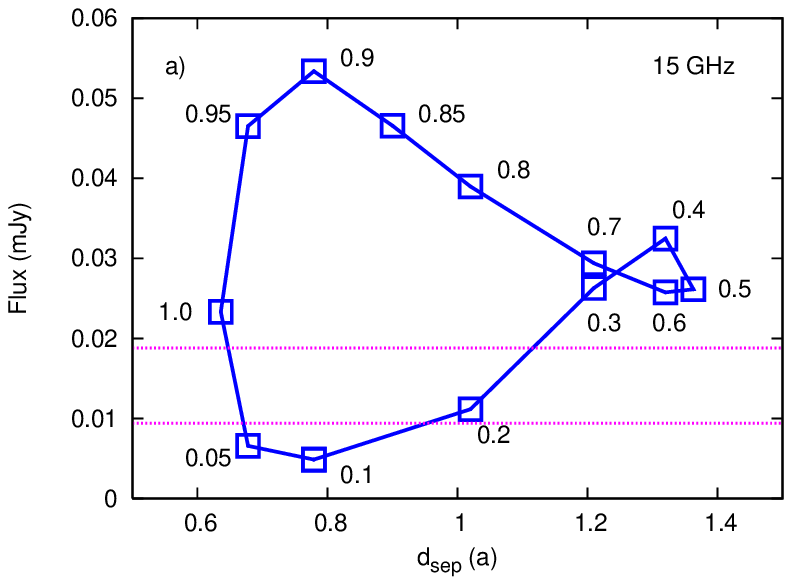,width=5.67cm}
\psfig{figure=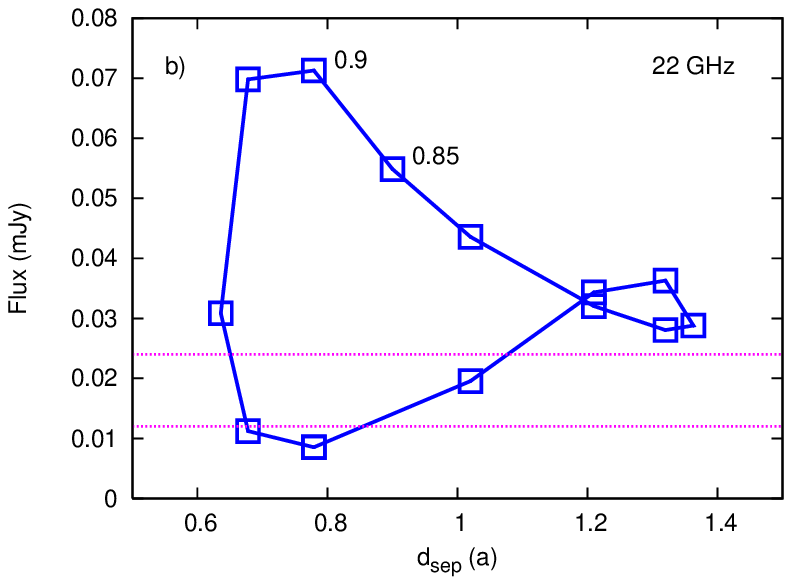,width=5.67cm}
\psfig{figure=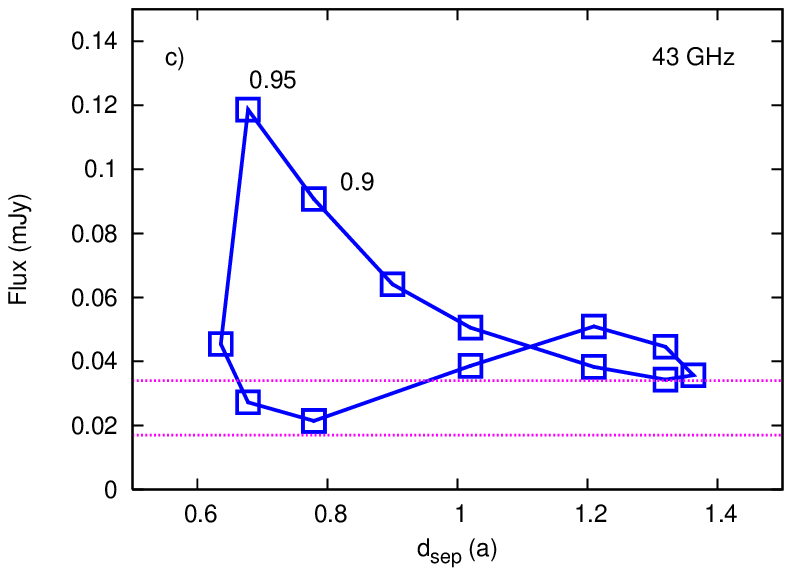,width=5.67cm}
\psfig{figure=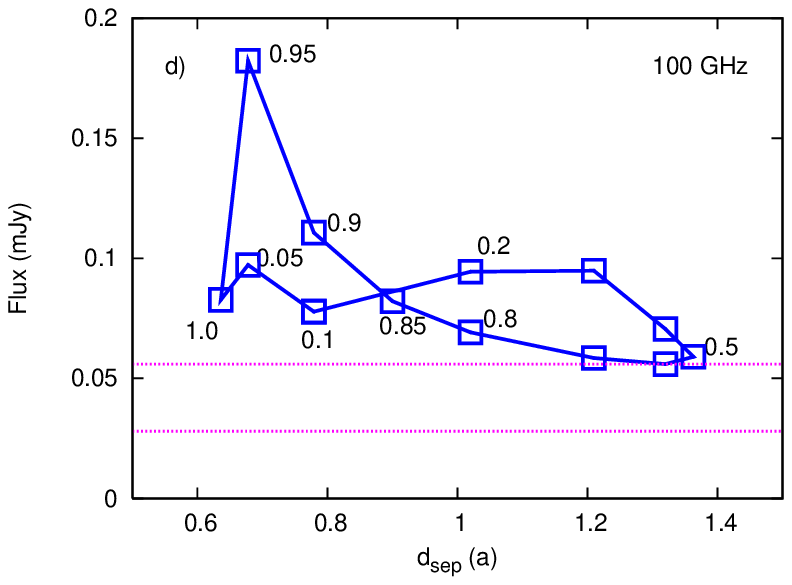,width=5.67cm}
\psfig{figure=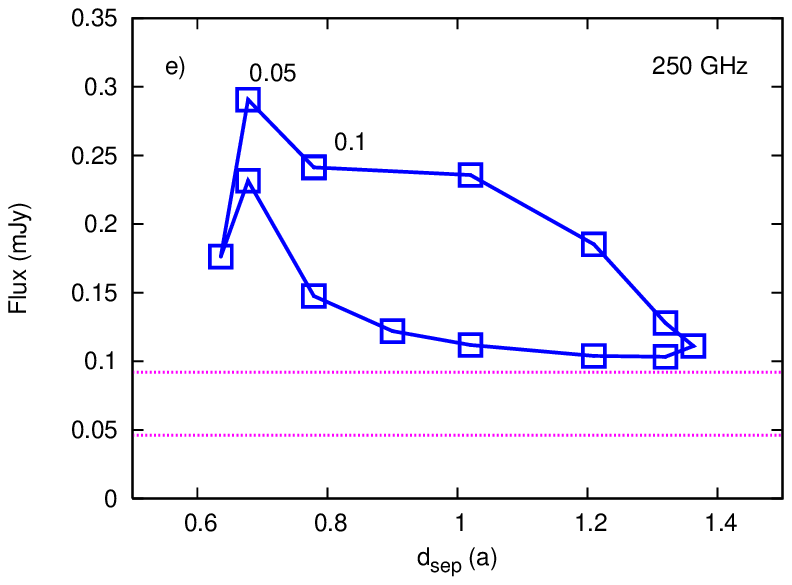,width=5.67cm}
\psfig{figure=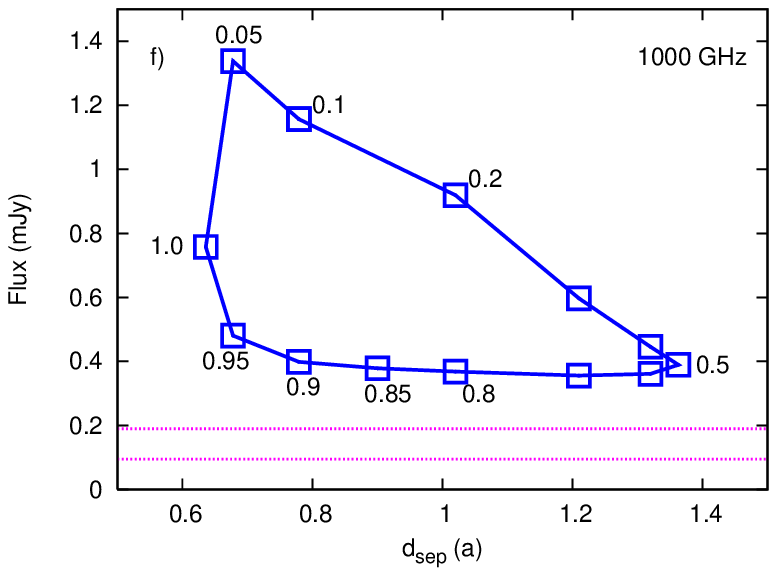,width=5.67cm}
\caption[]{Hystersis of the thermal free-free emission as a function
of orbital separation for a variety of frequencies ranging from
15\,GHz (a) to 1000\,GHz (f). The observer is located directly above
or below the orbital plane at $i=0^{\circ}$. The phases of some of the
points are marked. The lower horizontal line shows the theoretical
flux calculated from the terminal speed wind of a single O6V star. The
upper horizontal line shows twice this value.}
\label{fig:cwb4_hysteresis}
\end{figure*}

\subsubsection{Model cwb4}
Unlike the previous models which all had circular orbits, model cwb4
simulates a CWB with an eccentric orbit ($e=0.\overline{36}$).  The
intrinsic emission now varies with phase, whereas it was constant in
the circular orbit models cwb1$-$cwb3.
Fig.~\ref{fig:cwb4_radio_image} shows intensity images of the
free-free emission from model cwb4 at 1000\,GHz for an observer
directly above the orbital plane. The images show striking variations
in their brightness and morphology as a function of orbital phase,
reflecting the dramatic changes in the WCR during the orbit (see
Paper~I for full details of the hydrodynamics).

Discussion of this figure begins, most easily, at orbital phase 0.5
(apastron).  At this phase we see the emission regions of the two
winds (located very close to the stellar surfaces due to the high
frequency), plus between them emission from the hot WCR. There are
also small but very bright regions of emission within and close to the
WCR. These are very dense, cold, clumps of previously shocked gas
which formed during the previous periastron passage when the WCR
became highly radiative. Most of the dense, cold, gas which formed
during this time has now been cleared out of the inner parts of the
system.  The dense clump and its tail visible in the top left of the
image are actually located outside of the WCR, since the high inertia
of the clumps causes them to move on almost ballistic trajectories,
while the hot gas in the WCR responds more rapidly to the changing
positions of the stars (see Paper~I for more details).

The process of clearing out the remaining dense clumps continues as
the stars advance in orbital phase. By phase 0.7, no clumps are left
within the central regions of the system. At this point the stars are
moving in a slow waltz towards each other. The density within the WCR
steadily increases as the stellar separation reduces, resulting in
brighter emission from the WCR, so that by phase 0.95 the apex of the
WCR is brighter than the unshocked winds themselves, despite the high
frequency! The spatial distribution of the emission from the WCR also
reflects the WCR's increasing curvature as periastron is approached.

By periastron the gas in the densest parts of the WCR in model cwb4
has undergone substantial radiative cooling. Cool ($T \approx
10^{4}$\,K), dense gas forms in the WCR between the stars and for some
distance downstream (see Figs.~11 and~15 of Paper~I).  This gas is
optically thick to frequencies exceeding 1000\,GHz.  The enhancement
in the density of this gas relative to that of the neighbouring
unshocked winds means that its emission coefficient is significantly
higher, as indicated by the images shown in
Fig.~\ref{fig:cwb4_radio_image}. Further downstream, the plasma in the
WCR remains hotter, with some gas at temperatures exceeding
$10^{7}$\,K (see Fig.~15(e) in Paper~I). This plasma is optically thin
to emission above $\nu \gtsimm 5$\,GHz, and its emission coefficient
is much smaller than that of the optically thick cold gas nearer the
apex of the WCR. The wide variation in plasma density and temperature
throughout the WCR demonstrates that models of the emission from such
systems {\em must} be based on time-dependent hydrodynamical
calculations.

Even more mass within the WCR has cooled back to the wind temperature
($10^{4}$\,K) by phase 0.05. Newly shocked gas in the WCR becomes
largely adiabatic by phase 0.2, as the stars continue to separate and
the pre-shock velocity of the winds increases. No new clumps form
after this phase until the next periastron passage. Instead, the slow
process of clearing the existing clumps out of the system begins.

It is likely that the majority of the emission between orbital phase
$0.7-1.05$ is captured by the numerical grid, but it is clear that
significant flux losses occur from the simulation between phase
$0.05-0.3$ as optically thick clumps leave the finite boundaries of
the grid. This effect is likely to be greatest at the lower
frequencies since the clumps stay optically thicker for greater
outflow distances, and should be kept in mind when examining the
remaining figures in this section.

Fig.~\ref{fig:cwb4_ffradio_lc} displays lightcurves of the free-free
emission from model cwb4. Since the typical temperature of the plasma
in the WCR varies between ``cold'' near periastron, and ``hot'' for
the rest of the orbit (excepting the dense clumps formed within the
WCR), we have chosen to separately identify the emission from cold ($T
< 10^{5}$\,K) and hot ($T \geq 10^{5}$\,K) gas, rather than from the
unshocked winds and the WCR.  At times when the WCR is hot (such as at
apastron), this leads to the same identification of gas as in the
previous sections. However, during the small range in phases around
periastron, our analysis groups the cold gas within the WCR with the
unshocked gas from the winds, rather than with the hotter parts of the
WCR on the leading arms and downstream. Hence, the ``cold'' and
``hot'' components cannot be directly compared to the ``winds'' and
``WCR'' components from model cwb1, although as noted above it is
still reasonable to compare to model cwb2 at other phases. Note that
Fig.~\ref{fig:cwb4_ffradio_lc} shows {\em observed} fluxes, so, for
example, the contribution of hot material to the flux includes
absorption from overlying cold material in the circumstellar
environment.
 
Fig.~\ref{fig:cwb4_ffradio_lc} shows that the emission typically
reaches a maximum around periastron, as was already obvious from the
images at $\nu=1000$\,GHz and $i=0^{\circ}$, presented in
Fig.~\ref{fig:cwb4_radio_image}. The varying circumstellar absorption
due to the passage of the stars in front of the WCR complicates the
interpretation of the lightcurves when the observer is in the orbital
plane. Therefore, we shall concentrate on the lightcurve for an
observer at $i=0^{\circ}$.  At 43\,GHz the emission peaks just prior
to periastron, while at 1000\,GHz the maximum occurs just after
periastron. There are two reasons for this behaviour. One reason is
related to the changes with orbital phase of the optical depth of
sightlines through the WCR, and is discussed in more detail below. A
minimum centered on phase 0.1 also occurs in the 43\,GHz lightcurve -
this is primarily due to the high optical depth of the cold dense
plasma formed in the WCR during the periastron passage. Most of the
emission from this material is therefore absorbed. By phase 0.3 many
parts of the WCR become optically thin, and the emission from
background material starts to become visible. However, the emissivity
of this material is also dropping, and this becomes the dominant
factor affecting the observed flux at subsequent phases. The 43\,GHz
emission reaches a minimum around phase $0.5-0.6$, and then
subsequently increases as the stars approach each other, due to the
increase in density of the hot plasma in the WCR.

In contrast, the 1000\,GHz lightcurve of the total emission
(Fig.~\ref{fig:cwb4_ffradio_lc}j) does not display a minimum at phase
0.1. This is partly because some parts of the WCR remain optically
thin at this phase (although a small reduction in the emission from
``hot'' plasma relative to the trend is nevertheless visible in
Fig.~\ref{fig:cwb4_ffradio_lc}l). It is also partly because some of
the emission at 1000\,GHz is from the unshocked stellar winds (which
is responsible for some of the emission seen in
Fig.~\ref{fig:cwb4_ffradio_lc}k), and the stars are near quadrature at
phase 0.1.

Fig.~\ref{fig:cwb4_radio_spectrum} shows radio-to-sub-mm spectra from
model cwb4 as a function of orbital phase for an observer in the
orbital plane at $\phi=0^{\circ}$.  The spectrum from the WCR is
optically thin between phase 0.5 and 0.9 over the displayed frequency
range.  Comparing against the spectrum from model cwb2 in
Fig.~\ref{fig:cwb2_radio_spectrum}(a), we see that both the
``hot/WCR'' and the ``cold/winds'' flux are in good agreement at high
frequencies. However, there is less emission from model cwb4 than from
model cwb2 at low frequencies. At 10\,GHz, the ``winds'' flux from
model cwb2 is about three times greater than the ``cold'' flux from
model cwb4, while the ``WCR'' flux is more than an order of magnitude
greater than the ``hot'' flux.  This lack of emission from model cwb4
at the lower frequencies reflects the smaller grid compared to that of
model cwb2.

Since a large fraction of the emission from the WCR is optically thick
at periastron, it is not surprising to find that the spectral index of
the ``hot'' emission at phase 1.0 becomes positive like the unshocked
winds ($\alpha_{\rm WCR} \approx +0.75$ between 15 and 100\,GHz - see
Fig.~\ref{fig:cwb4_radio_spectrum_index}b). In fact, the ``hot''
spectrum is remarkably straight at periastron, with very little
curvature (see Figs.~\ref{fig:cwb4_radio_spectrum}a and~k, and
Fig.~\ref{fig:cwb4_radio_spectrum_index}b).

Due to the dramatic cooling at periastron, by phase 0.1 the majority
of the gas within the WCR {\em is now cold}.  This results in a much
lower flux from the ``hot'' component of the emission, which is now
mostly low density gas on the leading edges of the WCR and a narrow
strip of rapidly cooling gas near the apex and trailing edges of the
WCR. Conversely, the emission from the ``cold'' component is
significantly bolstered by the substantial amount of gas which has
cooled within the WCR, being $2.5\times$ more luminous at 100\,GHz
than the emission at periastron.

The change in viewing angle into the system also causes the low
frequency emission to drop (see e.g. panel l of
Fig.~\ref{fig:cwb4_radio_spectrum}). A gradual recovery occurs in the
low frequency emission up to phase 0.3, after which it declines again.
These changes are largely driven by changes to the optical depth to
the ``hot'' flux, which shows more variability at low frequencies than
the ``cold'' flux.  The level of variability is expected to be smaller
in calculations where the hydrodynamical grid extends to larger radii.

Although some parts of the WCR remain hot even at periastron (for
instance the leading edge of the shocks), the majority of the mass
within the WCR is optically thick at phases 0.0 and 0.1.  At the
highest frequencies considered, the WCR is only marginally optically
thick at phase 0.1, with the optical depth through its surface
occasionally exceeding unity, but more typically being less than
this. The WCR becomes increasingly optically thin thereafter, as the
stars continue their separation.

Just as the flux from the hot part of the WCR varies with phase, so
does the contribution from the unshocked winds plus cold parts of the
WCR. At phase 0.5 this part of the emission has a straight slope with
$\alpha_{\rm cold} \approx +1.0$ between 200 and 1000\,GHz. The flux
and spectral slope are relatively steady up to phase 0.9. At
periastron, the first gas in the WCR to cool back down to $10^{4}\,$K
creates a noticeable ``bump'' on the cold gas spectrum at $\nu \gtsimm
70$\,GHz (see Fig.~\ref{fig:cwb4_radio_spectrum_index}b). 
The extra emission becomes more extended in its range of
frequencies as more and more gas passing through the WCR cools back to
the temperature of the unshocked winds, and reaches its maximum
contribution at phase 0.1, with $6\times$ as much flux from cold gas
compared to phase 0.5 at 200\,GHz.  Thereafter the bump declines as
the cold plasma in the WCR is ablated and/or expelled out of the
system, so that the ``cold'' spectrum is almost entirely generated by
emission from the unshocked winds by apastron.

These changes to the spectra also affect the spectral index of
the emission, as shown in Fig.~\ref{fig:cwb4_radio_spectrum_index},
with the most rapid changes to $\alpha$ occuring around periastron 
passage. At phase 0.9, the spectral index of the total emission
increases from $+0.26$ at
55\,GHz, to $+0.95$ at 1\,THz, due to the steady increase of flux from
the cold unshocked winds relative to the hot WCR. At periastron, the
increasing opacity of the hot WCR leads to a significant increase in
$\alpha_{\rm hot}$ to $\approx +0.8$ between 20 and 400\,GHz. The 
enhancement to the emission around 100\,GHz due to the formation of
cold gas increases $\alpha_{\rm cold}$ to nearly $+1.8$ at 100\,GHz.
The spectral index of the total emission is in the range $+0.9$ to
$+1.3$ between 20 and 600\,GHz. By phase 0.1, $\alpha$ 
monotonically decreases from low to high frequencies. However, the
computed values of the spectral index at low frequencies 
shown in Fig.~\ref{fig:cwb4_radio_spectrum_index}(c)
are unlikely to be quantitatively correct, given the known underestimate
of the flux due to the finite grid size.

Fig.~\ref{fig:cwb4_hysteresis} shows the variation of the total
thermal flux with orbital separation for an observer with
$i=0^{\circ}$.  
Bearing in mind that our simulations likely underestimate the true
flux from this system between phase $0.05-0.3$, it nevertheless
reveals that there is a strong hystersis to the emission at all
frequencies. It is also interesting that at low
frequencies the emission is stronger as the stars approach periastron,
whereas at high frequencies the emission is stronger as the stars
recede from periastron. This is because at low frequencies the winds
are very opaque and the emission most easily escapes when the WCR is
full of hot plasma, which is the case in the second half of the orbit.
In contrast, at high frequencies the winds are always transparent, and
stronger intrinsic emission from the cold, dense, plasma which exists
in the WCR following periastron passage can be observed.

Fig.~\ref{fig:cwb4_hysteresis} also shows the theoretical flux from a
single O6V wind which accelerates instantaneously to its terminal
velocity (lower horizontal line). The upper horizontal line shows
twice this value. We have already seen from many of the lightcurves in
this paper that at the lower frequencies the models at times
significantly underestimate the true flux due to the finite extent of
the numerical grid, and we again see this here. At higher frequencies
(e.g. $\nu \gtsimm 100$\,GHz in model cwb4) this is no longer an
issue, at least over orbital phases $\approx 0.5-1.05$.

The flux at 43\,GHz from model cwb4 typically exceeds twice the
theoretical single star flux, although it is below this value at
phases 0.05 and 0.1 when the stars orbit deep into each other's wind -
at these times the optically thick contribution from the WCR is not
able to offset the loss of flux from the unshocked winds due to the
presence of the WCR (however, particularly at phase 0.1, we expect the
true flux to be significantly underestimated due to optically thick,
cool, clumps flowing through the grid boundaries).  By 250\,GHz, the
total flux from model cwb4 always exceeds twice the expected single
star value. This is again because the characteristic region of emission is
now close enough to the stars that the WCR no longer ``cuts'' out
significant parts of this region. At phase 0.5 the extra emission from
the WCR, which is mostly from hot gas, is negligible (see
Fig.~\ref{fig:cwb4_radio_spectrum}f). However, as the stars approach
periastron the emission from the WCR becomes significant as the plasma
within it increases in density and decreases in temperature, causing
the flux rise shown in Fig.~\ref{fig:cwb4_hysteresis}(e). The dip in
the flux at periastron is caused by the foreground wind occulting the
apex of the WCR where the brightest emission arises.

\citet{Pittard:2006} noted that the {\em thermal} radio flux from the
WCR in an adiabatic system scales as $d_{\rm sep}^{-1}$, so that we
would expect a variation of just over a factor of two from our
model. However, the variations from model cwb4 are obviously much greater.  For
example, Fig.~\ref{fig:cwb4_ffradio_lc}(l) shows that for an observer
with $i=0^{\circ}$, the flux increases by a factor of 3.4 between
phases 0.6 and 0.9, compared to a $d_{\rm sep}^{-1}$ increase of 1.7.
This is because the pre-shock velocities change as the stellar
separation varies, affecting the post-shock temperature, whereas the
analysis in \citet{Pittard:2006} is only valid in the scale-free limit
where the winds collide at constant velocity.

Finally, we note that the brightness of the WCR emission is highly
dependent on its density. In models cwb1 and cwb4 the cooled postshock
gas becomes very dense, partly because of the assumption that there is
no significant magnetic pressure which would otherwise support this
gas against collapse.  If magnetic pressure were to become
significant, the densities in the WCR would be limited and the
emissivity of the WCR would be reduced.

\begin{figure*}
\psfig{figure=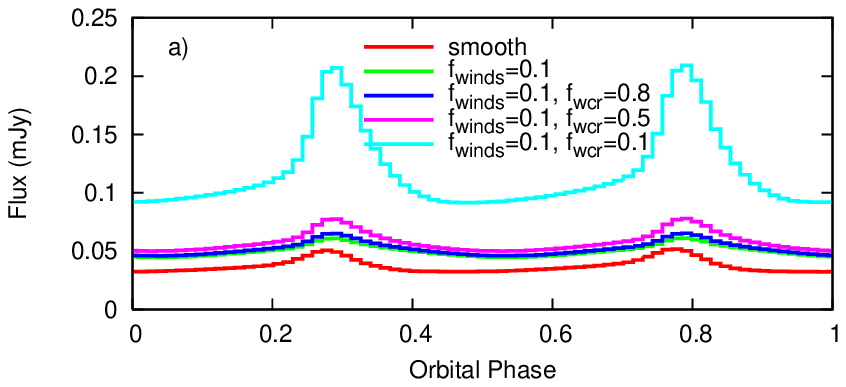,width=5.67cm}
\psfig{figure=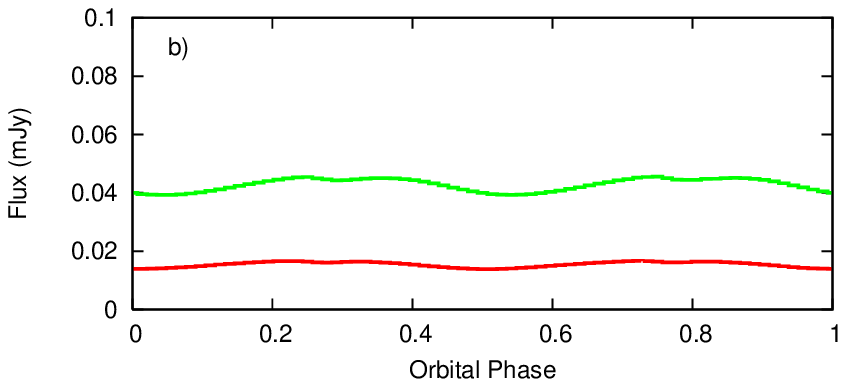,width=5.67cm}
\psfig{figure=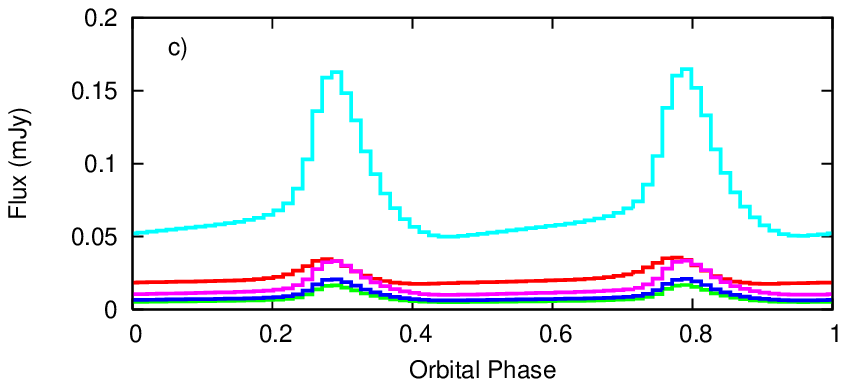,width=5.67cm}
\caption[]{Comparison of 43\,GHz lightcurves from model cwb2 with
smooth and clumpy winds for an observer with $i = 90^{\circ}$ and
$\phi=0^{\circ}$. The stars are at conjunction at phases 0.0 and 0.5,
and quadrature at phases 0.25 and 0.75. Note that while the flux from
clumpy unshocked winds increases as $f_{\rm winds}^{-2/3}$, the flux
from a clumpy WCR increases as $f_{\rm WCR}^{-1}$, since the former is
optically thick, while the latter is optically thin. Panel a) plots
the total flux, panel b) the flux from the unshocked winds, and
panel c) the flux from the WCR. Note that the y-axes (flux scales)
are different in each panel.}
\label{fig:cwb2_ffradio_lc_clumpff0.1}
\end{figure*}

\begin{figure*}
\psfig{figure=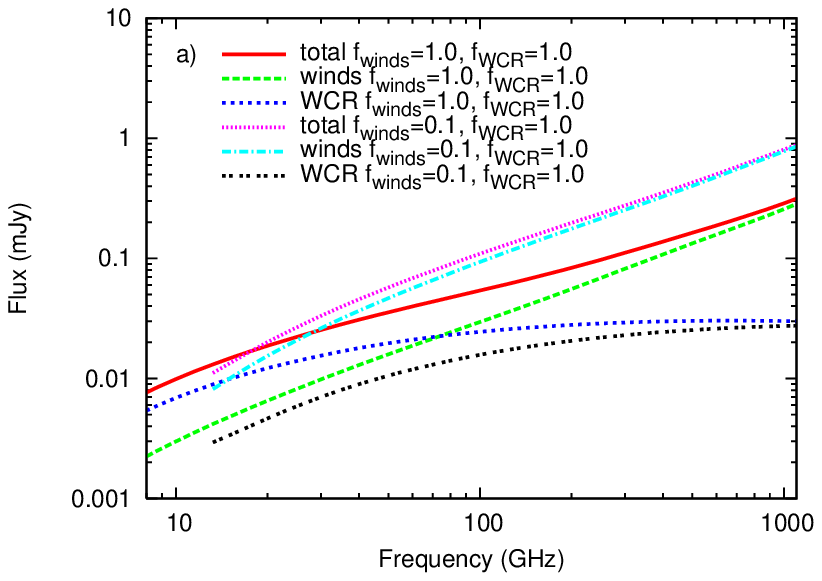,width=5.67cm}
\psfig{figure=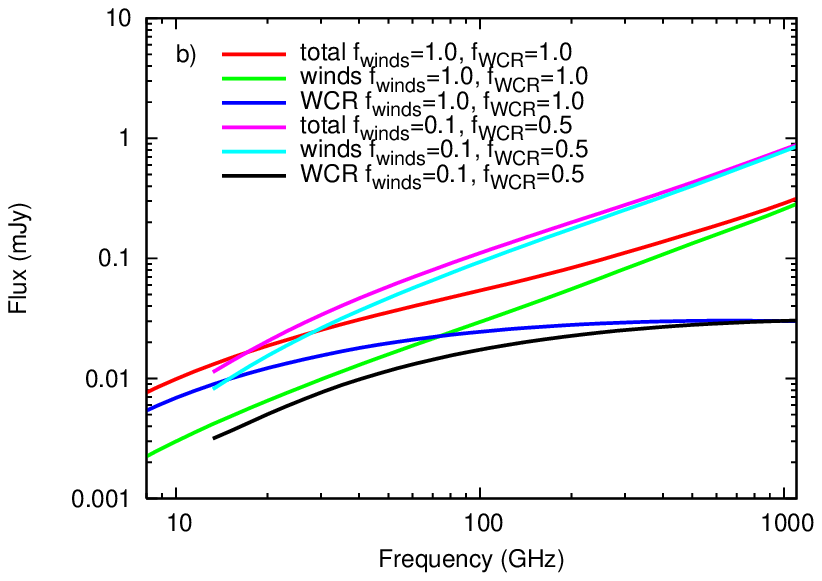,width=5.67cm}
\psfig{figure=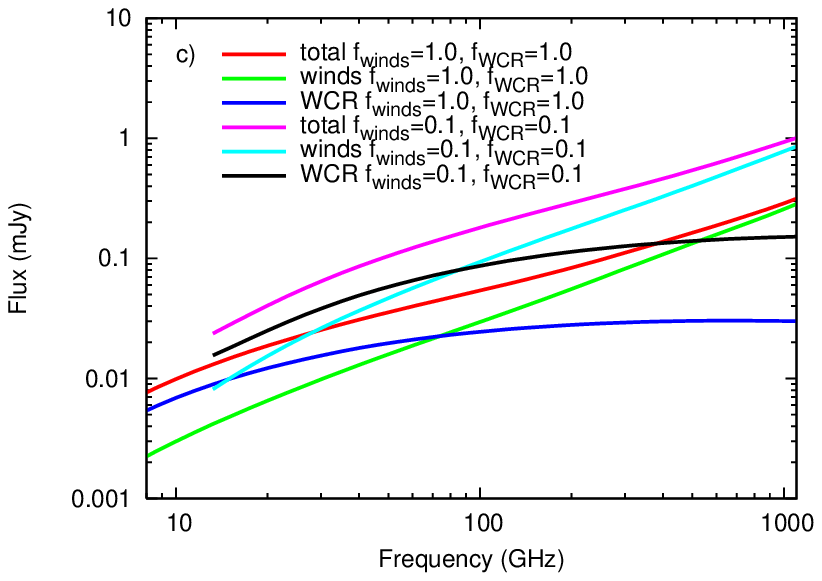,width=5.67cm}
\caption[]{Free-free radio spectra from model cwb2 as a function of
the clumping within the unshocked winds and the WCR, for
$i=90^{\circ}$ and $\phi=0^{\circ}$. a) Comparison of the smooth winds
and WCR case ($f_{\rm winds}=f_{\rm WCR}=1.0$) with clumpy winds but a
smooth WCR ($f_{\rm winds}=0.1$, $f_{\rm WCR}=1.0$).  b) As a), but
now the WCR is somewhat clumpy too, although the WCR remains much
smoother than the winds ($f_{\rm winds}=0.1$, $f_{\rm WCR}=0.5$).  c)
As b), except the clumping within the WCR remains as severe as in the
winds ($f_{\rm winds}=f_{\rm WCR}=0.1$).  Note that in the smooth
(clumpy) winds case the fluxes are underestimated for $\nu \ltsimm
20$\,(50)\,GHz, because of the finite extent of the numerical grid.}
\label{fig:cwb2_radio_spectrum_clumpff0.1}
\end{figure*}

\subsubsection{Clumpy winds}
\label{sec:clumpy}
Figs.~\ref{fig:cwb2_ffradio_lc_clumpff0.1}
and~\ref{fig:cwb2_radio_spectrum_clumpff0.1} show how the lightcurves
and spectra from model cwb2 are modified if it is assumed that the
unshocked winds (and at times also the shocked gas in the WCR) are clumpy
(note that the mass-loss rates remain the same).
Fig.~\ref{fig:cwb2_ffradio_lc_clumpff0.1}(b) shows how clumping
with a volume filling factor $f_{\rm winds}=0.1$ increases the
flux from the unshocked winds compared to the smooth winds case.
The flux increases by a factor of 2.8, which is a little below the
expected increase of 4.64 from the relation $S_{\nu} \propto \Mdot^{4/3}
\propto f^{-2/3}$, and is likely due to the greater loss of flux from the
numerical grid because of its finite size. 

Fig.~\ref{fig:cwb2_ffradio_lc_clumpff0.1}(c) shows how the flux from
the WCR varies with clumping. Comparing the smooth to the $f_{\rm
winds}=0.1$ results, we see that the larger optical depth unity
surfaces in the winds of the latter calculation lead to greater
attenuation of the WCR flux. Hence the observable flux from the WCR
can be affected by changes to the clumping in the unshocked winds,
even though in both of these calculations the WCR is assumed to be
smooth and the {\em intrinsic} emission from the WCR is identical.

\citet{Pittard:2007} showed that an adiabatic WCR rapidly smoothes out
clumps if their density contrast and size is not too high, so that the
destruction timescale of the clump is shorter than the dynamical
timescale for material within the WCR to flow out of the system.
However, this work also revealed that the plasma in the WCR contains a
multitude of weak shocks, and is far from being perfectly smooth. To
account for this, we have also computed calculations where there is
some ``residual clumping'' within the WCR.
Fig.~\ref{fig:cwb2_ffradio_lc_clumpff0.1}(c) shows that the flux from
the WCR gradually increases as the WCR is made more clumpy.  When
$f_{\rm WCR} = 0.5$, the observed flux from the WCR at quadrature is
roughly equal to that from the smooth winds case, with the extra
absorption from the unshocked winds (with $f_{\rm winds}=0.1$) being
offset by extra intrinsic emission from the WCR.

At the extreme limit of there being no ``smoothing'' effect by the WCR
(i.e. $f_{\rm WCR} = f_{\rm winds} < 1$), the observed flux from a
clumpy WCR can dramatically exceed its smooth counterpart, as shown in
Fig.~\ref{fig:cwb2_ffradio_lc_clumpff0.1}(c) when $f_{\rm WCR} =
f_{\rm winds}=0.1$. However, we regard this possibility as extremely
unlikely in systems with an adiabatic WCR, and more typically would
expect $f_{\rm WCR} >> f_{\rm winds}$ when $f_{\rm winds} << 1$.
Indeed, the level of the continuum in the theoretical X-ray spectra
shown in Fig.~3 of \citet{Pittard:2007} shows that the intrinsic flux
increases by only 20 per cent\footnote{It was this very behaviour
which led \citet{Pittard:2007} to note that ``clumping-independent''
measurements of the mass-loss could potentially be obtained from the
X-ray flux.}, equivalent to the WCR smoothing the clumping from
$f_{\rm winds} \approx 0.1$ to $f_{\rm WCR} \approx 0.8$.

Fig.~\ref{fig:cwb2_radio_spectrum_clumpff0.1} shows the spectra
obtained from calculations where the winds are significantly
clumped ($f_{\rm winds} = 0.1$) and the degree of clumping within
the WCR is varied between $f_{\rm WCR}=1.0$ (i.e. a smooth WCR) and
$f_{\rm WCR} = f_{\rm winds} = 0.1$ (i.e. the WCR has no smoothing
effect on the clumps). In Fig.~\ref{fig:cwb2_radio_spectrum_clumpff0.1}(a)
we see that clumping increases the free-free flux from the winds, at the
expense of increased absorption of the free-free flux from the WCR
(particularly at the lower frequencies). Increasing the clumping
within the WCR has little noticeable effect at first (see 
Fig.~\ref{fig:cwb2_radio_spectrum_clumpff0.1}b), but large enhancements
in the flux are apparent when the clumping within the WCR is extreme
(see Fig.~\ref{fig:cwb2_radio_spectrum_clumpff0.1}c). In fact, in such
cases the increase in flux from the WCR exceeds the increase seen 
from the winds, due to the differences in optical depth within
these regions.

Figs.~\ref{fig:cwb2_ffradio_lc_clumpff0.1}
and~\ref{fig:cwb2_radio_spectrum_clumpff0.1} demonstrate that even
with considerable clumping in the winds, free-free emission from the
WCR can still escape and create variability with orbital phase.

\section{Comparison with observations}
\label{sec:discussion}
\subsection{O-star binaries}
The parameters in model cwb1 are similar to those in 
HD\,215835 (DH\,Cep), HD\,165052, and HD\,159176. Currently, none of these
systems is detected in the radio or sub-mm bands, though
\citet{Bieging:1989} note 6\,cm upper limits of 0.27\,mJy for
HD\,165052, and 1.05\,mJy for HD\,159176.  Since we do not capture all
of the flux from model cwb1 (particularly at the lower frequencies
investigated), a direct comparison is not possible.  However, we can
obtain a very conservative upper limit to the 5\,GHz flux from model
cwb1 by extrapolating the 1000\,GHz flux to 5\,GHz with a spectral
index of $+0.6$ (the true spectral index may be significantly steeper
than this). This yields a 5\,GHz flux of $\approx 0.06$\,mJy, which is
still consistent with the observational upper limits noted above, even
if moderate clumping was added to our models.

However, these systems, and others like them, should be detected by
the next generation of interferometers. For instance, the EVLA will
have a point source sensitivity better than 1\,$\mu$Jy between
$2-40$\,GHz (1$\sigma$ rms in $\approx 9$\,hrs). 
e-MERLIN's specifications are almost as good.  The Square
Kilometre Array (SKA), which will operate over $0.1-25$\,GHz, will
probe even deeper. ALMA's band 3 ($84-116$\,GHz) also has a continuum
sensitivity of 0.06\,mJy, and should be just sensitive enough to
detect these systems. Obviously the chances for detection improve if the winds
are clumped. 

Model cwb2 is similar to HD\,93161A, and also shares some similarities
with Plaskett's star (HD\,47129). \citet{Benaglia:2004} failed to
detect HD\,93161A with ATCA at 3 and 6\,cm. Observations centered on
HD\,93129A (about 3 arcmin from HD\,93161A) reveal that at the 
position of HD\,93161A the rms noise is $\leq 0.2$\,mJy/beam at 3\,cm, 
and $\leq 0.25$\,mJy/beam at 6\,cm, giving $3\sigma$ upper limits
of 0.6\,mJy and 0.75\,mJy, respectively. These limits are consistent
with the fluxes predicted from model cwb2. On the other hand,
Plaskett's star was detected at 5\,GHz by \citet{Persi:1988},
who reported a flux of 0.2\,mJy. This flux is higher than we see
from model cwb2, and may indicate significant clumping within the
winds of this system. The fact that Plaskett's star is at a
distance of 1.5\,kpc (rather than the 1\,kpc of the models in this
work) makes this difference even worse.

Model cwb4 bears some similarities to some well-known O+O systems with
eccentric orbits, including HD\,152248, HD\,93205, HD\,93403,
Cyg\,OB2\,\#8A, and $\iota$\,Ori. Of these systems, HD\,152248,
HD\,93205, and HD\,93403 are currently not detected in the
radio. HD\,152248 fell within the field of view of ATCA observations
of Sco\,OB1 \citep{SetiaGunawan:2003}, and $3\sigma$ upper limits (at
both 3\,cm and 6\,cm) are 0.24\,mJy (Stevens, priv. communication,
2009).
In contrast, Cyg\,OB2\,\#8A is a bright, non-thermal, and strongly
variable radio source \citep{Bieging:1989}. Following the discovery of
this system's binarity \citep{DeBecker:2004}, \citet{Blomme:2005}
showed that the radio flux density undergoes phase-locked
variations\footnote{Other O+O systems with strongly varying
non-thermal radio emission are HD\,168112 \citep{Blomme:2005b},
HD\,167971 \citep{Blomme:2007}, and Cyg\,OB2\#9
\citep{VanLoo:2008}.}. The non-thermal emission from our models will
be investigated in a future paper.

Finally, $\iota$\,Orionis (HD\,37043) was detected by Howarth \& Brown (1991)
at 3.6\,cm, with a flux of $0.046\pm0.015$\,mJy \citep[see][]{Lamers:1993}.
Since it was observed only at one frequency it is not known if this
emission is predominantly thermal or non-thermal.
Moving the system to a distance of 1\,kpc yields a flux of 
$\approx 0.01$\,mJy, comparable to the predicted 8\,GHz flux 
from model cwb2, and similar to the flux from model cwb4 at
certain phases.

\subsection{WR and LBV systems}
Although this work has focussed on O+O systems, {\em thermal} radio
emission is also seen from WR and LBV stars
\citep[e.g.][]{Abbott:1986,Leitherer:1997,Dougherty:2008}.  However,
there are few accounts in the literature on the variability of such
sources\footnote{Of course, the variability of many non-thermal sources 
is much better studied.}. For instance, despite a large multi-frequency
monitoring campaign of $\gamma$\,Velorum, little evidence for
variation was found (Sean Dougherty, priv. communication, 2009).

Radio variability at the $10-30$ per cent level, has, however, been
recently reported from four WN9 stars in the Arches cluster
\citep[sources AR1, AR3, AR4, and AR8:][]{Lang:2005}. All but AR1 have
a positive spectral index ($\alpha \gtsimm +0.6$; observational
uncertainities do not allow an accurate measurement of the spectral
index of AR1). \citet{Lang:2005} suggested the variability might
indicate the presence of a time-variable non-thermal component, and/or
changes to the underlying mass-loss rate or velocity of the wind.
However, as this work demonstrates, an alternative explanation is that
these sources are binaries, and that the variability is caused by
changes to the circumstellar opacity along sightlines into the system
as the stars orbit each other, or due to a non-zero orbital
eccentricity. This view is supported by the fact that AR1 and AR4
\citep*[identified as sources A1N and A1S by][]{Wang:2006} have strong
and hard X-ray emission with a prominent Fe\,K line at 6.7\,keV, all
of which are characteristics of colliding wind binaries\footnote{On
the other hand, source AR6 clearly shows a non-thermal radio spectrum
\citep*{Lang:2001,Lang:2005}, and since it is associated with
strong X-ray emission \citep{Wang:2006}, it is also likely to be a
colliding wind binary.}.

Three point-like radio sources were also detected in the Quintuplet
cluster \citep[QR6, QR8, and QR9:][]{Lang:2005}. While these all have
positive spectral indices indicative of a strong stellar wind, we have
demonstrated in this work that the bulk of the emission could actually
arise from a radiative, and optically thick, WCR. In Westerlund\,1,
six WR stars are detected in the radio, but only one has a thermal
spectrum (WR\,L; Simon Clark, priv. communication, 2009). Note, however,
that WR\,L looks like it might be a CWB from it's X-ray properties 
\citep{Clark:2008}.

Variable thermal emission with $\alpha \gtsimm +0.6$ 
has also been detected from WR\,89, WR\,113,
and WR\,138 \citep{Montes:2009}. For WR\,89, the spectral index of the 
emission, $\alpha \gtsimm +0.6$, while for WR\,113, $\alpha \gtsimm +0.9$,
and $\alpha \gtsimm +0.7$ for WR\,138. WR\,89 is classified as a visual
binary \citep{vanderHucht:2001}, while WR\,113 is a WC8+O8-9 system
with an orbital period of 29.7\,d \citep{Niemela:1999}. WR\,138 is a
spectroscopic binary with a period of 1538\,d \citep{Annuk:1990},
though this system may also harbour a close companion with a short
period of order days \citep{Lamontagne:1982,Moffat:1986}.  It is
likely that the observed variability is related to the binarity of
these systems. \citet{Montes:2009} also note that the radio emission
from WR\,98a and WR\,104 (two of the so-called ``pinwheel'' systems)
is variable and the spectral index is positive, but less than the
value expected from a single wind (i.e. $\alpha < +0.6$).  These
systems could harbour non-thermal emission (i.e. be ``composite''
systems), or the emission could be purely thermal, combining 
optically thick emission from the winds
with optically thin emission from the WCR.

It is also interesting to note that the broad maxima seen in the
lightcurves of model cwb1 are similar to those seen in the radio
lightcurve of the LBV $\eta$\,Carinae \citep{Duncan:2003}.  The latter
is usually explained in terms of time-variable ionization of dense gas
within the spoked equatorial disc seen in the optical \citep[see
also][]{Kashi:2007}.  The spatial extent of the brightest part of the
radio emission spans a width of 1 to several arcseconds, depending on
the orbital phase.  At a distance of 2.3\,kpc, this corresponds to
thousands of AU.  However, 3-dimensional simulations of the structure
of the WCR reveals density enhancements on similar spatial scales
\citep{Parkin:2009,Gull:2009}. Thus, it is possible that at least some
of the variability could be due to time variations in the properties
and orientation of the optically thick shocked wind of the primary LBV
star.
Furthermore, $\eta$\,Car showed a sharp peak in the 7\,mm flux density
in 2003.5 (and again in 2009.0, Abraham, private communication, 2009),
coinciding with periastron passage \citep{Abraham:2005a}.  This spike
has been interpreted as the shocked plasma from the secondary wind
becoming optically thick at high temperatures
\citep{Abraham:2005b}. It is interesting that we also see a sharp
maximum in the flux from model cwb4 near periastron. However, for this
we have a different interpretation.  At low frequencies the spike in
model cwb4 occurs before periastron, when the density and thus
emissivity of the WCR is increasing. However, the WCR is optically
thin at this time. In contrast, at high frequencies ($\nu \gtsimm
250$\,GHz), the greatest flux from model cwb4 occurs after
periastron. At such frequencies, the majority of the unshocked winds
is optically thin, allowing the bright emission from the cold dense
clumps created during periastron passage to be seen.  It is clear that
further modelling of $\eta$\,Car is necessary, in order to better
understand the observed flux variations, on both short and long
timescales, from this fascinating system.

\section{Summary and conclusions}
\label{sec:summary}
This work examines the {\em thermal} radio-to-sub-mm emission from
colliding wind binaries, focussing in this first instance on
short-period O+O binaries (the potential {\em non-thermal} emission
will be studied in due course).  We investigate how the {\em thermal}
emission is influenced by the presence of a nearby companion star, and
the extent to which variability with orbital phase occurs.  The
emission is calculated from grids of density and temperature generated
from the 3D hydrodynamical models presented in \citet{Pittard:2009a}.
The models span a variety of different physical conditions, including
radiative and adiabatic wind-wind collision regions in systems with
circular orbits, and also a system with an eccentric orbit where the
WCR repeatedly cycles from radiative to adiabatic and back to
radiative again.

We find distinct differences in the spectra and lightcurves from
systems with radiative and adiabatic WCR's. In the former (model
cwb1), the WCR is optically thick to wavelengths as short as sub-mm,
and radiates as a dense sheet of emission. The WCR dominates the total
emission from the system, and only begins to show signs of becoming
optically thin at $\nu \gtsimm 1000$\,GHz. The flux (which has a
positive spectral index) can be over an order of magnitude greater
than from the wind of a single star, and highlights potentially
serious consequences for observational estimates of mass-loss rates
where sources are assumed to be single stars, but are in fact
binaries. The flux is highest just before quadrature, when the inner
regions of the WCR are viewed almost face on (due to the large
aberration of the WCR), and have the largest
projected extent. The lightcurve maxima are broad, with minima
occuring slightly before conjunction. For an observer in
the orbital plane, the total flux (from the unshocked winds plus the
WCR) varies by a factor of two or so. Although there is some loss of
flux from the models, the spectral index of the emission is likely to
be steeper than +0.6, and varies with orbital phase and the
orientation of the observer. Synthetic images for an observer directly
above the orbital plane display an ``S''-shaped region of emission
which traces the location of the WCR. 

In contrast, the emission from adiabatic WCR's is optically thin, and
most easily escapes the system when sight-lines to the observer are
themselves through the WCR. In model cwb2 the WCR dominates the
emission at $\nu \ltsimm 50-100$\,GHz, with free-free absorption
creating a low frequency turnover typically at $\nu \sim 10$\,GHz.
The emission from the WCR typically contributes only 10 per cent of
the total flux at 1000\,GHz (0.3\,mm).  The lightcurves show
relatively narrow maxima at phases just after quadrature. The
broad, and relatively flat minima inbetween the maxima indicate phases
where there is considerable absorption of the emission from the
WCR. There is a gradual drop in the degree of variability with
frequency as the circumstellar environment becomes less opaque and
increasingly transparent. At 1000\,GHz the vast majority of the WCR
emission escapes the system. Between 100 and 250\,GHz, the lightcurves
change morphology.  At $\nu \gtsimm 250$\,GHz, the maxima become
almost completely flat-topped, and the minima are now caused by
occultation of a narrow emission region near the surface of each
star. The spectral index is steepest at low frequencies (where
absorption of emission from the WCR is important) and at high
frequencies (where the emission is almost entirely from deep within
the acceleration region of the individual winds). Images of the emission
for an observer in the orbital plane reveal an intertwined 
``double-helix'' showing limb-brightened parts of the WCR, against which
the foreground wind is sometimes silhouetted. At other viewing angles
(e.g. $i=30^{\circ}$) an ``S''-shaped region of emission is again
seen.

Adiabatic systems where one of the winds is weaker than the other
generally display similar properties, the main differences being
slight asymmetries to the lightcurves (e.g. differing heights to the
maxima), and a change from a ``double-helix'' emission structure to a
single limb-brightened structure emitted by the relatively dense
trailing edge of the trailing arm of the WCR.
 
One of the more surprising findings in this work is the strength of
the flux variations with orbital phase from a system with an eccentric
orbit (model cwb4), which exceed an order of magnitude from the WCR),
and may be almost as strong from the system as a whole. This is caused
by dramatic changes to the physical properties of the WCR, which is
highly radiative at periastron, but adiabatic at apastron, and changes
from optically thick to optically thin, respectively.  This creates a
strong frequency-dependent hystersis to the emission around the orbit.
At low frequencies the emission is stronger as the stars approach
periastron, due to the low opacity through the hot WCR at this
time. The flux variations easily exceed the $d_{\rm sep}^{-1}$ scaling
proposed by \citet{Pittard:2006}, which is invalidated by changes to
the pre-shock wind speeds as the stars orbit deep within the
acceleration zone of each other's wind. In contrast, at high
frequencies the emission is stronger as the stars recede from
periastron, because of the high intrinsic emission from dense, cold,
post-shock gas created during periastron passage.  Even larger
variations are seen when the observer is in the orbital plane,
indicating that changes in the circumstellar absorption also play an
important role.

An examination of the effects of clumping reveal that in addition to
higher fluxes, clumping can also effect the variability of radio
lightcurves. In the case of strong clumping in the unshocked winds but
a smooth WCR, the variability decreases. However, if the clumps are
not smoothed by the WCR (a case which we consider rather unlikely),
the variations of the flux with orbital phase can be enhanced compared to 
the smooth winds case.

We find that the fluxes from our models are consistent with the mostly
upper limits which exist to date from the O+O systems which they most
resemble. On the other hand, there are many instances where radio
variability is observed from WR sources with positive spectral
indices, some with $\alpha \gtsimm +0.6$. Many of these sources are
likely to be binaries, with the variation linked to changes in the
absorption through the circumstellar material, as the orientation of
the system to the observer changes, or perhaps due to changes in the
separation of the stars. However, we also caution that
systems with positive spectral indices are not necessarily thermal -
instead, one could be witnessing the absorption of a non-thermal
component. This possibility must be carefully considered, especially
if the spectral index is calculated at relatively low frequencies
(e.g. between 5 and 8\,GHz).  To safely distinguish between thermal
and non-thermal emission in close, spatially unresolved 
systems where the spectral index is
positive may require flux measurements over a wide frequency range,
additional observations at X-ray and $\gamma$-ray energies, and
detailed comparisons with theoretical models.

It is difficult to predict how the fluxes from the models
investigated in this work will scale to other systems (e.g. WR+O systems),
since there are so many key variables, such as whether the
WCR is optically thick or thin, the recent thermal history of 
the plasma in the WCR, the relative sizes of the radio photospheres in
each wind to the stellar separation, the orientation of the system, etc.
In many ways this work, in revealing some of these complications, 
demonstrates that it may not always be valid to apply some of the scaling
laws noted in \citet{Dougherty:2003} and \citet{Pittard:2006}.
In this respect there is a clear need for further numerical
simulations of CWBs with a range of key parameters (mass-loss rates,
stellar separations, etc.). Such simulations will aid the 
study and interpretation of the resulting emission from these
fascinating systems.

The future commissioning of more sensitive telescopes, such as the
EVLA, e-MERLIN, ALMA, and SKA, should result in detections of
short-period O+O systems, thus allowing the predictions of this work
to be tested. Although the synthetic images shown in this work are of
an angular scale which is too small to be resolved with EVLA or ALMA,
planned upgrades to the VLBA offer the exciting prospect of spatially
resolving the emission from tight CWBs. The EVLA, ALMA, and other
facilities may also resolve some of the wider O+O (and WR+O) CWBs.

\section*{acknowledgements}
I would like to thank the referee for a helpful and timely report,
Sean Dougherty and Ross Parkin for some of the inspiration behind this
research, Ian Stevens for use of his modified CAK code, and Ian, Paula
Benaglia, Simon Clark, and Perry Williams for providing some
information on the current radio observations of early-type stars. I
would also like to thank the Royal Society for a University Research
Fellowship.

\label{lastpage}


\begin{thebibliography}{99}
\bibitem[\protect\citeauthoryear{Abbott et al.}{1980}]{Abbott:1980}
Abbott D.~C., Bieging J.~H., Churchwell E., Cassinelli J.~P., 1980, ApJ, 238, 196 
\bibitem[\protect\citeauthoryear{Abbott, Bieging \& Churchwell}{Abbott et al.}{1981}]{Abbott:1981}
Abbott D.~C., Bieging J.~H., Churchwell E., 1981, ApJ, 250, 645
\bibitem[\protect\citeauthoryear{Abbott et al.}{1986}]{Abbott:1986}
Abbott D.~C., Bieging J.~H., Churchwell E., Torres A.~V., 1986, ApJ, 303, 239
\bibitem[\protect\citeauthoryear{Abraham et al.}{2005a}]{Abraham:2005a}
Abraham Z., et al., 2005a, A\&A, 437, 977
\bibitem[\protect\citeauthoryear{Abraham et al.}{2005b}]{Abraham:2005b}
Abraham Z., Falceta-Gon\c{c}alves D., Dominici T., Caproni A., Jatenco-Pereira V., 2005b, MNRAS, 364, 922
\bibitem[\protect\citeauthoryear{Altenhoff, Thum \& Wendker}{Altenhoff et al.}{1994}]{Altenhoff:1994}
Altenhoff W.~J., Thum C., Wendker H.~J., 1994, A\&A, 281, 161
\bibitem[\protect\citeauthoryear{Annuk}{1990}]{Annuk:1990}
Annuk K., 1990, Acta Astronomica, 40, 267
\bibitem[\protect\citeauthoryear{Arias et al.}{2002}]{Arias:2002}
Arias J.~I., Morrell N.~I., Barb\'{a} R.~H., Bosch G.~L., Grosso M., Corcoran M., 2002, A\&A, 333, 202	
\bibitem[\protect\citeauthoryear{Bagnuolo et al.}{2001}]{Bagnuolo:2001}
Bagnuolo W.~G. Jr., Riddle R.~L., Gies D.~R., Barry D.~J., 2001, ApJ, 554, 362
\bibitem[\protect\citeauthoryear{Benaglia}{2010}]{Benaglia:2010}
Benaglia P., 2010, ASP Conf. Ser., ``High Energy Phenomena in Massive Stars'' (arXiv:0904.0533)
\bibitem[\protect\citeauthoryear{Benaglia, Cappa \& Koribalski}{Benaglia et al.}{2001}]{Benaglia:2001}
Benaglia P., Cappa C.~E., Koribalski B.~S., 2001, A\&A, 372, 952
\bibitem[\protect\citeauthoryear{Benaglia \& Koribalski}{2004}]{Benaglia:2004}
Benaglia P., Koribalski B.~S., 2004, A\&A, 416, 171
\bibitem[\protect\citeauthoryear{Bertout et al.}{1985}]{Bertout:1985}
Bertout C., Leitherer C., Stahl O., Wolf B., 1985, A\&A, 144, 87
\bibitem[\protect\citeauthoryear{Bieging, Abbott \& Churchwell}{Bieging et al.}{1989}]{Bieging:1989}
Bieging J.~H., Abbott D.~C., Churchwell E., 1989, ApJ, 340, 518
\bibitem[\protect\citeauthoryear{Blomme}{2005}]{Blomme:2005}
Blomme R., 2005, in Rauw G., Naz\'{e} Y., Blomme R., Gosset R., eds, Proc. JENAM 2005, Massive Stars and High-Energy Emission in OB Associations. IAGL, Li\`{e}ge, p.\,45
\bibitem[\protect\citeauthoryear{Blomme et al.}{2007}]{Blomme:2007}
Blomme R., De Becker M., Runacres M.~C., Van Loo S., Setia Gunawan D.~Y.~A., 2007, A\&A, 464, 701
\bibitem[\protect\citeauthoryear{Blomme et al.}{2002}]{Blomme:2002}
Blomme R., Prinja R.~K., Runacres M.~C., Colley S., 2002, A\&A, 382, 921
\bibitem[\protect\citeauthoryear{Blomme et al.}{2005}]{Blomme:2005b}
Blomme R., Van Loo S., De Becker M., Rauw G., Runacres M.~C., Setia Gunawan D.~Y.~A., Chapman J.~M., 2005, A\&A, 436, 1033
\bibitem[\protect\citeauthoryear{Castor, Abbott \&\ Klein}{1975}]{Castor:1975} 
Castor J.~I, Abbott D.~C., \&\ Klein R.~I., 1975, ApJ, 195, 157 (CAK)
\bibitem[\protect\citeauthoryear{Clark et al.}{2008}]{Clark:2008}
Clark J.~S., Muno M.~P., Negueruela I., Dougherty S.~M., Crowther P.~A., Goodwin S.~P., de Grijs R., 2008, A\&A, 477, 147
\bibitem[\protect\citeauthoryear{Contreras et al.}{1996}]{Contreras:1996}
Contreras M.~E., Rodr\'{\i}guez L.~F., G\'{o}mez Y., Vel\'{a}zquez A., 1996, ApJ, 469, 329
\bibitem[\protect\citeauthoryear{De Becker}{2007}]{DeBecker:2007}
De Becker M., 2007, A\&ARv, 14, 171
\bibitem[\protect\citeauthoryear{De Becker, Rauw \& Manfroid}{2004}]{DeBecker:2004}
De Becker M., Rauw G., Manfroid J., 2004, A\&A, 424, L39 
\bibitem[\protect\citeauthoryear{De Becker et al.}{2004b}]{DeBecker:2004b}
De Becker M., Rauw G., Pittard J.~M., Antokhin I.~I., Stevens I.~R., Gosset E., Owocki S.~P., 2004b, A\&A, 416, 221
\bibitem[\protect\citeauthoryear{De Becker et al.}{2006}]{DeBecker:2006}
De Becker M., et al., 2006, MNRAS, 371, 1280
\bibitem[\protect\citeauthoryear{Dougherty \& Clark}{2008}]{Dougherty:2008}
Dougherty S.~M., Clark J.~S., 2008, in P.~Benaglia, G.~Bosch, C.~Cappa, eds, ``Massive Stars: Fundamental Parameters and Circumstellar Interactions'', RevMexAA, 33, 68
\bibitem[\protect\citeauthoryear{Dougherty, Taylor \& Waters}{Dougherty et al.}{1991}]{Dougherty:1991}
Dougherty S.~M., Taylor A.~R., Waters L.~B.~F.~M., 1991, A\&A, 248, 175
\bibitem[\protect\citeauthoryear{Dougherty et al.}{2003}]{Dougherty:2003}
Dougherty S.~M., Pittard J.~M., Kasian L., Coker R.~F., Williams P.~M., Lloyd H.~M., 2003, A\&A, 409, 217
\bibitem[\protect\citeauthoryear{Dougherty \& Williams}{2000}]{Dougherty:2000}
Dougherty S.~M., Williams P.~M., 2000, MNRAS, 319, 1005
\bibitem[\protect\citeauthoryear{Duncan \& White}{2003}]{Duncan:2003}
Duncan R.~A., White S.~M., 2003, MNRAS, 338, 425
\bibitem[\protect\citeauthoryear{Gull et al.}{2009}]{Gull:2009}
Gull T.~R., et al., 2009, MNRAS, 396, 1308
\bibitem[\protect\citeauthoryear{Hummer}{1988}]{Hummer:1988}
Hummer D.~G., 1988, ApJ, 327, 477
\bibitem[\protect\citeauthoryear{Kashi \& Soker}{2007}]{Kashi:2007}
Kashi A., Soker N., 2007, MNRAS, 378, 1609
\bibitem[\protect\citeauthoryear{Lamers \& Leitherer}{1993}]{Lamers:1993}
Lamers H.~J.~G.~L.~M., Leitherer C., 1993, ApJ, 412, 771
\bibitem[\protect\citeauthoryear{Lamontagne et al.}{1982}]{Lamontagne:1982}
Lamontagne R., Koenigsberger G., Seggewiss W., Moffat A.~F.~J., 1982, ApJ, 253, 230
\bibitem[\protect\citeauthoryear{Lang, Goss \& Rodr\'{i}guez}{Lang et al.}{2001}]{Lang:2001}
Lang C.~C., Goss W.~M., Rodr\'{i}guez L.~F., 2001, ApJ, 551, L143
\bibitem[\protect\citeauthoryear{Lang et al.}{2005}]{Lang:2005}
Lang C.~C., Johnson K.~E., Goss W.~M., Rodr\'{i}guez L.~F., 2005, AJ, 130, 2185
\bibitem[\protect\citeauthoryear{Leitherer, Chapman \& Koribalski}{Leitherer et al.}{1995}]{Leitherer:1995}
Leitherer C., Chapman J.~M., Koribalski B., 1995, ApJ, 450, 289
\bibitem[\protect\citeauthoryear{Leitherer, Chapman \& Koribalski}{Leitherer et al.}{1997}]{Leitherer:1997}
Leitherer C., Chapman J.~M., Koribalski B., 1997, ApJ, 481, 898
\bibitem[\protect\citeauthoryear{Leitherer \& Robert}{1991}]{Leitherer:1991}
Leitherer C., Robert C., 1991, ApJ, 377, 629 
\bibitem[\protect\citeauthoryear{Linder et al.}{2006}]{Linder:2006}
Linder N., Rauw G., Pollock A.~M.~T., Stevens I.~R., 2006, MNRAS, 370, 1623
\bibitem[\protect\citeauthoryear{Linder et al.}{2007}]{Linder:2007}
Linder N., Rauw G., Sana H., De Becker M., Gosset E., 2007, A\&A, 474, 193
\bibitem[\protect\citeauthoryear{Linder et al.}{2008}]{Linder:2008}
Linder N., Rauw G., Martins F., Sana H., De Becker M., Gosset E., 2008, A\&A, 489, 713
\bibitem[\protect\citeauthoryear{Moffat \& Shara}{1986}]{Moffat:1986}
Moffat A.~F.~J., Shara M.~M., 1986, AJ, 92, 952
\bibitem[\protect\citeauthoryear{Montes et al.}{2009}]{Montes:2009}
Montes G., P\'{e}rez-Torres M.~A., Alberdi A., Gonz\'{a}lez R.~F., 2009, astro-ph 0810.5026
\bibitem[\protect\citeauthoryear{Morrell et al.}{2001}]{Morrell:2001}
Morrell N.~I., et al., 2001, MNRAS, 326, 85
\bibitem[\protect\citeauthoryear{Naz\'{e} et al.}{2005}]{Naze:2005}
Naz\'{e} Y., Antokhin I.~I., Sana H., Gosset E., Rauw G., 2005, MNRAS, 359, 688
\bibitem[\protect\citeauthoryear{Niemela et al.}{1999}]{Niemela:1999}
Niemela V.~S., Gamen R., Morrell N.~I., Gim\'{e}nez Ben\'{\i}tez S., 1999, in K.~A.~van~der~Hucht, G.~Koenigsberger, P.~R.~J.~Eenens, eds, Proc. IAU Symp. No. 193, ``Wolf-Rayet Phenomena in Massive Stars and Starburst Galaxies'', p. 26 
\bibitem[\protect\citeauthoryear{Nugis, Crowther \& Willis}{Nugis et al.}{1998}]{Nugis:1998}
Nugis T., Crowther P., Willis A., 1998, A\&A, 333, 956
\bibitem[\protect\citeauthoryear{Panagia \& Felli}{1975}]{Panagia:1975}
Panagia N., Felli M., 1975, A\&A, 39, 1
\bibitem[\protect\citeauthoryear{Parkin et al.}{2009}]{Parkin:2009}
Parkin E.~R., Pittard J.~M., Corcoran M.~F., Hamaguchi K., Stevens I.~R., 2009, MNRAS, 394, 1758
\bibitem[\protect\citeauthoryear{Pauldrach, Puls \& Kudritzki}{Pauldrach et al.}{1986}]{Pauldrach:1986}
Pauldrach A.~W.~A., Puls J., Kudritzki R.~P., 1986, A\&A, 154, 86
\bibitem[\protect\citeauthoryear{Persi et al.}{1988}]{Persi:1988}
Persi P., et al., 1988, ASSL, 142, 227
\bibitem[\protect\citeauthoryear{Pittard}{2007}]{Pittard:2007}
Pittard J.~M., 2007, ApJ, 660, L141
\bibitem[\protect\citeauthoryear{Pittard}{2009a}]{Pittard:2009a}
Pittard J.~M., 2009a, MNRAS, 396, 1743 (Paper~I)
\bibitem[\protect\citeauthoryear{Pittard}{2009b}]{Pittard:2009b}
Pittard J.~M., 2009b, MNRAS, in preparation (Paper~III)
\bibitem[\protect\citeauthoryear{Pittard et al.}{2006}]{Pittard:2006}
Pittard J.~M., Dougherty S.~M., Coker R.~F., O'Connor E., Bolingbroke N.~J., 2006, A\&A, 446, 1001
\bibitem[\protect\citeauthoryear{Pittard \& Dougherty}{2006}]{Pittard:2006b}
Pittard J.~M., Dougherty S.~M., 2006, MNRAS, 372, 801
\bibitem[\protect\citeauthoryear{Pittard \& Stevens}{1997}]{Pittard:1997}
Pittard J.~M., Stevens I.~R., 1997, MNRAS, 292, 298
\bibitem[\protect\citeauthoryear{Puls, Vink \& Najarro}{Puls et al.}{2008}]{Puls:2008}
Puls J., Vink J.~S., Najarro F., 2008, Astron. Astrophys. Rev., 16, 209 
\bibitem[\protect\citeauthoryear{Rauw et al.}{2002}]{Rauw:2002}
Rauw G., Vreux J.-M., Stevens I.~R., Gosset E., Sana H., Jamar C., Mason K.~O., 2002, A\&A, 388, 552
\bibitem[\protect\citeauthoryear{Runacres \& Blomme}{1996}]{Runacres:1996}
Runacres M.~C., Blomme R., 1996, A\&A, 309, 544
\bibitem[\protect\citeauthoryear{Rybicki \& Lightman}{1979}]{Rybicki:1979}
Rybicki G.~B., Lightman A.~P., 1979, Radiative processes in astrophysics (New York, Wiley-Interscience, 1979.~393 p.)
\bibitem[\protect\citeauthoryear{Sana et al.}{2004}]{Sana:2004}
Sana H., Stevens I.~R., Gosset E., Rauw G., Vreux J.-M., 2004, MNRAS, 350, 809
\bibitem[\protect\citeauthoryear{Schmid-Burgk}{1982}]{Schmid-Burgk:1982}
Schmid-Burgk J., 1982, A\&A, 108, 169
\bibitem[\protect\citeauthoryear{Schnerr et al.}{2007}]{Schnerr:2007}
Schnerr R.~S., Rygl K.~L.~J., van der Horst A.~J., Oosterloo T.~A., Miller-Jones J.~C.~A., Henrichs H.~F., Spoelstra T.~A.~Th., Foley A.~R., 2007, A\&A, 470, 1105
\bibitem[\protect\citeauthoryear{Scuderi et al.}{1998}]{Scuderi:1998}
Scuderi S., Panagia N., Stanghellini C., Trigilio C., Umana G., 1998, A\&A, 332, 251
\bibitem[\protect\citeauthoryear{Setia Gunawan et al.}{2003}]{SetiaGunawan:2003}
Setia Gunawan D.~Y.~A., Chapman J.~M., Stevens I.~R., Rauw G., Leitherer C., 2003, in K.~A.~van der Hucht, A.~Herrero, C.~Esteban, eds, Proc. IAU~Symp.~No.~212,``A Massive Star Odyssey, from Main Sequence to Supernova'', p.\,230 
\bibitem[\protect\citeauthoryear{Stevens}{1995}]{Stevens:1995}
Stevens I.~R., 1995, MNRAS, 277, 163
\bibitem[\protect\citeauthoryear{van der Hucht}{2001}]{vanderHucht:2001}
van der Hucht, K.~A., 2001, New Astronomy Review, 45, 135
\bibitem[\protect\citeauthoryear{Van Loo et al.}{2008}]{VanLoo:2008}
Van Loo S., Blomme R., Dougherty S.~M., Runacres M.~C., 2008, A\&A, 483, 585
\bibitem[\protect\citeauthoryear{Wang, Dong \& Lang}{Wang et al.}{2006}]{Wang:2006}
Wang Q.~D., Dong H., Lang C., 2006, MNRAS, 371, 38 
\bibitem[\protect\citeauthoryear{Williams et al.}{1990}]{Williams:1990}
Williams P.~M., van der Hucht K.~A., Sandell G., Th\'{e} P.~S., 1990, MNRAS, 244, 101
\bibitem[\protect\citeauthoryear{Wright \& Barlow}{1975}]{Wright:1975}
Wright A.~E., Barlow M.~J., 1975, MNRAS, 170, 41
\end{thebibliography}
\end{document}